\documentclass{vldb}

\usepackage{amsmath}
\usepackage{algorithm}
\usepackage[noend]{algorithmic}
\usepackage{cite}
\usepackage{graphicx}
\usepackage{multirow}
\usepackage{amssymb}
\usepackage{xcolor}
\usepackage{url}
\usepackage{xspace}
\usepackage{subfig}
\usepackage{relsize}
\usepackage{afterpage}
\usepackage{colortbl}
\usepackage{balance}
\usepackage{tabu}
\usepackage{enumitem}
\usepackage{times}
\usepackage{xspace}
\usepackage{ifthen}

\newif\ifarxiv
\arxivtrue

\newlength{\oldtextfloatsep}
\newlength{\oldfloatsep}

\newtheorem{defn}{Definition}

\newtheorem{ex}{Example}

\newcommand{\moveups}{\vspace*{-1mm}}
\newcommand{\moveup}{\vspace*{-2mm}}

\newcommand{\tabcaption}[1]{\vspace*{-3mm}\caption{#1}\vspace*{-4mm}}
\newcommand{\figcaption}[1]{\vspace*{-3mm}\caption{#1}\vspace*{-5mm}}

\newcommand{\imagewidth}{0.09\columnwidth}
\newcommand{\filterwidth}{0.90\linewidth}
\newcommand{\trianglewidth}{0.75\linewidth}

\newcommand{\name}{HD-Index\xspace}
\newcommand{\tree}{RDB-tree\xspace}
\newcommand{\trees}{RDB-trees\xspace}

\newcommand{\pivot}{SSS\xspace}

\newcommand*\samethanks[1][\value{footnote}]{\footnotemark[#1]}

\newcommand{\algocomment}{\hfill{}\ensuremath{\triangleright}}

\title{HD-Index: Pushing the Scalability-Accuracy Boundary for Approximate kNN Search in High-Dimensional Spaces}

\numberofauthors{1}

\author{
\alignauthor
Akhil Arora$^1$\thanks{The author was employed at American Express Big Data Labs during the submission of this work.} ~ Sakshi Sinha$^2$\thanks{The work was done while the authors were students at Indian Institute of Technology Kanpur, and they contributed equally.} ~ Piyush Kumar$^3$\samethanks ~ Arnab Bhattacharya$^4$\\
	$^1$\affaddr{EPFL, Lausanne, Switzerland}  ~  $^2$\affaddr{Fresh Gravity Inc., Pune, India}  ~  $^3$\affaddr{Visa Inc., Bangalore, India}\\
	$^4$\affaddr{Dept. of Computer Science and Engineering, Indian Institute of Technology, Kanpur, India}\\
	\email{arora.akhilcs@gmail.com ~ sakshi.sinha@freshgravity.com}
	\email{pkumarda@visa.com ~ arnabb@cse.iitk.ac.in}
}

\begin{document}

\setlength{\abovedisplayskip}{1pt}
\setlength{\belowdisplayskip}{1pt}

\maketitle

\begin{abstract}
	Nearest neighbor searching of large databases in high-dimensional spaces is
	inherently difficult due to the curse of dimensionality.  A flavor of
	approximation is, therefore, necessary to practically solve the problem of
	nearest neighbor search.  In this paper, we propose a novel yet simple
	indexing scheme, \emph{\name}, to solve the problem of \emph{approximate
	$k$-nearest neighbor queries} in massive high-dimensional databases.  \name
	consists of a set of novel hierarchical structures called \emph{\trees}
	built on Hilbert keys of database objects.  The leaves of the \trees store
	distances of database objects to reference objects, thereby allowing
	efficient pruning using distance filters.  In addition to triangular
	inequality, we also use Ptolemaic inequality to produce better lower
	bounds.  Experiments on massive (up to billion scale) high-dimensional (up
	to 1000+) datasets show that \name is \emph{effective}, \emph{efficient},
	and \emph{scalable}.
\end{abstract}


\section{Motivation}
\label{sec:intro}

\emph{Nearest neighbor (NN) search} is a fundamental problem with applications
in information retrieval, data mining, multimedia and scientific databases,
social networks, to name a few. Given a query point $q$, the NN problem is to
identify an object $o$ from a dataset of objects $\mathcal{D}$ such that $q$ is
closest to $o$ than any other object.  A useful extension is to identify
$k$-nearest neighbours (kNN), i.e., an ordered set of top-$k$ closest objects
$\{o_1, \ldots, o_k\}$ to $q$.

While there exist a host of scalable methods to efficiently answer NN queries
in low- and medium-dimensional spaces \cite{indexingbook, indexingbook1, rtree,
rstartree, mtree, xtree, jensen, iqtree, iminmax, pyramid}, they do not scale
gracefully with the increase in dimensionality of the dataset. Under the strict
requirement that the answer set returned should be \emph{completely accurate},
the problem of high dimensionality in similarity searches, commonly dreaded as
the ``\emph{curse of dimensionality}'' \cite{indexingbook}, renders these
methods useless in comparison to a simple linear scan of the entire database
\cite{vafile}.

In many practical applications, however, such a strict correctness requirement
can be sacrificed to mitigate the dimensionality curse and achieve practical
running times \cite{valle,c2lsh,srs}.  A web-scale search for similar images,
e.g., Google Images (\url{www.google.com/imghp}), Tineye
(\url{www.tineye.com}), RevIMG (\url{www.revimg.com}), perhaps best exemplifies
such an application.  Moreover, a single query is often just an intermediate
step and aggregations of results over many such queries is required for the
overall retrieval process.  For example, in image searching, generally multiple
image keypoints are searched.  The strict accuracy of a single image keypoint
search is, therefore, not so critical since the overall aggregation can
generally tolerate small errors.

With this relaxation, the NN and kNN problems are transformed into \emph{ANN
(approximate nearest neighbors)} and \emph{kANN} problems respectively, where
an error, controlled by a parameter $c$ that defines the \emph{approximation
ratio} (defined formally in Def.~\ref{def:ratio}), is allowed. The retrieved
object(s) $o' = \{o'_1,\ldots,o'_k\}$ is such that their distance from $q$ is
at most $c$ times the distance of the actual nearest object(s) $o =
\{o_1,\ldots,o_k\}$ respectively.

The locality sensitive hashing (LSH) family \cite{lsh} is one of the most
popular methods capable of efficiently answering kANN que\-ries in sub-linear
time.  However, these methods are not \emph{scalable} owing to the requirement
of large space for constructing the index that is super-linear in the size of
the dataset \cite{e2lsh}.  While there have been many recent improvements
\cite{lsb,c2lsh,sklsh} in the original LSH method with the aim of reducing the
indexing space, asymptotically they still require super-linear space and, thus,
do not scale to very large datasets.  A recent proposition, SRS \cite{srs},
addresses this problem by introducing a linear sized index while retaining high
efficiency and is the state-of-the-art in this class of techniques.

A notable practical limitation of these techniques with theoretical guarantees
is that the error is measured using the approximation ratio $c$.  Given that
the task is to identify an \emph{ordered} set of $k$ nearest objects to a query
point $q$, the \emph{rank} at which an object is retrieved is significant.
Since $c$ is characterized by the properties of the \emph{ratio of distances}
between points in a vector space, a low (close to $1$) approximation ratio does
not always guarantee similar relative ordering of the retrieved objects. This
problem worsens with increase in dimensionality $\nu$, as the distances tend to
lose their meaning, and become more and more similar to each other
\cite{indexingbook}.  Mathematically, $\lim\limits_{\nu \to
\infty}\frac{d_{max}}{d_{min}}=1$ where $d_{max}$ and $d_{min}$ are the maximum
and minimum pairwise distances between a finite set of $n$ points.  Hence, it
is not surprising that $c$ rapidly approaches $1$ for extremely
high-dimensional spaces (dimensionality $\geq 100$).  Thus, while $c$ might be
a good metric in low- and medium-dimensional spaces ($\lesssim 20$), it loses
its effectiveness with increasing number of dimensions.

Thus, for assessing the quality of kANN search, especially in high-dimensional
spaces, a metric that gives importance to the \emph{retrieval rank} is
warranted.  \emph{Mean Average Precision (MAP)}, which is widely used and
accepted by the information retrieval community \cite{irbook} for evaluation of
\emph{ranked retrieval} tasks, fits perfectly.  MAP@k (defined formally in
Def.~\ref{def:map}) takes into account the relative ordering of the retrieved
objects by imparting higher weights to objects returned at ranks closer to
their original rank. Additionally, it has been shown to possess good
discrimination and stability \cite{irbook}.  Thus, MAP should be the target for
evaluating the quality.

\begin{figure}[t]
\moveup
\moveup
\centering
	\subfloat[Dataset: SIFT10K]
	{
		\includegraphics[width=0.42\linewidth]{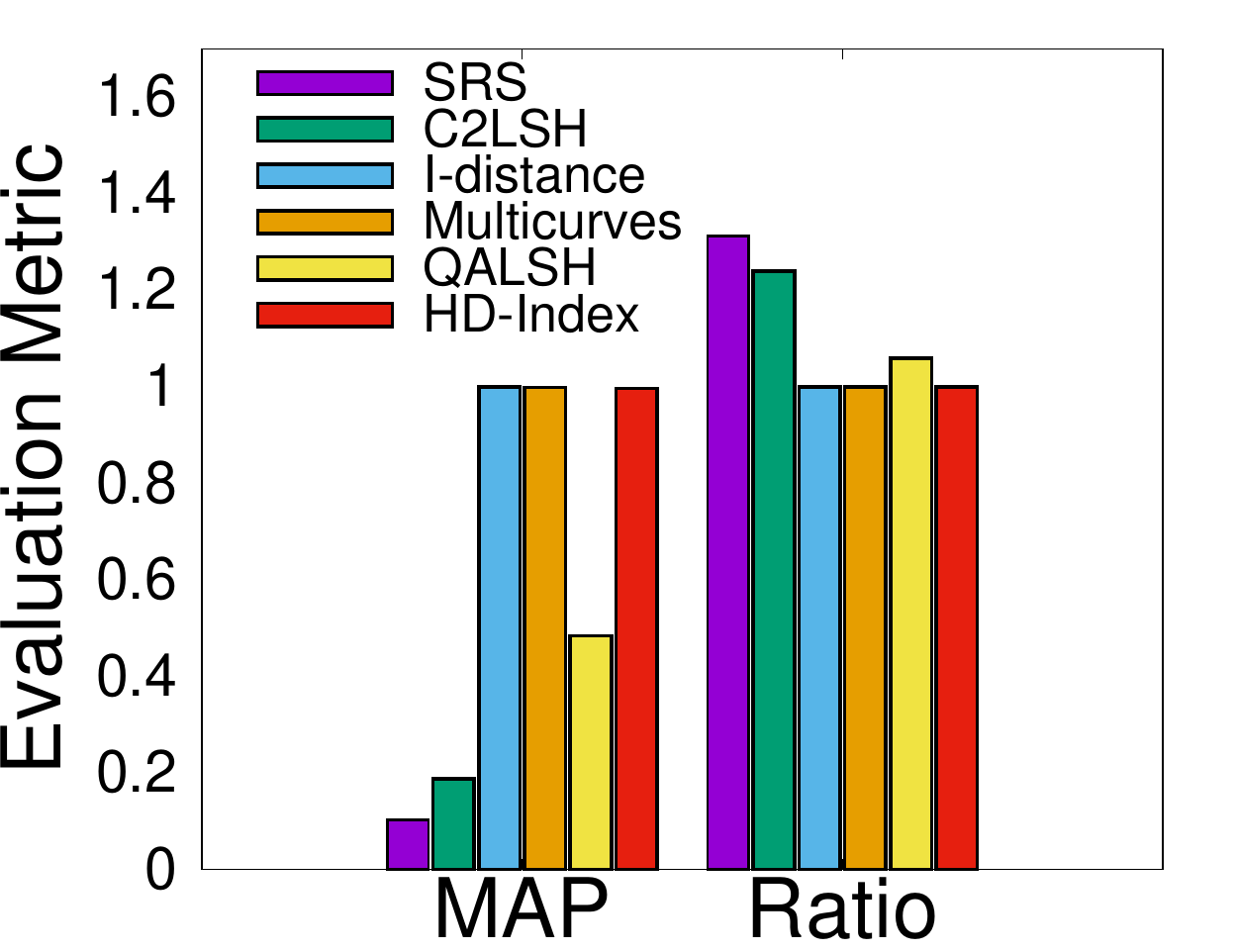}
		\label{fig:motivate_map_vs_ratio_sift10k}
	}
	\subfloat[Dataset: Audio]
	{
		\includegraphics[width=0.42\linewidth]{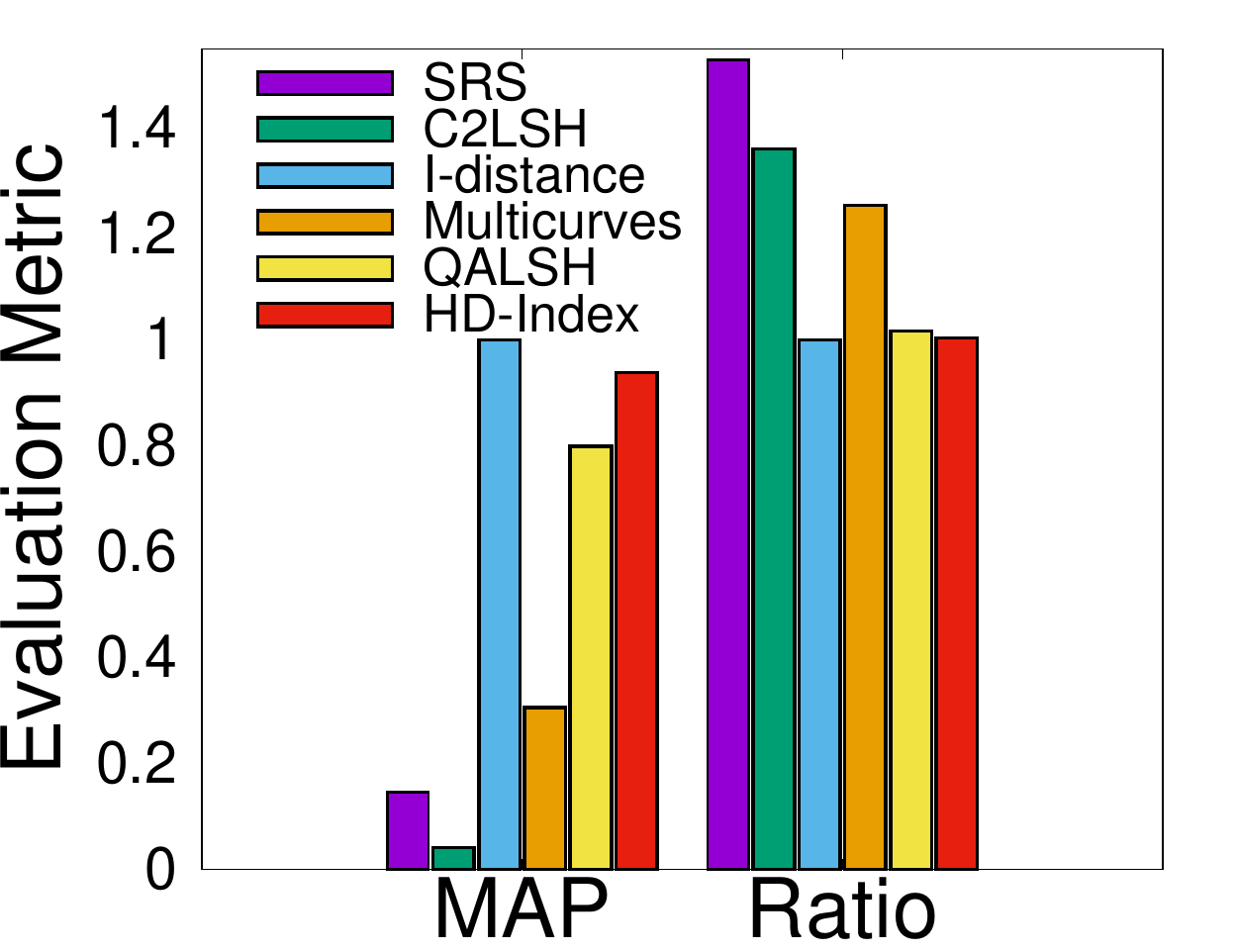}
		\label{fig:motivate_map_vs_ratio_audio}
	}
	\figcaption{\textbf{Comparing MAP with approximation ratio ($k=10$).}}
	\label{fig:motivate_map_vs_ratio}
	\moveups
\end{figure}

To highlight the empirical difference between the two metrics, we plot both
MAP@10 and approximation ratio (for $k=10$) for different techniques and real
datasets (details in Sec.~\ref{sec:exp}) in
Fig.~\ref{fig:motivate_map_vs_ratio} (these and other figures look better in
color).  It is clear that good approximation ratios can lead to poor MAP values
(e.g., in Fig.~\ref{fig:motivate_map_vs_ratio_sift10k}, C2LSH and SRS have
ratios $1.24$ and $1.31$ respectively with MAP values of $0.18$ and $0.10$
respectively).  Moreover, the relative orders may be flipped (such as in
Fig.~\ref{fig:motivate_map_vs_ratio_audio} where C2LSH offers better ratio but
poorer MAP than SRS).  On the whole, this shows that MAP cannot be improved by
just having better approximation ratios.

To the best of our knowledge, there is no method that can guarantee a bound on
the MAP.  Thus, the existing techniques that provide bounds on approximation
ratio cannot be deemed to be advantageous in practice for the task of kANN in
high-dimensional spaces.

We next analyze another class of techniques that rely on the use of
space-filling curves (such as Hilbert or Z-order) \cite{space} for indexing
high-dimensional spaces \cite{valle,mult1,ruiz,liao,lawder}. Although devoid of
theoretical guarantees, these methods portray good retrieval efficiency and
quality. The main idea is to construct single/multiple single-dimensional
ordering(s) of the high-dimensional points, wh\-ich are then indexed using a
standard structure such as B+-tree for efficient querying. These structures are
based on the guarantee that points closer in the transformed space are
necessarily close in the original space as well.  The vice versa, however, is
not true and serves as the source of inaccuracy.

The leaf node of a B+-tree can either store the entire object which improves
efficiency while hampering scalability, or store a pointer to the database
object which improves scalability at the cost of efficiency.  Hence, these
methods do \emph{not} scale gracefully to high-dimensional datasets due to
their large space requirements as only a few objects can fit in a node.

In light of the above discussion, the need is to design a scheme that
congregates the \emph{best of both the worlds} (one-dimensional representation
of space-filling curves with hierarchical pruning of B+-trees with leaves
modified as explained later) to ensure \emph{both} efficiency and scalability.

To this end, in this paper, we propose a hybrid index structure, \textsc{\name
(High-Dimensional Index)}.  We focus on solving the kANN query for large
\emph{disk-resident} datasets residing in \emph{extremely high-dimensional}
spaces.

We first divide the $\nu$-dimensional space into disjoint partitions. For each
lower-dimensional partitioned space, we construct a hierarchical tree structure
using the Hilbert space-filling curve orderings.  We then propose a novel
design called \textsc{\trees (Reference Distance B+-trees)}, where instead of
object descriptors or pointers, the distances to a fixed set of reference
objects are stored in the leaves.  While querying, in addition to distance
approximation using triangular inequality, we incorporate the use of a more
involved lower bounding filter known as the \emph{Ptolemaic inequality}
\cite{ptolemy}.  Instead of using a single reference object, it approximates
the distance using a pair of reference objects.  Although costlier to compute,
it yields tighter lower bounds.  The novel design of \tree leaves allows
application of these filtering techniques using the reference objects easily.

\noindent
\textbf{Contributions.} In sum, our contributions in this paper are:

\begin{itemize}[nosep,leftmargin=*]

	\item We propose a novel, scalable, practical, and yet simple indexing
		scheme for large extremely high-dimensional Euclidean spaces, called
		\emph{\name} (Sec.~\ref{sec:construction}).

	\item We design a novel structure called \emph{\tree} with leaves designed
		to store distances to reference objects for catering to extremely
		high-dimensional spaces (Sec.~\ref{sec:bptree}).

	\item We devise an efficient and effective querying algorithm
		(Sec.~\ref{sec:query}) that leverages lower-bounding filters to prune
		the search space.

	\item We advocate the use of mean average precision or MAP
		(Def.~\ref{def:map}) as a better metric to benchmark techniques in
		terms of quality of the retrieved top-$k$ objects, instead of the
		approximation ratio, especially in high-dimensional spaces.  We
		motivate the need of MAP in real-world applications through an actual
		information retrieval application, that of image search
		(Sec.~\ref{subsec:image_search}).

	\item Through an in-depth empirical analysis (Sec.~\ref{sec:exp}), we show
		that \name scales gracefully to massive high-dimensional datasets on
		commodity hardware, and provides speed-ups in running time of up to
		orders of magnitude, while providing significantly better quality
		results than the state-of-the-art techniques.

\end{itemize}

\section{Background}

\subsection{Preliminaries}
\label{sec:problem}

We consider a dataset $\mathcal{D}$ with $n=|\mathcal{D}|$ $\nu$-dimensional
points spanning a vector space.  Given a query $q = (q^1, \dots, q^{\nu})$, and
the Euclidean distance function $d$ ($L_2$ norm), the \emph{k-nearest neighbor
(kNN)} query is to retrieve the set $T \subset \mathcal{D}, |T| = k$ of $k$
nearest neighbors of $q$ from $\mathcal{D}$ such that $\forall o_i \in T,
\forall o_j \notin T, d(q,o_i) < d(q,o_j)$.  The \emph{approximate k-nearest
neighbor (kANN)} problem relaxes the strict requirement and aims to retrieve a
set $A$ of size $k$ that contains \emph{close} neighbors of $q$.

We next describe two \emph{quality metrics} to assess the quality of the
retrieved kANN answer set.  Suppose the \emph{true} $k$ nearest neighbors of
$q$ are $T_k = \{o_1, \dots, o_k\}$ \emph{in order}, i.e., $o_1$ is closest to
$q$, then $o_2$, and so on.  Assume that the \emph{approximate} $k$ nearest
neighbors returned are $A_k = \{o'_1, \dots, o'_k\}$, again, \emph{in order}.

\begin{defn}[Approximation Ratio]
\label{def:ratio}
The \emph{approximation ratio} $c \geq 1$ is simply the average of the ratio of
the distances:
\begin{align}
	\label{eq:ratio}
	c = \frac{1}{k} . \sum_{i = 1}^k \frac{d(q,o'_i)}{d(q,o_i)}
\end{align}
\end{defn}

\begin{defn}[Average Precision]
\label{def:ap}
Consider the $i^\text{th}$ returned item $o'_i$.  If $o'_i$ is \emph{not}
relevant, i.e., it does not figure in the true set at all in any rank, its
\emph{precision} is $0$.  Otherwise, the recall at position $i$ is captured.  If
out of $i$ items $o'_1, \dots, o'_i$, $j$ of them figure in the true set, then
the recall is $(j / i)$.
The average of these values for ranks $1$ to $k$ is the \emph{average
precision} at $k$, denoted by $AP@k$:
\begin{align}
	\label{eq:ap}
	AP@k = \frac{1}{k} . \sum_{i = 1}^{k} \left[ I_{o'_i \in T_k} \cdot (j / i) \right]
\end{align}
$I$ is the indicator function: $I = 1$ if $o'_i \in T_k$; and $0$ otherwise.
\end{defn}

The AP function, thus, prefers correct answers at \emph{higher} ranks.

\begin{defn}[Mean Average Precision]
\label{def:map}
The \emph{mean average precision}, denoted by
$MAP@k$, is the mean over $AP@k$ values for $Q$ queries:
\begin{align}
	\label{eq:map}
	MAP@k = \frac{1}{Q} . \sum_{\forall i = 1}^{Q} AP@k(q_i)
\end{align}
\end{defn}

AP and, thus, MAP lie in the range $[0,1]$.

\begin{ex}
	Suppose the true ranked order for a query is $\{o_1,$ $o_2, o_3\}$.
	Consider $A_1 = \{o_4, o_3, o_2\}$.  Since $o_4$ is not relevant, its
	precision is $0$.  Next, for rank $2$, only $o_3$ is relevant.  Thus, the
	precision is $1/2$.  Similarly, the precision at $o_2$ is $2/3$.  The AP,
	therefore, is $(0 + 1/2 + 2/3) / 3 = 0.39$.
	
	The AP for the same set of returned elements but in a different ranked
	order, $A_2 = \{o_3, o_2, o_4\}$, is $(1 + 1 + 0) / 3 = 0.67$.
	
	Hence, the MAP for $A_1$ and $A_2$ is $(0.39 + 0.67) / 2 = 0.53$.
\end{ex}

\subsection{Related Work}
\label{sec:work}

The problem of nearest neighbor searches, more popularly known as the \emph{kNN
query}, has been investigated by the database community for decades.  The basic
technique of scanning the entire data does not scale for large disk-resident
databases and, consequently, a host of indexing techniques (e.g., R-tree
\cite{rtree} and M-tree families \cite{mtree}) have been designed to
efficiently perform the kNN query \cite{indexingbook,indexingbook1}.

\subsubsection{Hierarchical Structures}

The R-tree family of structures builds a hierarchy of data objects with the
leaf pages containing the actual data points.  In each level, the MBR (minimum
bounding rectangle) corresponding to the internal node covers all its children
nodes.  The branching factor indicates how many children nodes can be packed
within one disk page.  The MBRs can be hyper-rectangular (R-tree \cite{rtree},
R*-tree \cite{rstartree}) or hyper-spherical (SS-tree \cite{sstree}) or mixed
(SR-tree \cite{srtree}).

Unfortunately enough, for datasets having a large dimensionality, the infamous
``curse of dimensionality'' \cite{indexingbook} renders almost all these
hierarchical indexing techniques inefficient and linear scan becomes the only
feasible solution \cite{vafile}.  Structures such as the VA-file \cite{vafile}
have, therefore, been designed to compress the data and perform the unavoidable
linear scan faster.  Techniques to particularly handle the problem of high
dimensionality have also been proposed (e.g., the pyramid technique
\cite{pyramid}, iMinMax \cite{iminmax}, iDistance \cite{idistance}, etc.).
These structures map a high-dimensional point to a single number and then index
that using a B+-tree \cite{btree}.  A specialized hierarchical structure X-tree
\cite{xtree}, on the other hand, adjusts itself with growing dimensionality.
In the extreme case, it resembles a simple flat file and the searching degrades
to that of a linear scan.  To beat the curse of dimensionality, compression
through quantization has also been employed.  A-tree \cite{atree} uses a fixed
number of bits per dimension to quantize the data, thereby enabling packing of
more children nodes in an internal node.  IQ-tree \cite{iqtree} uses different
number of quantization bits depending on the number of points in a data MBR and
is restricted to only two levels of hierarchy before it indexes the leaf pages
directly.  Both of them in the end, however, produce the exact correct answer.
A survey of such high-dimensional techniques, especially in the context of
multimedia databases, is available at \cite{mmsurvey}.

Hierarchical index structures of the M-tree family \cite{mtree} exploit the
property of triangular inequality of metric distances.  These structures do not
need the vector space representation of points and work directly with the
distance between two points.  Nevertheless, at high dimensions, the Euclidean
distance itself tends to lose its meaning \cite{indexingbook} and these
structures fail as well.

\begin{figure*}[t]
\moveup
\centering
\includegraphics[width=0.99\textwidth]{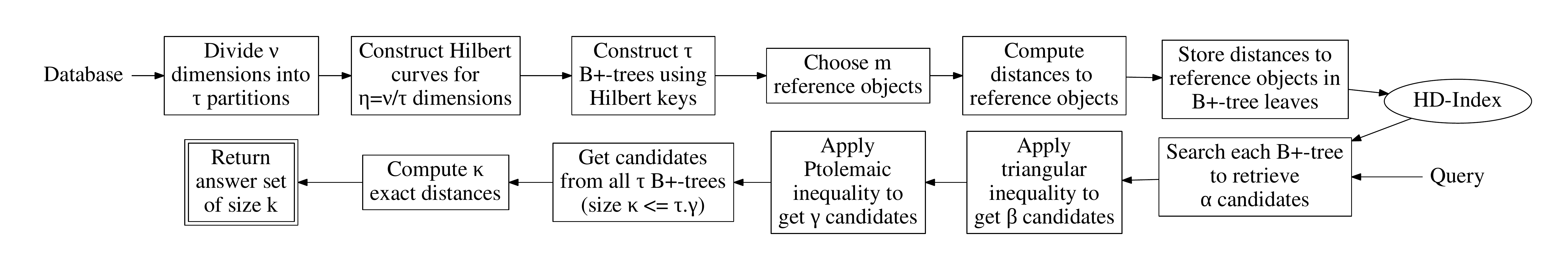}
\moveup
\moveup
\figcaption{\textbf{Overview of our approach.}}
\label{fig:overview}
\vspace*{2mm}
\end{figure*}

\subsubsection{Reference Object-Based Structures}

Instead of hierarchical tree-based structures, another important line of
research has been the use of reference objects or landmarks or
pivots \cite{sss_dyn,pivot2}.  Certain database points are marked as reference
objects and distance of all other points from one or more of these reference
objects are pre-computed and stored.  Once the query arrives, its distance to
the reference objects are computed.  Using the triangular inequality (based on
the assumption that the underlying distance measure is a metric), certain points
can be pruned off without actually computing their distance to the query.  AESA
\cite{aesaold,aesa} and LAESA \cite{laesa} use the above generic scheme.

Marin et al. \cite{marin} employed clustering and the cluster representatives
were used as reference objects.  Amato et al. \cite{amato} proposed inverted
distance mappings from a set of representative points based on the premise that
in a metric space, if two objects are close to each other, then the relative
ordering of the data points from these two objects would also be similar.  Chen
et al.  \cite{jensen} used representative objects to obtain a vector of
distances.  The one-dimensional ordering of these vectors were then indexed
using a B+-tree.

The choice of representative points has been governed by various factors such
as maximizing distance to the already chosen set \cite{brisaboa, sss_dyn},
maximizing the minimum distance among themselves \cite{dasgupta}, location on
the boundary of the dataset \cite{jr}, whether they are outliers
\cite{bustos2}, using a nearest neighbor distance loss function \cite{hennig},
minimizing the correlation among themselves \cite{leuken}, or through a sample
of queries \cite{venkat}.  Chen et al. \cite{jensen} included a nice survey of
methods to choose representative objects.

\subsubsection{Space-Filling Curves}

Space-filling curves \cite{space,indexingbook} provide an intuitive way of
approximating the underlying data space by partitioning it into grid cells.  A
one-dimensional ordering is then imposed on the grid cells such that the
resultant curve covers each and every grid cell \emph{exactly once}.  The
sequence of an \emph{occupied} grid cell according to the curve is its
one-dimensional index.  The multi-dimensional points are indexed by the
one-dimensional index of the grid cell it occupies.  The basic property of
these curves ensure that when two points are close in the one-dimensional
ordering, they are necessarily close in the original high-dimensional space.
The reverse may not be true, though, due to the ``boundary'' effects where the
curve visits a neighboring cell after a long time.  Among the different types
of space-filling curves, the Hilbert curve is considered to be the most
appropriate for indexing purposes \cite{packinghilbert}.

To diminish the boundary effects, the use of multiple space-filling curves have
been incorporated.  Deterministic translation vectors were used by \cite{liao}
whereas \cite{shep} used random transformation or scaling of data points.
Valle et al. \cite{valle} proposed an alternative way where different subsets
of the dimensions are responsible for each curve.  This helps to reduce the
dimensionality of the space handled by each curve.  Different B+-trees, each
corresponding to a particular Hilbert curve, were built.  When a query point
arrives, the nearest neighbor searches in the different B+-trees aggregate the
points.  Finally, the approximate nearest neighbors are chosen from this
aggregated set.

\subsubsection{Locality-Sensitive Hashing}

The locality-sensitive hashing (LSH) family \cite{lsh} of randomized algorithms
adopt a completely different approach.  The database points are
probabilistically hashed to multiple hash tables.  The hash functions follow
two crucial properties: (1) if two points are close enough, they are very
likely to fall in the same hash bucket, and (2) if two points are far-off, they
are quite unlikely to fall in the same hash bucket.  At run time, the query
point is also hashed, and the points falling in the same hash buckets
corresponding to each hash table form the candidates.  The final approximate
nearest neighbor set is chosen from these candidates.

The basic LSH scheme \cite{lsh} was extended for use in Euclidean spaces by
E2LSH \cite{e2lsh}.  C2LSH \cite{c2lsh} was specifically designed on top of
E2LSH to work with high-dimensional spaces.  LSB-forest \cite{lsb} builds
multiple trees to adjust to the NN distance.  Sun et al. devised SRS \cite{srs}
to have a small index footprint so that the entire index structure can fit in
the main memory.

Recently, a query-aware data-dependent LSH scheme, QALSH, has been proposed
\cite{qalsh}.  The bucket boundaries are not decided apriori, but only after
the query arrives in relation to the position of the query.  As a result,
accuracy improves.

Although the LSH family \cite{lsh} provides nice theoretical guarantees in
terms of running time, space and approximation ratio, however, importantly
enough, they provide no guarantees on MAP.

\begin{table}[t]
\centering
\scalebox{0.85}{
\begin{tabular}{cl}
\hline
{\bf Symbol} & {\bf Definition} \\
\hline
\hline
$\mathcal{D}$ & Database \\
$n = |\mathcal{D}|$ & Size of database \\
$o$ & Database object \\
$\nu$ & Dimensionality of each object \\
\hline
$\mathcal{P}$ & Partition of dimensions \\
$\tau = |\mathcal{P}|$ & Total number of partitions \\
\hline
$\mathcal{H}$ & Set of Hilbert curves corresponding to $\mathcal{P}$ \\
$\eta = \nu / \tau$ & Dimensionality of each Hilbert curve \\
$\omega$ & Order of Hilbert curve \\
\hline
$\mathcal{B}$ & \trees corresponding to $\mathcal{H}$ \\
$\Omega$ & Order of \tree leaf \\
$B$ & Disk page size \\
\hline
$\mathcal{R}$ & Set of reference objects \\
$m = |\mathcal{R}|$ & Number of reference objects \\
\hline
$q$ & Query object \\
\hline
$\alpha$ & Number of candidates after \tree search \\
$\beta$ & Number of candidates after triangular inequality \\
$\gamma$ & Number of candidates after Ptolemaic inequality \\
\hline
$\mathcal{C}$ & Final candidate set \\
$\kappa = |\mathcal{C}|$ & Size of final candidate set \\
\hline
$\mathcal{A}$ & Final answer set \\
$k = |\mathcal{A}|$ & Number of nearest neighbors returned \\
\hline
\end{tabular}
}
\tabcaption{\textbf{Symbols used in this paper.}}
\label{tab:symbols}
\end{table}

\subsubsection{In-memory Indexing Techniques}

Flann \cite{flann} uses an ensemble of randomized KD-forests \cite{randkd} and
priority K-means trees to perform fast approximate nearest neighbor search. Its
key novelty lies in the automatic selection of the best method in the ensemble,
and allowing identification of optimal parameters for a given dataset.
Recently, Houle et al. \cite{rct} proposed a probabilistic index structure
called the Rank Cover tree (RCT) that entirely avoids the use of numerical
pruning techniques like triangle inequality etc., and instead uses rank-based
pruning. This alleviates the curse of dimensionality to a large extent.

Proximity graph-based kANN methods \cite{graph1,graph2,graph4,graph5} have
gained popularity, out of which, the hierarchical level based navigable small
world method HNSW \cite{hnsw} has been particularly successful.

Quantization based methods aim at significantly reducing the size of the index
structure, thereby facilitating \emph{in-memory} querying of approximate
$k$-nearest neighbors. The product quantization (PQ) method \cite{pq} divides
the feature space into disjoint subspaces and quantize them independently.
Each high-dimensional data point is represented using just a short code
composed of the subspace quantization indices.  This facilitates efficient
distance estimation. The Cartesian K-means (CK-Means) \cite{ckmeans} and
optimized product quantization (OPQ) \cite{opq} methods were proposed later as
improvements and a generalization of PQ respectively.

Despite their attempt to reduce the index structure size, the techniques
discussed above find difficulty in scaling to massive high-dimensional databases
on commodity hardware, and instead require machines possessing a humongous
amount of main memory.

There have been many other indexing schemes built specifically for ANN queries
\cite{covertree, medrank, sash, navigating, spilltree, sklsh, mplsh}.

\subsubsection{Representative Methods}

For comparison, we choose the following representative methods: iDistance
\cite{idistance} (as an exact searching technique), Multicurves \cite{valle}
(hierarchical index using space-filling curves), SRS \cite{srs} (low memory
footprint), C2LSH \cite{c2lsh} (low running time), QALSH \cite{qalsh}
(data-dependent hashing with high quality), OPQ \cite{opq} (product
quantization), and HNSW \cite{hnsw} (graph-based method) since they claim to
outperform the other competing methods in terms of either running time, or
space overhead, or quality.

\pagebreak
\section{HD-Index Construction}
\label{sec:construction}

\begin{algorithm}[t]
	\caption{\name Construction}
	\label{algo:index}
	\textbf{Input:} Dataset $\mathcal{D}$ of size $n$ and dimensionality $\nu$ \\
	\textbf{Output:} Index \name
	\begin{algorithmic}[1]
	{\scriptsize
		\STATE $\mathcal{R} \gets$ set of reference objects \algocomment Sec.~\ref{sec:reference}
		\STATE $rdist \gets$ dist($\forall \mathcal{R}_i, \forall \mathcal{D}_j$)
		\STATE $\mathcal{P} \gets$ partition of dimensions \algocomment equal and contiguous
		\STATE $\tau = |\mathcal{P}|$; $\eta = \nu / \tau$
		\FOR {$i \gets 1$ to $\tau$}
			\STATE $\mathcal{P}_i \gets$ dimensions $i . \eta$ to $(i+1) . \eta - 1$
			\STATE $\mathcal{H}_i \gets$ Hilbert($\mathcal{P}_i, \omega$) \algocomment Sec.~\ref{sec:hilbert}
			\STATE Insert $\mathcal{D}_j$ with key $HK(\mathcal{D}_j)$ into \tree $\mathcal{B}_i$ \algocomment Sec.~\ref{sec:bptree}
			\FOR {$\forall \mathcal{D}_j$ in each leaf of $\mathcal{B}_i$}
			    \STATE Store $rdist(\mathcal{D}_j, \forall \mathcal{R}_i)$
			\ENDFOR
		\ENDFOR
		}
	\end{algorithmic}
\end{algorithm}

In this section, we describe in detail the construction of our index structure,
\emph{\name} (Algo.~\ref{algo:index}).  Fig.~\ref{fig:overview} shows the
overall scheme of our method.  We use the terminology listed in
Table~\ref{tab:symbols}.

\begin{figure*}[t]
	\centering
	\moveup
	\moveups
	\subfloat[Dim 1-2]
	{
		\includegraphics[width=0.30\columnwidth]{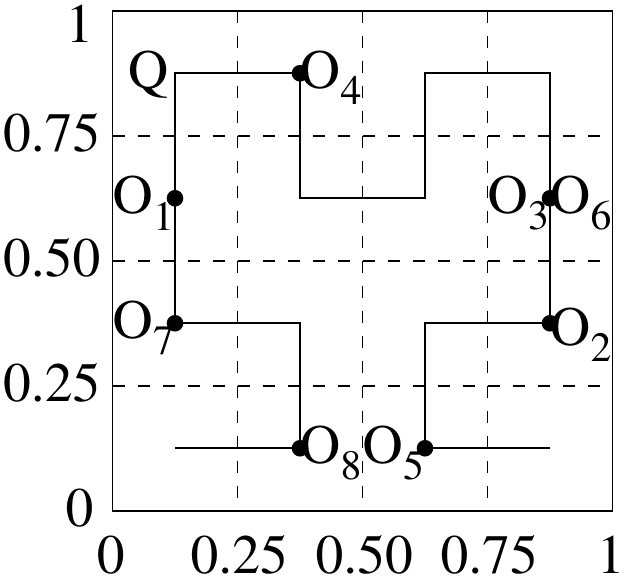}
		\label{fig:sf1}
	}
	\quad
	\subfloat[Dim 3-4]
	{
		\includegraphics[width=0.30\columnwidth]{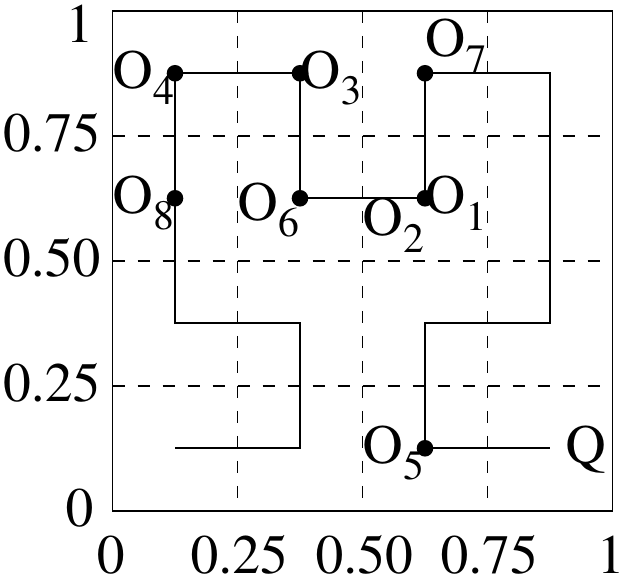}
		\label{fig:sf2}
	}
	\qquad
	\subfloat[Modified \trees.]
	{
		\includegraphics[width=1.25\columnwidth]{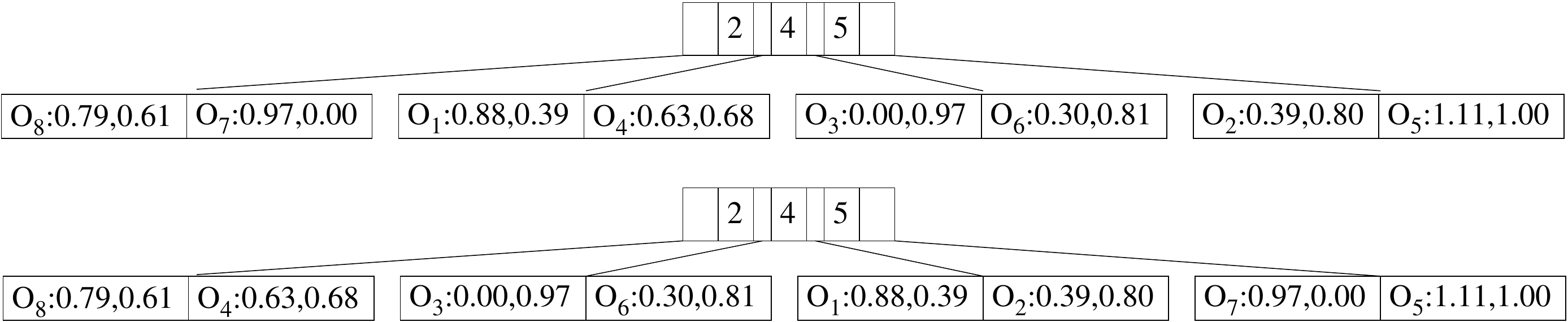}
		\label{fig:btrees}
	}
	\figcaption{\textbf{Hilbert keys and resulting \trees for the example in
	Table~\ref{tab:example}.}}
	\label{fig:example}
	\vspace*{2mm}
\end{figure*}

Since we are specifically targeting datasets in high-dimensional vector spaces,
we first divide the entire dimensionality into smaller partitions.  A Hilbert
curve is then passed through each partition.  The resultant one-dimensional
values are indexed using a novel structure called \emph{\tree}.  While the
upper levels of a \tree resembles that of a B+-tree, the leaves are modified.

We next explain each of the fundamental steps in detail.

\subsection{Space-Filling Curves}
\label{sec:hilbert}

The total number of dimensions $\nu$ is first divided into $\tau$ disjoint and
equal partitions.  Although the partitioning can be non-equal and/or
overlapping as well, we choose an equal and contiguous partitioning scheme due
to two reasons: (1) it is simpler, and (2) the other partitioning schemes do
not give any tangible benefit since we assume that the dimensions are
independent.  The claim is empirically verified in Sec.~\ref{sec:partitioning}.

For each partition, a Hilbert curve, catering to $\eta = \nu / \tau$
dimensions, is constructed using the Butz algorithm \cite{butz}.  The
\emph{order} of a Hilbert curve specifies the number of bits used to describe
every dimension.  Thus, if the order is $\omega$, each dimension is divided
into $2^{\omega}$ equal grid partitions.  The single-dimensional value of an
object in the Hilbert curve ordering is called its \emph{Hilbert key}.

As an example, consider the $4$-dimensional dataset of $8$ objects shown in
Table~\ref{tab:example}.  The dimensions are broken into two partitions, 1-2 and
3-4.  The Hilbert curve for each of these partitions is shown in
Fig.~\ref{fig:sf1} and Fig.~\ref{fig:sf2}.  The Hilbert keys for the objects are
obtained by considering the sorted order along the curve, and are shown in the
last two columns of Table~\ref{tab:example}.

Multiple Hilbert curves are beneficial in not only reducing the dimensionality
but also tackling the ``boundary'' effect.  It is unlikely that nearby objects
will have far-off Hilbert keys in \emph{all} the partitions.  For example, in
Fig.~\ref{fig:sf1}, $O_7$ and $O_1$ having adjacent Hilbert keys are nearby in
space, while objects $O_8$ and $O_4$, despite being close in space, have
far-off Hilbert keys.  However, in Fig.~\ref{fig:sf2}, $O_8$ and $O_4$ have
adjacent Hilbert keys.  Objects $O_7$ and $O_1$ continue to be close in both
the cases.  It is enough for an object that is a nearest neighbor to have a
close Hilbert key in only \emph{one} partition to be chosen as a candidate.

\begin{table}[t]
	\centering
	\scalebox{0.85}{
	\begin{tabular}{ccccccc}
		\hline
		\textbf{Object} & \bf Dim 1 & \bf Dim 2 & \bf Dim 3 & \bf Dim 4 & \bf HK 1 & \bf HK 2 \\
		\hline
		$O_1$ & 0.20 & 0.74 & 0.68 & 0.73 & 3 & 5 \\
		$O_2$ & 0.84 & 0.34 & 0.49 & 0.81 & 6 & 5 \\
		$O_3$ & 0.97 & 0.64 & 0.32 & 0.93 & 5 & 3 \\
		$O_4$ & 0.42 & 0.86 & 0.12 & 0.82 & 4 & 2 \\
		$O_5$ & 0.62 & 0.09 & 0.56 & 0.07 & 7 & 7 \\
		$O_6$ & 0.84 & 0.59 & 0.49 & 0.73 & 5 & 4 \\
		$O_7$ & 0.05 & 0.43 & 0.52 & 0.82 & 2 & 6 \\
		$O_8$ & 0.40 & 0.24 & 0.10 & 0.64 & 1 & 1 \\
		\hline
		$Q$   & 0.18 & 0.87 & 0.76 & 0.23 & - & - \\
		\hline
	\end{tabular}
	}
	\tabcaption{\textbf{Running example ($O_3, O_7$ are reference objects).}}
	\label{tab:example}
\end{table}

\subsection{\trees}
\label{sec:bptree}

Corresponding to the $\tau$ Hilbert curves, $\tau$ \trees are next constructed
using the Hilbert keys of the objects.

The reasons for not adapting the standard B+-tree leaf structure are as
follows.  If only the pointers to the actual objects are stored in the leaf,
then accessing $\alpha$ objects require $\alpha$ random disk accesses.  The
number of such random disk accesses can be reduced by storing the entire object
descriptor (and not a pointer to it) in the leaf itself.  However, in high
dimensions, the number of objects that can be stored in the leaf is extremely
low.  For example, assuming a page size of 4\,KB, only $4$ objects of
dimensionality $128$ can fit in a page, where each dimension is of $8$ bytes.
Hence, none of these designs is suitable for high-dimensional spaces.

We, therefore, modify the leaves to store distances to \emph{reference
objects}.  Reference objects are specially designated objects from the dataset
that are used to quickly obtain approximate distances to the query.  We explain
the querying process using reference objects in Sec.~\ref{sec:query} and how
they are chosen in Sec.~\ref{sec:reference}.  In this section, we assume that
the set of reference objects $\mathcal{R}$ of size $m$ is known.

For each object, the distance to all the $m$ reference objects are stored as
part of its description.  Hence, as long as $m < \nu$, where $\nu$ is the
dimensionality of the object, more number of objects can fit in a page of \tree
than a B+-tree (assuming that the storage requirement for a distance value is
the same as that for a dimension).

In Sec.~\ref{sec:exp}, we show that even for extremely large datasets, a
reference set size of $10$ suffices.  Thus, the reduction in number of random
disk accesses required to retrieve the same number of candidate objects for a
$128$-dimensional space is almost $13$ times.

\begin{table}[t]
\centering
\scalebox{0.82}{
\begin{tabular}{ccccccc}
\hline
\multirow{2}{*}{\textbf{Dataset}} & \textbf{Dimen-} & \textbf{Hilbert} & \textbf{Dimensions} & \textbf{Reference} & \textbf{\tree} \\
& \textbf{sions, $\nu$} & \textbf{order, $\omega$} & \textbf{per curve, $\eta$} & \textbf{objects, $m$} & \textbf{leaf order, $\Omega$} \\
\hline
SIFTn & 128 & 8 & 16 & 10 & 63 \\
Yorck & 128 & 32 & 16 & 10 & 36 \\
SUN & 512 & 32 & 64 & 10 & 13 \\
Audio & 192 & 32 & 24 & 10 & 28 \\
Enron & 1369 & 16 & 37 & 10 & 18 \\
Glove & 100 & 32 & 10 & 10 & 40 \\
\hline
\end{tabular}
}
\tabcaption{\textbf{\tree leaf order (page size = 4 KB).}}
\label{tab:leaves}
\moveups
\end{table}

The \emph{order} of the \tree leaves is computed as follows.  Storing
the Hilbert key of an object requires $\eta. (\omega / 8)$ bytes since for each
of the $\eta$ dimensions, $\omega / 8$ bytes describe the key.  The distances to
$m$ reference objects requires $4.m$ bytes of storage.  The pointer to the
complete object descriptor consumes a further $8$ bytes.  Thus, if there are
$\Omega$ objects in a \tree, its storage requirement is $(\eta. (\omega / 8) +
4.m + 8) . \Omega$ bytes.  Adding the overheads of left and right pointers ($8$
bytes each) to the sibling leaves and an indicator byte to indicate that it is a
leaf, if the page size is $B$ bytes,
\begin{align}
	\label{eq:leaf}
	(\eta \cdot (\omega / 8) + 4 \cdot m + 8) \cdot \Omega + 16 + 1 \leq B
\end{align}
The order $\Omega$ is the \emph{largest} integer that satisfies
Eq.~\eqref{eq:leaf}.  Table~\ref{tab:leaves} lists the \tree leaf orders for
the datasets we use.

Assuming that the reference objects chosen are $O_3$ and $O_7$, the \trees
constructed using the Hilbert keys are shown in Fig.~\ref{fig:btrees}.  The
leaves contain actual distances to $O_3$ and $O_7$ respectively.

\subsection{Reference Object Selection}
\label{sec:reference}

In order to avoid the actual distance computation between an object and the
query at run time, \emph{reference objects} are used to approximate the distance
and quickly filter out objects.  We describe the details of how the distance is
approximated later in Sec~\ref{sec:query}.  In this section, we describe the
desirable properties of the set of reference objects $\mathcal{R}$ and how they
are chosen.

The nature of the distance approximation is such that the closer a reference
object is to the query, the better is the approximation.  Hence, the set of
reference objects should be well spread in the data space.  This ensures that no
matter where the query is located, there is at least one reference object that
is not too far away from it and, hence, the approximation of distances to other
objects using this reference object is good.
The above objective assumes that any reference object can be used to
approximate the distance for any other object.  In other words, the purview of
every reference object is the entire dataset.  Thus, the distance from an
object to all the reference objects must be pre-computed and stored.

Considering the above issues, a good choice for reference object selection is
the \emph{sparse spatial selection (SSS)} algorithm \cite{pivot2}.  The method
first estimates the largest distance, $d_{max}$, between any two objects in the
database using the following heuristic.  A random object is selected and its
farthest neighbor is found.  This neighbor is then selected and the process of
finding the farthest object is repeated.  The process continues till
convergence or for a fixed number of iterations.

Once $d_{max}$ is estimated, an object is chosen randomly as the first
reference object.  In every iteration, the dataset is scanned till an object
whose distance to \emph{all} the previously selected reference objects is
greater than a fraction $f$ of $d_{max}$ is found.  This object is added to the
set $\mathcal{R}$.  The process is continued till $|\mathcal{R}| = m$.

The \emph{SSS dynamic (SSS-Dyn)} \cite{sss_dyn} method extends the above by
continuing the process beyond the first $m$ iterations.  For every new object
that satisfies the fraction $f$ condition, it chooses a victim from the current
set.  The victim is the one that contributes the least towards approximating the
distance between a set of (pre-selected and fixed) object pairs.  If the
contribution of the new object is better than the victim, it replaces the victim
as a reference object.  This goes on till there are no more objects that satisfy
the $f$ condition.  The choice of the method is discussed in
Sec.~\ref{sec:refobj}.

The number of reference objects, $m$, determines the amount of pre-processing
and storage required for each database object.  More importantly, since the
\tree leaves store the distances to these $m$ reference objects, it determines
the \emph{order} of the \trees at the leaf level as shown in
Eq.~\eqref{eq:leaf}.

The offline index building phase is completed with the construction of the
$\tau$ \trees.  We refer to this union of the $\tau$ \trees, each storing
$n$ Hilbert key values of the database points from $\eta$-dimensional
sub-spaces, with the leaves storing the distances to $m$ reference objects, as
the \emph{\name (High-Dimensional Index)}.

\subsection{Tuning of Parameters}

The order $\omega$ of the Hilbert curve should be such that there is no large
loss in the representation after the quantization using $\omega$ bits.
Therefore, it depends on the domain from which the individual dimensions of the
descriptors draw their value.  Table~\ref{tab:datasets} shows the details of
the datasets including their domain sizes while Table~\ref{tab:leaves} lists
the Hilbert curve orders and other parameters.

If the number of partitions $\tau$ is more, then the larger number of Hilbert
curves increases the number of \tree searches, thereby increasing the overall
searching time.  However, simultaneously, $\eta$ decreases which results in
better approximation by the Hilbert curve, thereby increasing the accuracy.
Hence, the value of $\tau$ (or equivalently, $\eta$) provides a handle to
balance the trade-off between running time and accuracy.  We discuss the tuning
of these parameters experimentally in Sec.~\ref{sec:exp}.

From our experiments, we observed that the effect of varying $f$ on the overall
quality and efficiency of the reference object selection method is minimal.
Hence, we chose $f = 0.3$.

\subsection{Analysis}

\subsubsection{Time and Space Complexities}

We first analyze the index construction time of \name.

Estimating $d_{max}$ for reference object selection requires $O(n)$ time since
the iterations are continued for at most a (constant) threshold number of
times.  In each iteration, a potential reference object is checked against all
the $O(m)$ existing reference objects.  Assuming a worst case of $O(n)$
potential reference objects, the total time for $m$ iterations is $O(m^2.n)$.

Computing the distances between the objects and the references having
dimensionality $\nu$ consumes $O(m.n.\nu)$ time.

Mapping a single point to a Hilbert key of order $\omega$ catering to a
dimensionality of $\eta$ requires $O(\omega.\eta)$ time.  Thus, the time for $n$
objects is $O(\omega.\eta.n)$.  The \tree construction requires
$O(n.\log_{\theta} n)$ time where $\theta$ is the branching factor of the
internal nodes.
Since there are $\tau = \nu / \eta$ \trees, the total time for this phase is
$O((\omega.\eta.n + n.\log_{\theta} n).\tau) = O(\nu.\omega.n +
\tau.n.\log_{\theta} n)$.

The total construction time of \name is, therefore, $O(m^2.n + m.n.\nu +
\nu.\omega.n + \tau.n.\log_{\theta} n)$.

We next analyze the space requirements for \name.  The original data objects
require $O(n.\nu)$ storage.  In each of the $\tau$ \trees and for each of the
database objects, the following are stored: (1)~Hilbert key composed from
$\eta$ dimensions, (2)~pointer to the original object, and (3)~distances to $m$
reference objects.  This requires a total of $O(\tau.n.(\eta + O(1) + m)) =
O(n.\nu + \tau.n.m)$ space.  The rest of the \tree requires $O(n.\eta)$ space
to store the $O(n)$ internal nodes consisting of Hilbert keys.  Hence, the
total space requirement of \name is $O(n.\nu + n.m.\tau)$.

\subsubsection{Discussion}

In practice, $\omega$, $\tau$, the height of the \tree ($\log_{\theta} n$), and
$m \ll n$ are small constants (as verified experimentally in Sec.~\ref{sec:exp}
as well).  Hence, the construction time and space complexities of \name are
essentially $O(n.\nu)$, which is the size of the input.  This \emph{linear}
scaling factor allows \name to scale gracefully to extremely large datasets and
high dimensionality.

\subsection{Handling Updates}

In this section, we describe how updates are handled.  We only discuss
insertions since deletions can be handled by simply marking the object as
``deleted'' and not returning it as an answer.  Since B+-trees are naturally
update-friendly, inserting new points in a \tree requires a very small amount of
computation: the Hilbert key of the new object and its distances to reference
objects.

Although the set of reference objects may change with insertions, since the
performance with even random reference objects are quite close in quality to a
carefully chosen set (Sec.~\ref{sec:refobj}), we do \emph{not} re-compute the
set $\mathcal{R}$ with every new insertion.  Also, the number of updates, if
any, are generally insignificant with respect to the total number of objects
and, hence, there is little merit in updating the reference set.

\section{Querying}
\label{sec:query}

We next describe the kNN search procedure for a query object $q$ residing in the
same $\nu$-dimensional Euclidean space (Algo.~\ref{algo:query}).

\begin{algorithm}[t]
	\caption{kNN Query using \name}
	\label{algo:query}
	\textbf{Input:} Query $q$, Number of nearest neighbors $k$ \\
	\textbf{Output:} Approximate kNN $\mathcal{A}$ 
	\begin{algorithmic}[1]
		{\scriptsize
		\FOR {$i \gets 1$ to $\tau$}
			\STATE $q_i \gets$ dimensions of $q$ according to $\mathcal{P}_i$
			\STATE $k_i \gets$ Hilbert key of $q_i$
			\STATE $C^1_i \gets \alpha$ nearest objects of $k_i$ from $\mathcal{B}_i$
			\FOR {$j \gets 1$ to $\alpha$}
				\STATE Compute $\tilde{d}_{\triangle}(q,C^1_i(j))$ using Eq.~\eqref{eq:triangle}
			\ENDFOR
			\STATE $C^2_i \gets \beta$ nearest objects from $\tilde{d}_{\triangle}(q,C^1_i(j))$
			\FOR {$j \gets 1$ to $\beta$}
				\STATE Compute $\tilde{d}_{\square}(q,C^2_i(j))$ using Eq.~\eqref{eq:ptolemy}
			\ENDFOR
			\STATE $C^2_i \gets \gamma$ nearest objects from $\tilde{d}_{\square}(q,C^2_i(j))$
		\ENDFOR
		\STATE $\mathcal{C} \gets \cup_{i = 1}^{\tau} C^2_i$
		\FOR {$j \gets 1$ to $\kappa = |\mathcal{C}|$}
			\STATE Access object $\mathcal{C}(j)$
			\STATE Compute actual distance $d(q,\mathcal{C}(j))$
		\ENDFOR
		\STATE $\mathcal{A} \gets k$ nearest objects from $d(q,\mathcal{C}(j))$
		\RETURN {$\mathcal{A}$}
		}
	\end{algorithmic}
\end{algorithm}

The querying process follows three broad steps: (i)~retrieving candidate objects
from the \trees using the Hilbert keys, (ii)~refining the candidate set by
applying distance approximation inequalities, and (iii)~retrieving the $k$
approximate nearest neighbors by computing the actual distances from the
candidates.

We next describe each of them in detail.

\subsection{Retrieving Candidates from \trees}
\label{sec:candidates}

The $\nu$ dimensions of the query $q$ are divided into $\tau$ groups according
to the same partitioning scheme $\mathcal{P}$ as the dataset.

For each of these $\tau$ partitions, the corresponding \tree $\mathcal{B}_i$
is searched with the Hilbert key of $q$ associated with the dimensions
$\mathcal{P}_i$.  The nearest $\alpha$ candidate objects are returned.  These
objects are close in the space represented by $\mathcal{P}_i$ and, therefore,
have the potential to be close in the entire $\nu$-dimensional space as well.

Returning to our example in Fig.~\ref{fig:example}, assuming $\alpha = 2$, the
first \tree returns $O_1, O_4$ while the second one returns $O_5, O_7$.

\subsection{Refining Candidates using Approximation}
\label{sec:refining}

The $\alpha$ objects are then successively refined by applying two distance
approximation schemes---\emph{triangular inequality} and \emph{Ptolemaic
inequality}---using the reference objects.
As a pre-requisite, the distance of query $q$ to all the $m$
reference objects are computed.

The triangular inequality using the $i^\text{th}$ reference object
$\mathcal{R}_i$ lower bounds the distance between the query $q$ and an object
$o$ as follows:
\begin{align*}
	d(q,o) \geq \tilde{d}_{\triangle}(q,o)_{i} =
		|d(q,\mathcal{R}_i) - d(o,\mathcal{R}_i)|
\end{align*}

Using $m$ reference objects produces $m$ lower bounds for an object $o$.  The
best lower bound is the one that is \emph{maximal} and is, therefore, considered
as the \emph{approximate} distance of $o$ from $q$:
\begin{align}
	\label{eq:triangle}
	\tilde{d}_{\triangle}(q,o) =
	\max_{\forall \mathcal{R}_i \in \mathcal{R}} \tilde{d}_{\triangle}(q,o)_{i}
\end{align}

Using this approximate distance, the $\alpha$ candidates are sorted and the
$\beta$ nearest ones to $q$ are chosen as the next set of candidates for
the application of the next distance approximation, the Ptolemaic inequality
\cite{ptolemy}, which is respected by the Euclidean distance.

The Ptolemaic inequality uses two reference objects $\mathcal{R}_i$ and
$\mathcal{R}_j$ to lower bound the distance:
\begin{align*}
	d(q,o) \geq \tilde{d}_{\square}(q,o)_{i,j} =
		\frac{ |d(q,\mathcal{R}_i) . d(o,\mathcal{R}_j)
			- d(q,\mathcal{R}_j) . d(o,\mathcal{R}_i)| }
			{ d(\mathcal{R}_i,\mathcal{R}_j) }
\end{align*}

The best (i.e., \emph{maximal}) lower bound among the $m \choose 2$ possibilities
is considered as the \emph{approximate} distance:
\begin{align}
	\label{eq:ptolemy}
	\tilde{d}_{\square}(q,o) =
		\max_{\forall \mathcal{R}_i, \mathcal{R}_j \in \mathcal{R}} \tilde{d}_{\square}(q,o)_{i,j}
\end{align}

The nearest $\gamma$ candidates, chosen from the $\beta$ candidates using
Eq.~\eqref{eq:ptolemy}, forms the final candidate set from the \tree
$\mathcal{B}_i$.

The order of applying the inequalities is important as the approximation using
Ptolemaic inequality is \emph{costlier} but \emph{better} than the triangular
inequality.

The $\gamma$ candidates from each $\mathcal{B}_i$ are merged to form the final
candidate set $\mathcal{C}$.  The size of $\mathcal{C}$, denoted by $\kappa$,
is at least $\gamma$ (when all the partitions produce the same set) and at most
$\tau . \gamma$ (when the partitions produce completely disjoint sets).  Thus,
$\gamma \leq \kappa \leq \tau . \gamma$.

\subsection{Retrieving k Nearest Neighbors}

Finally, the complete descriptions of the candidate objects in $\mathcal{C}$ are
accessed using the corresponding object pointers from the \trees, and the
\emph{actual} distances from $q$ are computed.  The $k$ nearest objects from
$\mathcal{C}$ are returned as the kANN answer set $\mathcal{A}$.

\subsection{Analysis}

\subsubsection{Time and Space Complexities}

We first analyze the time complexity of the querying phase.

The cost of distance computation from $q$ to all the $m$ reference objects is
$O(m.\nu)$.

For each of the $\tau$ partitions, the following costs are paid.  Retrieval of
$\alpha$ candidates from a \tree requires $O(\log_{\theta} n + \alpha)$ time
where $\theta$ is the branching factor of the \tree internal nodes.  Computing
the triangular inequalities for $\alpha$ candidates using $m$ reference objects
takes $O(\alpha.m)$ time due to pre-processing.  By using a heap, the nearest
$\beta$ candidates are chosen in $O(\alpha. \log \beta)$ time.  Subjecting the
$\beta$ candidates to Ptolemaic inequalities requires $O(\beta.  m^2)$ time.
Finally, $\gamma$ candidates are extracted using a heap in $O(\beta. \log
\gamma)$ time.  The total cost over $\tau$ partitions is, therefore,
$O(\tau.(\log_{\theta} n + \alpha + \alpha.m + \alpha. \log \beta + \beta.m^2 +
\beta. \log \gamma))$.

Computing the actual distances for $\kappa$ candidates requires $O(\kappa.\nu)$
time.  Since $\kappa$ is at most $\tau.\gamma$, this phase runs in
$O(\tau.\gamma.\nu)$ time.

Adding up the costs of each phase gives the total time.

The \emph{extra} space overhead is that of $\alpha$ candidates.  Since the
\trees are processed sequentially, the extra space overhead remains
$O(\alpha)$ which is essentially a constant.

Since \name is a disk-based structure, we next analyze the number of random
disk accesses needed for a kANN query.  The computation of $m$ distances from
the query to the reference objects require $O(m)$ disk accesses.  However,
since $m \ll n$, the reference object set can be assumed to fit in memory and,
therefore, no I/O cost is incurred.  Assuming the order of a \tree leaf to be
$\Omega$, retrieving $\alpha$ objects requires $O(\log_{\theta} n + (\alpha /
\Omega))$ disk accesses.  The computation of the distance inequalities do
\emph{not} require any more disk access since the distances to the reference
objects are stored in the leaves themselves.  Computing $\kappa$ exact
distances from the candidates require $\kappa = O(\tau.\gamma)$ disk accesses.
Hence, the total number of disk accesses is $O(\tau.(\log_{\theta} n + (\alpha
/ \Omega) + \gamma))$.

\subsubsection{Discussion}

As shown in Sec.~\ref{sec:exp}, it suffices to have $\alpha \ll n$ for even
extremely large datasets.  Consequently, $\alpha$, $\beta$ and $\gamma$ are
essentially constants in the overall analysis.  Since $\omega$ and $m \ll n$
are also small constants, the running time of a kANN query can be summarized as
$O(\tau . (\log_{\theta} n + \nu))$, i.e., it is linear on the number of \tree
searches, the height of a \tree, and the dimensionality.  The extra space
consumed during the search is $O(\alpha)$, which is negligible.  The number of
disk accesses is \emph{linear} in the number of trees, the height of the trees,
and candidate size, thus allowing \name to scale gracefully to large datasets
and high-dimensional spaces.

\section{Experiments}
\label{sec:exp}

All the experiments were done using codes written in C++ on an Intel(R)
Core(TM) i7-4770 machine with 3.4 GHz CPU, 32 GB RAM, and 2 TB external 
disk space running Linux Ubuntu 12.04.  The results are summarized in
Sec.~\ref{sec:expsummary}.
\ifarxiv
\else
The reader is referred to the extended version \cite{full} of this paper for
additional results.
\fi

\noindent
{\bf Datasets:} We present results on large-scale high-dimensional
\emph{real-world} and \emph{benchmark} datasets, taken from various publicly
available sources \cite{yorck,sun,enron,glove,audio,data}, as described in
Table~\ref{tab:datasets}. Our datasets constitute a judicious mix of scale,
dimensionality, and domain types.

\begin{itemize}[nosep,leftmargin=*]
	\item The \emph{Audio} dataset comprises of audio features extracted
		using Marsyas library from the DARPA TIMIT speech database
		\cite{audio}.
	\item The \emph{SUN} dataset comprises of extremely high-dimensional GIST
		features of images \cite{sun}.
	\item The \emph{SIFT} datasets comprise of SIFT features \cite{data} at
		varying scale.
	\item The \emph{Yorck} dataset comprises of SURF features extracted from
		images in the Yorck project \cite{yorck}.
	\item The \emph{Enron} dataset comprises of feature vectors of bi-grams
		of collection of emails \cite{enron}.
	\item The \emph{Glove} dataset comprises of word feature vectors
		extracted from Tweets \cite{glove}.
\end{itemize}

\begin{table}[t]
\centering
\vspace*{1mm}
\scalebox{0.85}{
\begin{tabular}{lrrrrr}
\hline
\multirow{2}{*}{\textbf{Dataset}} & \multirow{2}{*}{\textbf{Type}} & \textbf{Dimen-} & \multirow{2}{*}{\textbf{\# Objects}} & \textbf{Domain} & \multirow{2}{*}{\textbf{\# Queries}} \\
& & \textbf{sionality} & & \textbf{of values} & \\
\hline
SIFT10K & Tiny & 128 & 10,000 & [0,255]	& 100\\
\hline
Audio & Small & 192 & 54,287 & [-1,1] & 10,000\\
SUN & Small & 512 & 80,006 & [0,1] & 100\\
\hline
SIFT1M & Medium & 128 & 1,000,000 & [0,255]	& 10,000\\
SIFT10M & Medium & 128 & 10,000,000 & [0,255] & 10,000\\
Yorck & Medium  & 128 & 15,120,935 & [-1,1]	& 1,254\\
\hline
SIFT100M & Large & 128 & 100,000,000 & [0,255] & 10,000\\
SIFT1B & Large & 128 & 999,494,170 & [0,255] & 10,000\\
\hline
Enron & Small & 1369 & 93,986 & [0,252429] & 1,000\\
Glove & Medium & 100 & 1,183,514 & [-10,10] & 10,000\\
\hline
\end{tabular}
}
\tabcaption{\textbf{Datasets.}}
\label{tab:datasets}
\moveup
\end{table}

Note that SIFT and Enron features are integers, while all other datasets
possess floating point feature values.  The Hilbert curve and \tree orders are
shown in Table~\ref{tab:leaves}.

Next, we state the \emph{methods}, their \emph{parameters} and the
\emph{evaluation metrics} used for obtaining the results.

\noindent
{\bf Methods:} We compare \name for effectiveness, efficiency and scalability
against a number of \emph{representative} state-of-the-art methods, namely,
Multicurves, C2LSH, SRS, QALSH, OPQ, and HNSW \cite{valle,c2lsh,srs,qalsh, opq,
hnsw}. In addition, we also consider iDistance \cite{idistance}, which is an
\emph{exact} method for kNN queries and will, therefore, always exhibit
\emph{perfect} quality.  For all these techniques, we adopt the C++
implementations made available by their authors.

\noindent
{\bf Parameters:} The disk page size for the machine on which the experiments
were run is $B=4096$ bytes. We set the number of nearest neighbors $k=100$.
The parameters for the competing techniques were set as follows: iDistance:
initial radius $r=0.01$, $\Delta r=0.01$; Multicurves: $\tau=8$, $\alpha=4096$;
C2LSH: approximation ratio $c=2$, interval size $w=1$, false-positive
percentage $\beta^{C2LSH}=100/n$, error probability $\delta =1/e$; QALSH:
approximation ratio $c=2$, false-positive percentage $\beta^{QALSH}=100/n$,
error probability $\delta =1/e$; SRS: algorithm SRS-12, approximation ratio
$c=2$, number of 2-stable random projections $m^{SRS}=6$, early termination
threshold $\tau^{SRS}=0.1809$, maximum percentage of examined points
$t=0.00242$; OPQ: number of subspaces $M=8$; HNSW: number of nearest neighbors
$M=10$.  The search parameters of OPQ and HNSW were set such that their MAP
values were close to those of \name.  Note that most of these parameters have
been set according to the recommendations by their respective authors.

\noindent
{\bf Evaluation Metrics:} We consider the following metrics:

\begin{itemize}[nosep,leftmargin=*]
	\item \textbf{Quality:} We adopt approximation ratio (Def.~\ref{def:ratio})
		and MAP@$k$ (Def.~\ref{def:map}) to evaluate the quality of the
		discussed methods.
	\item \textbf{Efficiency:} We evaluate the efficiency of the
		methods using the running time of the queries. For fairness,
		we turn off buffering and caching effects in all the experiments.
	\item \textbf{Scalability:} To evaluate the scalability,
		we measure (1) the size and (2) construction time of the index
		structures, and main memory consumption during (3) index construction
		and (4) querying.
\end{itemize}

All the reported values are averages over the entire query set.

\subsection{Comments}
\label{sec:comments}

All the datasets are pre-processed to remove the duplicate points, if any. We
used the set of queries provided with the Audio and the SIFT datasets for our
experiments. For the datasets where query sets were not provided, namely, SUN,
Yorck, Enron, and Glove, we reserved $100$, $1254$, $1000$, and $10000$ random
data points respectively as queries.  Since the implementations of some of the
techniques do not support floating point data, to enable them to run on datasets
with floating point feature values, the values are scaled by a sufficiently
large number.  This is as suggested by the corresponding authors of C2LSH and
SRS. 

Although it is not a requirement of the algorithms, the publicly available
implementations of C2LSH, QALSH, and iDistance load the entire dataset into main
memory during the index construction step. Thus, it crashed on our machine for
the two largest datasets, SIFT100M and SIFT1B, that require $\sim$52GB and
$\sim$520GB of RAM to be loaded respectively. Further, the underlying
implementation of R-trees used by SRS suffers from memory leaks.  Hence, to
enable SRS run on SIFT100M and SIFT1B, as suggested by the authors, these
datasets were divided into chunks, indexed separately and then combined together
into a single index for the whole dataset. This discussion highlights the
importance of purely disk-based methods, as even with today's hardware, methods
using very large amounts of main memory in a single machine to build and load
the index structure may not be deemed as scalable.

\subsection{Internal Parameter Tuning}
\label{sec:tuning}

Since the \name structure and its querying mechanism uses some parameters, the
first set of experiments is performed to determine a good range of their values.
This is a critical step.  The parameter tuning process follows an order such
that the most general parameter, i.e., the one which is independent of other
parameters, is tuned first and so on. In general, we follow the intuitive order
of first tuning the parameters specific to \name construction followed by those
used during querying.

\begin{figure*}[t]
\moveups
\centering
	\includegraphics[width=0.81\linewidth]{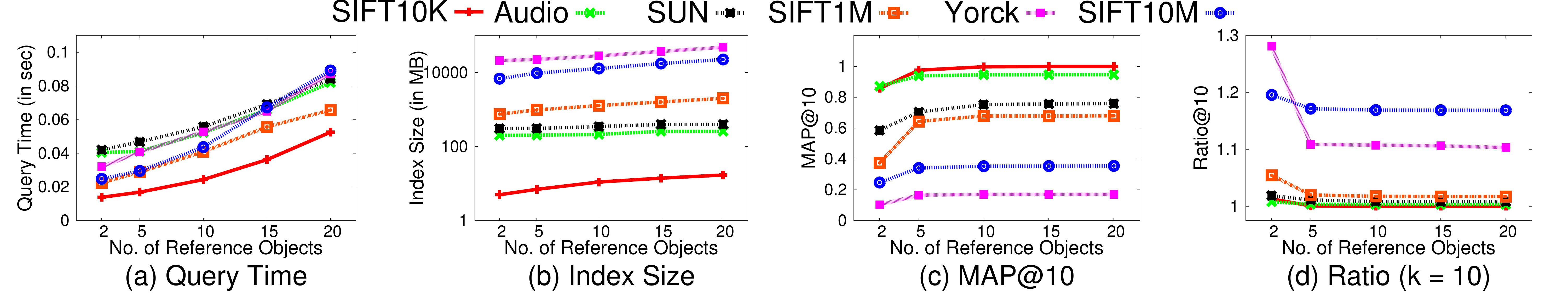}
	\\
	\includegraphics[width=0.81\linewidth]{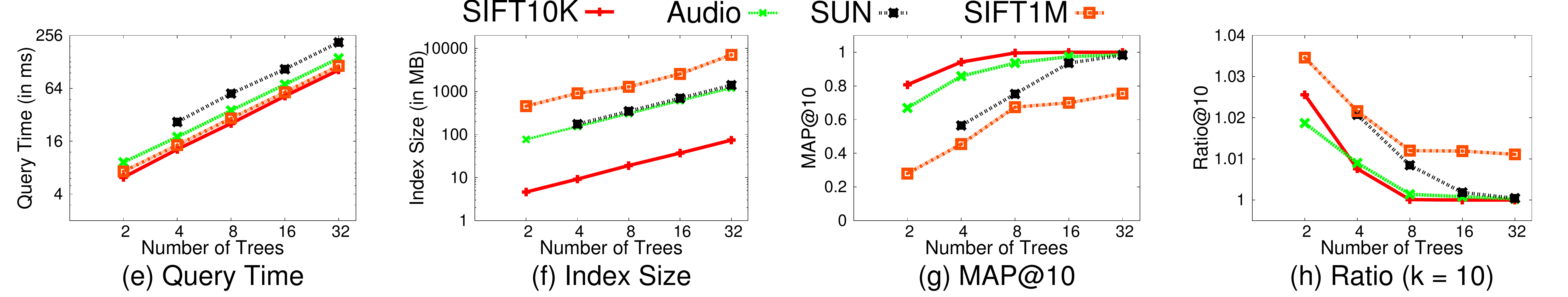}
	\figcaption{\textbf{Effect of varying number of reference objects, $m$, on
	(a) Query time, (b) Index size, (c) MAP@10, and (d) Ratio ($k=10$).  Effect
of varying number of \trees, $\tau$, on (e) Query time, (f) Index size, (g)
MAP@10, and (h) Ratio ($k=10$).}
}
\label{fig:varying_pivots_trees}
\end{figure*}

\begin{figure*}[t]
\centering
	\subfloat[Query Time (SIFT10K)]
	{
		\scalebox{0.24}{
			\includegraphics[width=\trianglewidth]{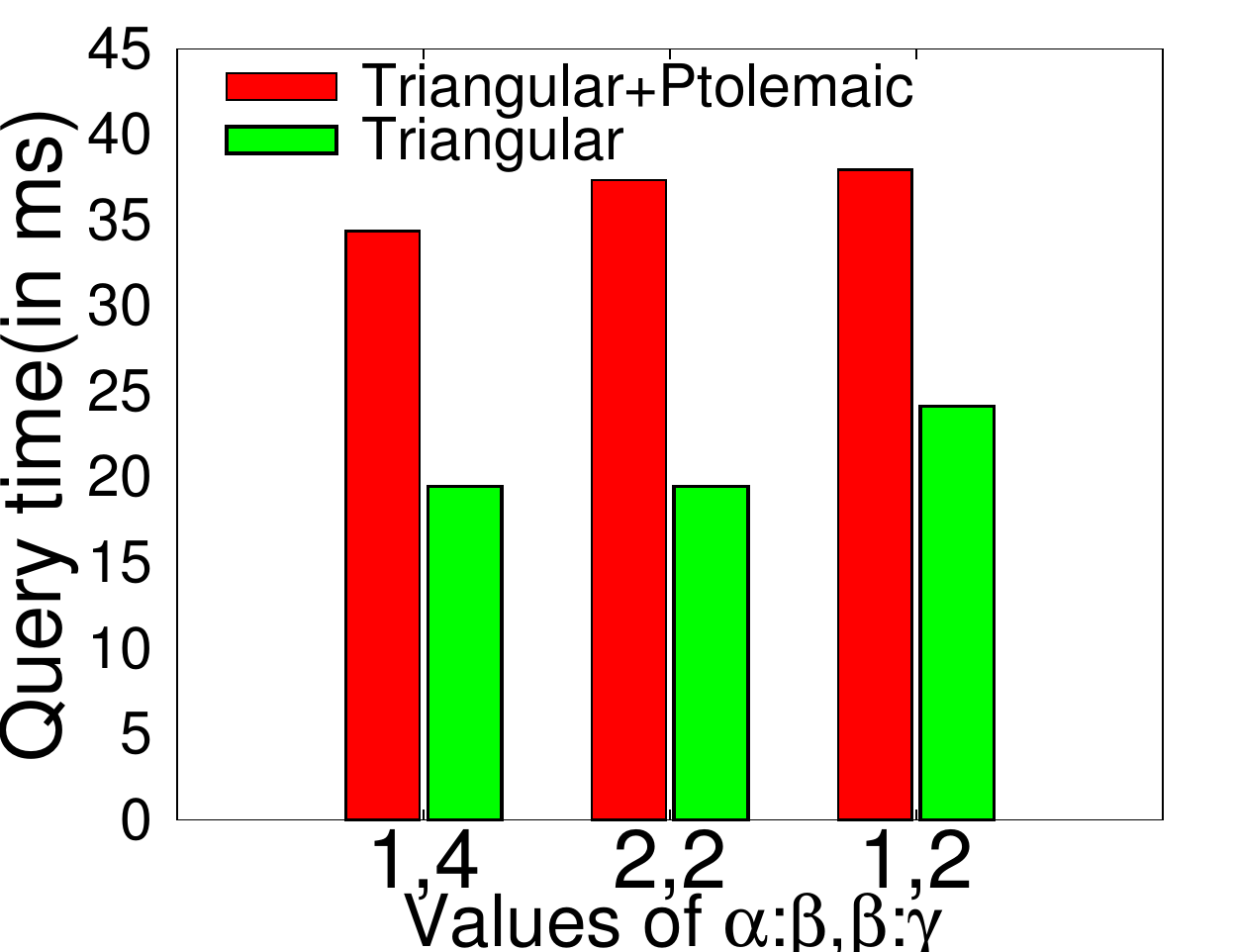}
		}
		\label{fig:ptolemaic1_time1}
	}
	\subfloat[Query Time (Audio)]
	{
		\scalebox{0.24}{
			\includegraphics[width=\trianglewidth]{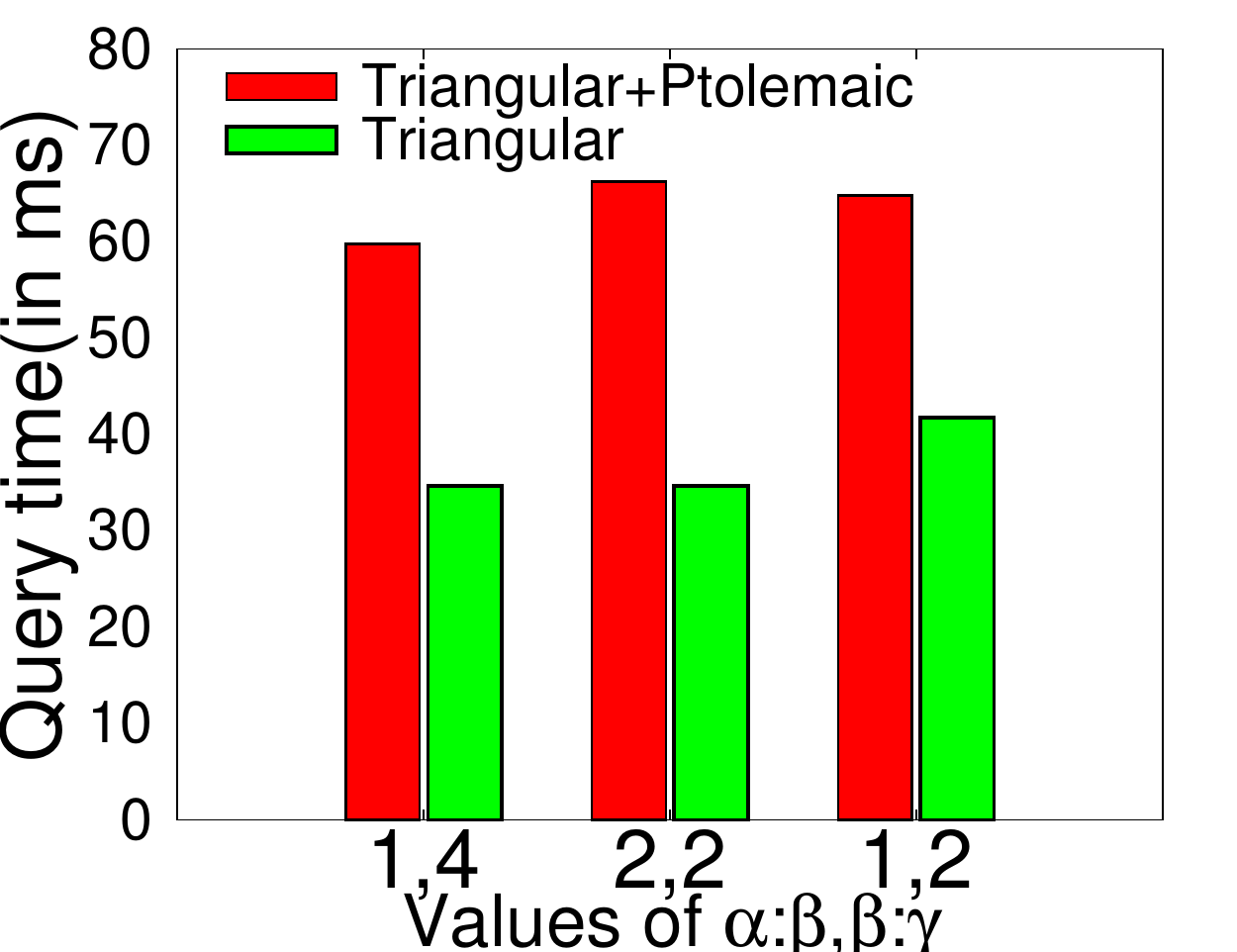}
		}
		\label{fig:ptolemaic1_time2}
	}
	\subfloat[Query Time (SUN)]
	{
		\scalebox{0.24}{
			\includegraphics[width=\trianglewidth]{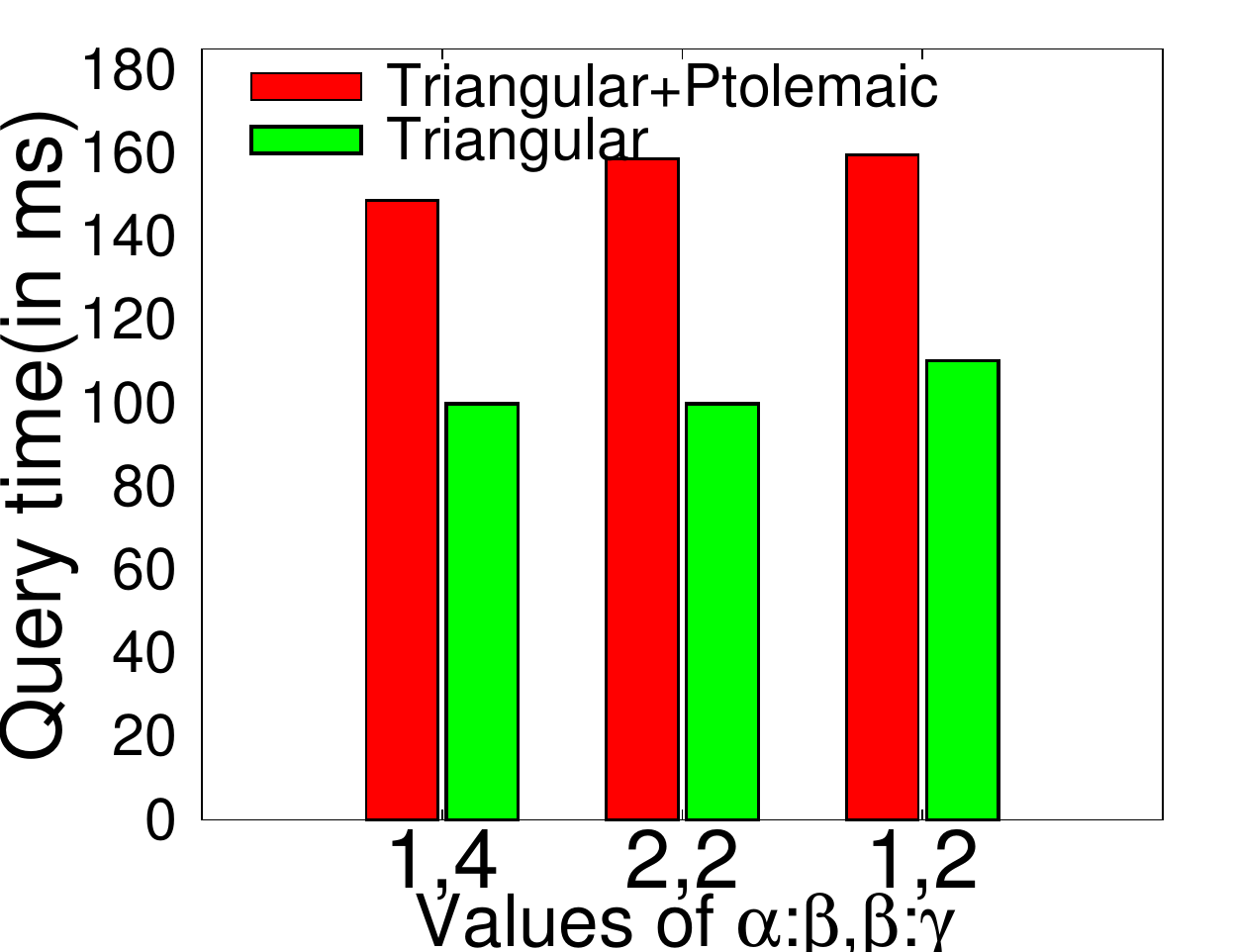}
		}
		\label{fig:ptolemaic1_time3}
	}
	\subfloat[Query Time (SIFT1M)]
	{
		\scalebox{0.24}{
			\includegraphics[width=\trianglewidth]{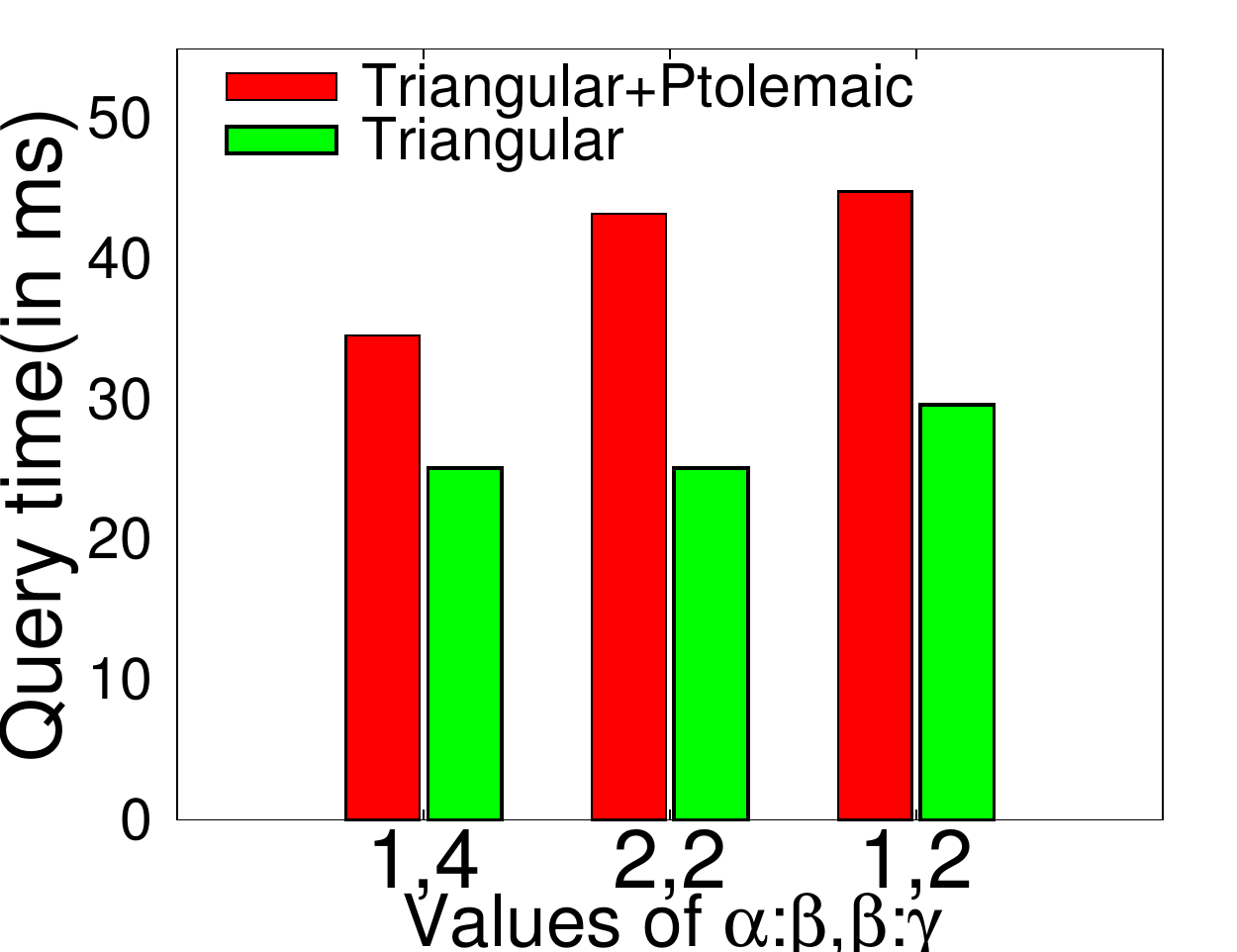}
		}
		\label{fig:ptolemaic1_time4}
	}
	\\
	\vspace{-2mm}	
	\subfloat[MAP@10 (SIFT10K)]
	{
		\scalebox{0.24}{
			\includegraphics[width=\trianglewidth]{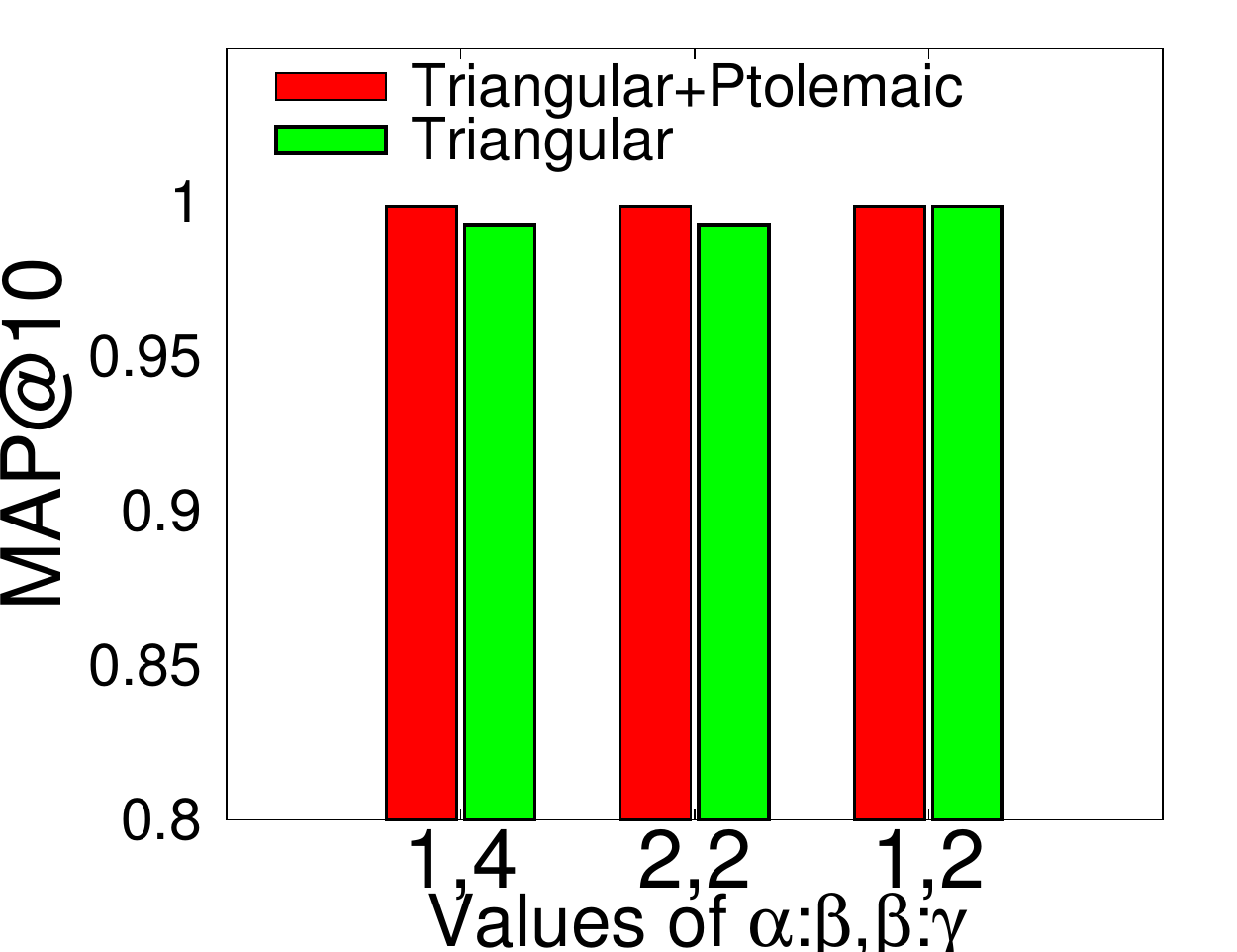}
		}
		\label{fig:ptolemaic1_map1}
	}
	\subfloat[MAP@10 (Audio)]
	{
		\scalebox{0.24}{
			\includegraphics[width=\trianglewidth]{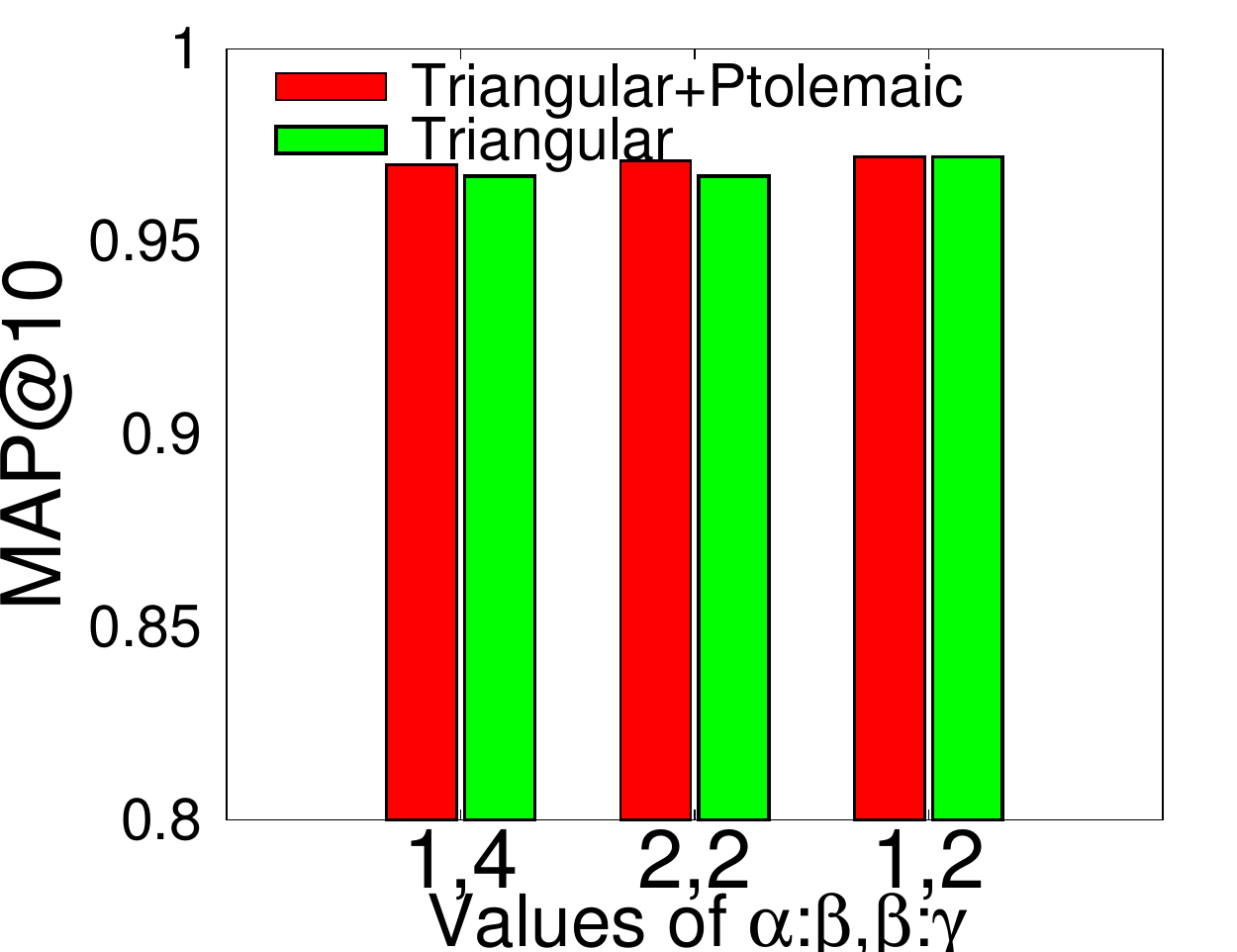}
		}
		\label{fig:ptolemaic1_map2}
	}
	\subfloat[MAP@10 (SUN)]
	{
		\scalebox{0.24}{
			\includegraphics[width=\trianglewidth]{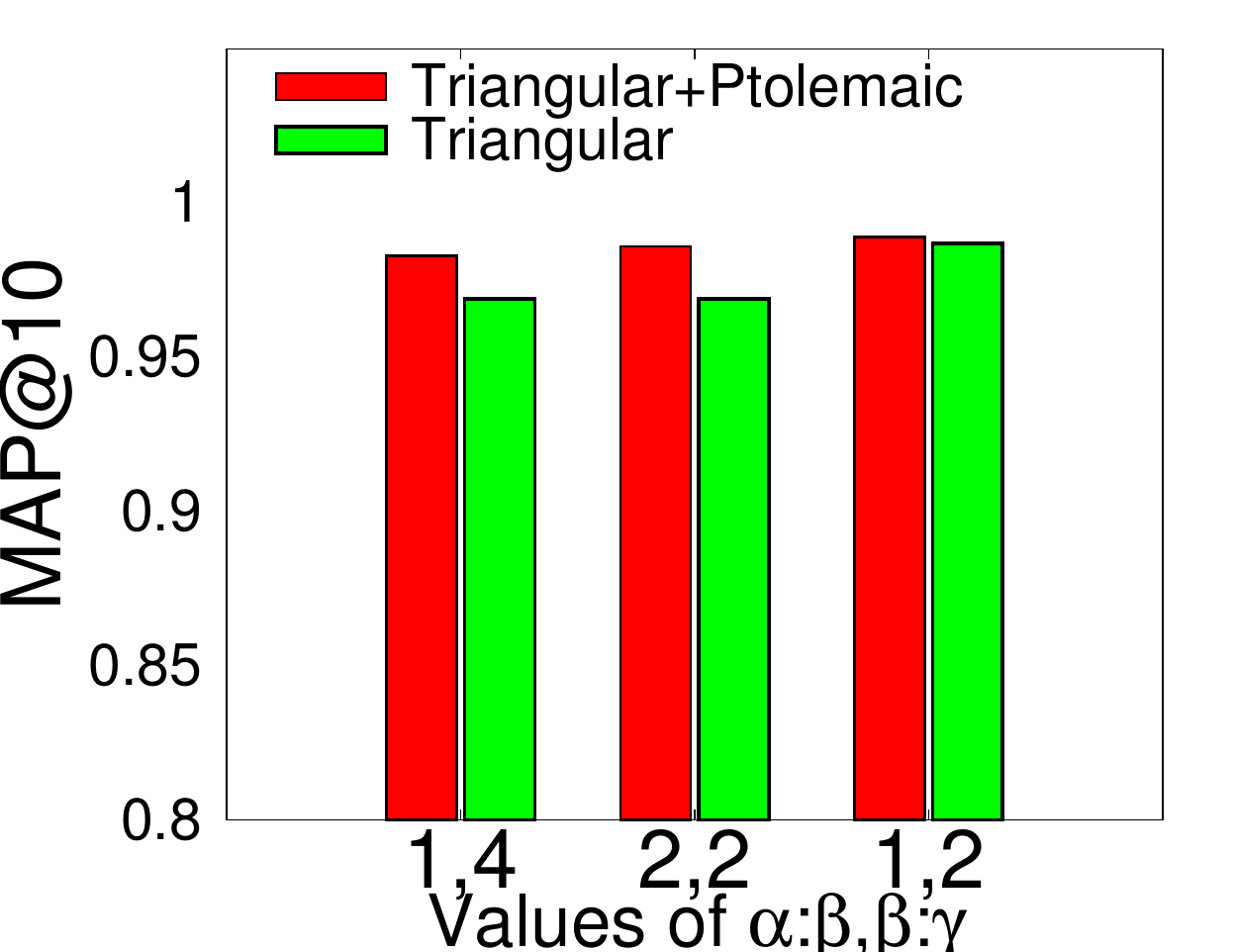}
		}
		\label{fig:ptolemaic1_map3}
	}
	\subfloat[MAP@10 (SIFT1M)]
	{
		\scalebox{0.24}{
			\includegraphics[width=\trianglewidth]{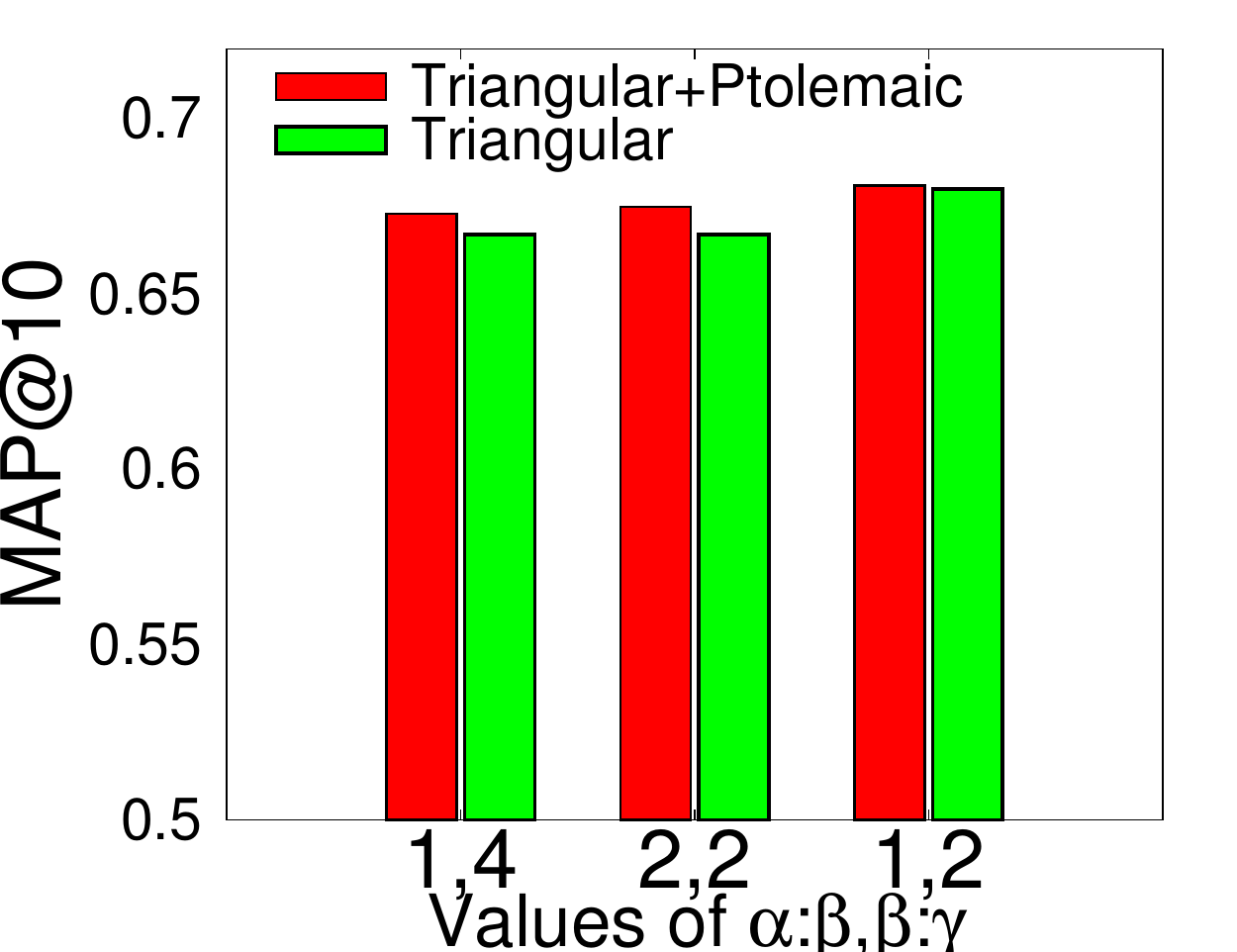}
		}
		\label{fig:ptolemaic1_map4}
	}
	\figcaption{\textbf{Effect of the filtering mechanism (either triangular
	inequality alone or a combination of triangular and Ptolemaic inequalities)
on query time (a-d) and MAP@10 (e-h) with $\alpha=4096$.} }
\label{fig:ptolemaic1}
\vspace*{1mm}
\end{figure*}

\subsubsection{Indexing: Subspace Partitioning Scheme}
\label{sec:partitioning}

During the index construction step, \name requires data to be partitioned into
different lower dimensional sub-spaces.  For ease of use, we sequentially
partition the feature space.  However, that is not necessary, since \name is not
dependent upon a particular partitioning scheme. To substantiate this claim, we
perform querying on $100$ indices constructed by randomly partitioning the full
space into multiple sub-spaces.  The MAP@10 values over the $100$ indices for
the different datasets are as follows: (a)~SIFT10K: $0.974 \pm 0.002$,
(b)~Audio: $0.934 \pm 0.013$, (c)~SUN: $0.750 \pm 0.010$, (d)~SIFT1M: $0.660 \pm
0.010$.  Considering the small standard deviations and comparable results of
MAP@10 for contiguous partitioning (as shown later in
Fig.~\ref{fig:ptolemaic1}e-Fig.~\ref{fig:ptolemaic1}h), we conclude that quality
does not depend significantly on the choice of partitioning scheme.

\subsubsection{Indexing: Reference Object Selection Method}
\label{sec:refobj}

As mentioned earlier in Sec.~\ref{sec:work}, there are multiple methods of
choosing the reference objects.  We have experimented with three algorithms: the
random method, where $m$ random objects are chosen as reference objects, \pivot
and SSS-Dyn.
Compared to the other sophisticated methods, even the random reference object
selection has comparable accuracy levels. (The MAP values are within 90\% of
those of \pivot).  The figures are shown in
\ifarxiv
Fig.~\ref{fig:ref_algo} in Appendix~\ref{app:reference}.
\else
Appendix~A in the extended version \cite{full}.
\fi

This indicates that the design of \name by itself is good enough to achieve a
high accuracy, and the choice of reference objects is not very critical.  The
\pivot method is faster and has similar MAP values as SSS-Dyn.  Also, the larger
the dataset, the lesser the difference between the methods.  Consequently, we
recommend the \pivot method as the reference object selection algorithm.

\subsubsection{Indexing: Number of Reference Objects}

The first important parameter is the number of reference objects, $m$. To
effectively tune $m$, we consider the top $6$ datasets of up to medium size from
Table~\ref{tab:datasets}.  Fig.~\ref{fig:varying_pivots_trees}a shows that the
query time increases in a sub-linear fashion with $m$. The size of the final
index (Fig.~\ref{fig:varying_pivots_trees}b) grows linearly with $m$ (y-axis in
log scale). However, for all the datasets, the quality metrics saturate after
$m=10$, with little or no further changes (Fig.~\ref{fig:varying_pivots_trees}c
and Fig.~\ref{fig:varying_pivots_trees}d).  Thus, we choose $m=10$ as the
recommended number of reference objects. For the rest of the experiments, we use
this value.

\subsubsection{Indexing: Number of \trees}

Next, we analyze the effect of the number of \trees $\tau$ on the different
evaluation metrics. We use the top $4$ datasets from Table~\ref{tab:datasets}.
Figs.~\ref{fig:varying_pivots_trees}e and~\ref{fig:varying_pivots_trees}f show
that, as expected, the running time and the size of the final index grow
linearly with increasing $\tau$, for all the datasets. However,
Figs.~\ref{fig:varying_pivots_trees}g and~\ref{fig:varying_pivots_trees}h
portray that quality saturates after a while. The MAP@10 values for all the
datasets, except SUN, saturates after $\tau=8$. A similar effect is observed
for the approximation ratio as well.  Thus, we choose $\tau = 8$ as the
recommended number of \trees.  However, for the SUN dataset, which has a very
high dimensionality of $\nu = 512$, choosing $\tau = 8$ makes each Hilbert
curve cater to a dimensionality of $\eta = 512 / 8 = 64$. Consequently, for
such a high dimensionality, the MAP performance suffers.  The MAP performance
increases significantly by using double the number of trees, i.e., $\tau = 16$.
Hence, while the default value is $\tau = 8$, for extremely high dimensional
datasets (500+), we recommend $\tau = 16$.  For datasets where the
dimensionality is not a power of $2$, we choose parameters that are as close as
possible to the recommended values.  Thus, for Enron, we use $\tau = 37$ trees,
each catering to $\eta = 37$ dimensions, while for Glove, we use $\tau = 10$
trees, each handling $\eta = 10$ dimensions.

\begin{figure*}[t]
\moveups
\centering
	\includegraphics[width=0.85\linewidth]{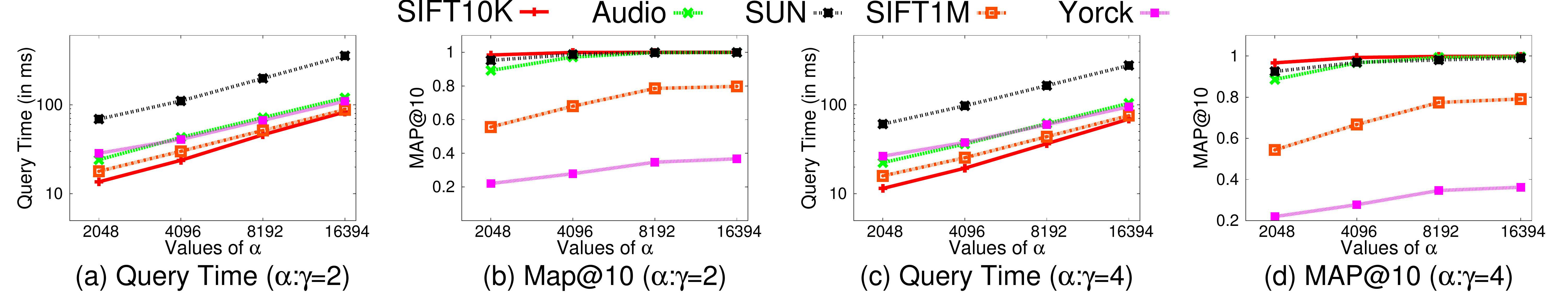}
	\\
	\vspace*{-2mm}
	\includegraphics[width=0.85\linewidth]{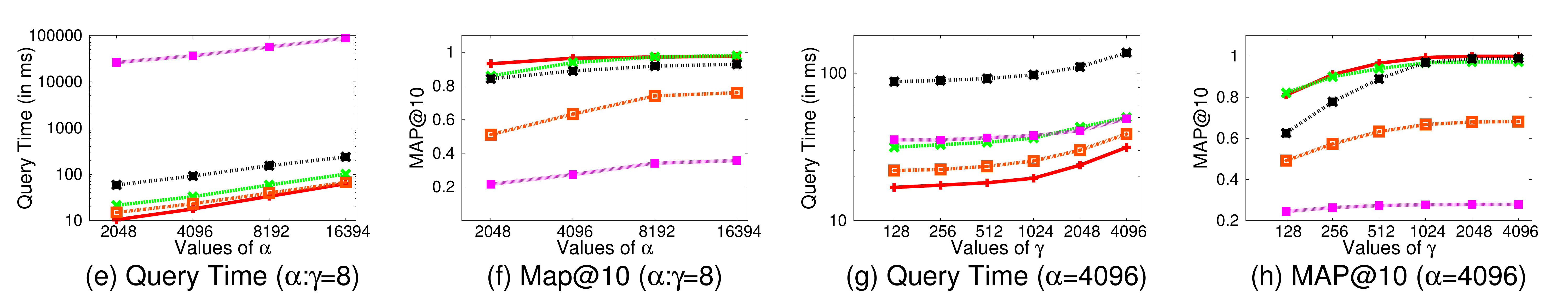}
	\figcaption{\textbf{Effect of varying $\alpha$ on query time and MAP@10
	with $\alpha/\gamma=2$ (a,b); with $\alpha/\gamma=4$ (c,d); and with
$\alpha/\gamma=8$ (e,f); and Effect of varying $\gamma$ on (g) query time and
(h) MAP@10 with $\alpha=4096$.}
}
\label{fig:filter_parameters}
\vspace*{1mm}
\end{figure*}

\begin{figure*}[t]
\centering
	\vspace*{1mm}
	\includegraphics[width=0.98\linewidth]{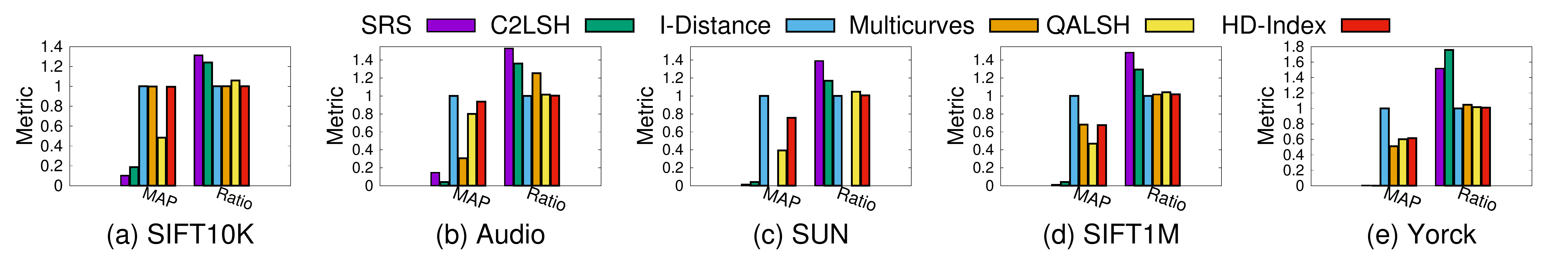}
	\vspace*{-1mm}
	\figcaption{\textbf{Comparison of MAP@10 and approximation ratio for $k=10$
	on different datasets.}
}
\label{fig:map_ratio}
\vspace*{1mm}
\end{figure*}

\subsubsection{Querying: Importance of Ptolemaic Inequality}

Having tuned the parameters specific to \name, we next analyze the effect of
parameters associated with the querying mechanism. We first evaluate the effect
of the lower bounding filters -- triangular and Ptolemaic inequalities.
Fig.~\ref{fig:ptolemaic1} presents a detailed analysis of using (1) just the
triangular inequality, i.e., $\beta = \gamma$ (as explained in
Sec.~\ref{sec:refining}), and (2) a combination of both triangular and Ptolemaic
inequalities, for an initial candidate set size of $\alpha=4096$ on different
datasets.
While it is clear from Figs.~\ref{fig:ptolemaic1_map1}-\ref{fig:ptolemaic1_map4}
that MAP@10 obtained using a combination of triangular and Ptolemaic inequality
filters is always better when compared to using the triangular inequality alone,
the gain is not significant under all scenarios. Notable gains are observed only
when we use $\alpha/\beta=1$ and $\beta/\gamma=4$ for the combined filtering
process, and $\alpha/\gamma=4$ for triangular inequality alone. This is because
tighter lower bounds using Ptolemaic inequality are more robust to large
reductions during the filtering process. On the other hand, it is evident from
Figs.~\ref{fig:ptolemaic1_time1}-~\ref{fig:ptolemaic1_time4} that the running
times obtained using the combined filters are consistently $1.5$--$2$ times
slower when compared to using the triangular inequality alone. A similar pattern
is followed for different $\alpha$ values as well (additional results are shown
in
\ifarxiv
Appendix~\ref{app:ptolemaic}).
\else
Appendix~B in the extended version \cite{full}).
\fi

In sum, although the use of Ptolemaic inequality results in better MAP, the
gain in quality is not significant in majority of the cases when compared to
the increase in query time, which almost doubles. Therefore, it is more prudent
to use the triangular inequality \emph{alone} to trade a little amount of
quality (loss in MAP) for a much higher query efficiency. Thus, we henceforth
choose triangular inequality alone as the recommended choice of lower bounding
filter.

As an alternate analysis, although the application of Ptolemaic inequality is
slower, importantly, it does \emph{not} affect the number of disk I/Os.  Since
all the $\alpha$ candidates are processed in main memory without the need to
access the disk, Ptolemaic inequality only requires more CPU time.  Thus, if the
performance metric of an application is the \emph{number of disk I/Os} (instead
of the wall clock running time), it is better to use Ptolemaic inequality since
it produces better quality results using the \emph{same} number of disk
accesses.

\subsubsection{Querying: Filter Parameters}

Since we do not use the Ptolemaic inequality, the intermediate filter parameter
$\beta$ is not needed any further.  We next fine tune the other filter
parameters $\alpha$ and $\gamma$ by plotting (Fig.~\ref{fig:filter_parameters})
MAP@10 and running time for the top $5$ datasets from Table~\ref{tab:datasets}.
Figs.~\ref{fig:filter_parameters}a,~\ref{fig:filter_parameters}c
and~\ref{fig:filter_parameters}e show that, as expected, the running time scales
linearly with increase in $\alpha$. However, Figs.~\ref{fig:filter_parameters}b,
~\ref{fig:filter_parameters}d and~\ref{fig:filter_parameters}f portray that the
quality saturates after a while (after $\alpha=4096$).  Thus, we choose
$\alpha=4096$ as the recommended value.

However, for SIFT1M and Yorck, $\alpha = 4096$ is not sufficient since these
datasets are extremely large.  MAP improves when double the value, i.e., $\alpha
= 8192$ is used and saturates beyond that.  Hence, for very large datasets, we
recommend $\alpha=8192$.

Having fixed $\alpha$, we analyze the MAP@10 and running times for all the
datasets with varying $\gamma$. Fig.~\ref{fig:filter_parameters}g shows that the
running time scales linearly with increase in $\gamma$.  However, once again,
MAP@10 saturates after $\gamma=1024$ for all the datasets. Thus, we choose
$\gamma=1024$, and $\alpha/\gamma=4$, as the recommended values.

\subsubsection{Robustness with Number of Nearest Neighbors}
\label{sec:k}

We also evaluate MAP@$k$ and running times of the techniques with varying $k$
on the tiny, small and medium datasets (results in
\ifarxiv
Appendix~\ref{app:k}).
\else
Appendix~C in the extended version \cite{full}).
\fi
While the running times of the other techniques increase with $k$, it remains
almost constant for \name and Multicurves.  This is mainly due to the design of
these structures where $\alpha$ candidates are selected and then refined to
return $k$ answers in the end. Since $\alpha > k$, the running times are not
affected by $k$.

\subsubsection{Limitations}
\label{sec:limitations}

It needs to be noted that the current parameter tuning of \name is almost
purely empirical.  In future, we would like to analyze the parameters from a
theoretical stand-point and understand what values can be reasonable for a
given dataset.

Also, if the dataset really exhibits ultra-high dimensionality (in order of
hundreds of thousands), then \name is unable to process the queries efficiently
since there are too many B+-trees.  However, on that note, we would also like to
clarify that to the best of our knowledge, there is no disk-based database
indexing method that can really handle such ultra-high dimensionality.  However,
our method can be easily parallelized and/or distributed with little
synchronization steps due to its nature of building and querying using multiple
independent \trees.

For small datasets, \name is not as efficient as some of the other techniques
(as explained in more detail in Sec.~\ref{subsec:comparison}).  However, as the
dataset size increases, it scales better.

\begin{figure*}[t]
\moveups
\centering
	\includegraphics[width=1.01\linewidth]{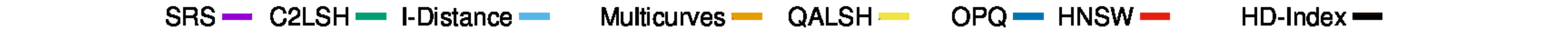}
	\\
	\vspace*{-3mm}	
	\includegraphics[width=1.01\linewidth]{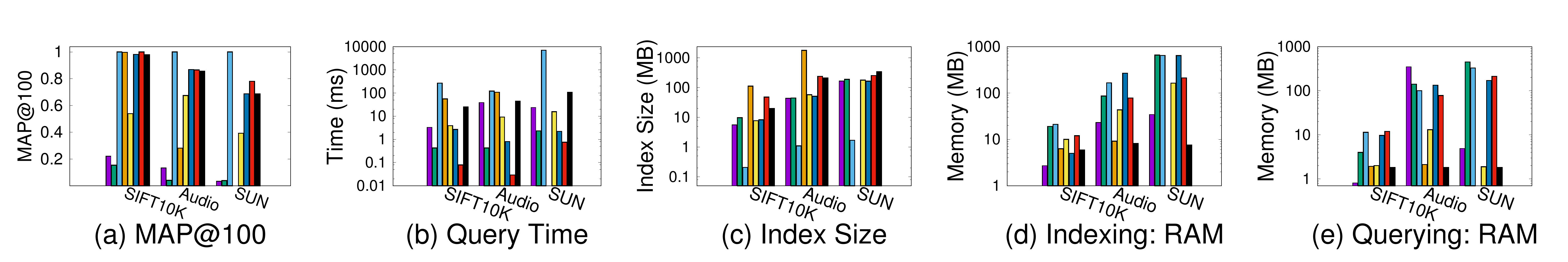}
	\\
	\vspace*{-3mm}	
	\includegraphics[width=1.01\linewidth]{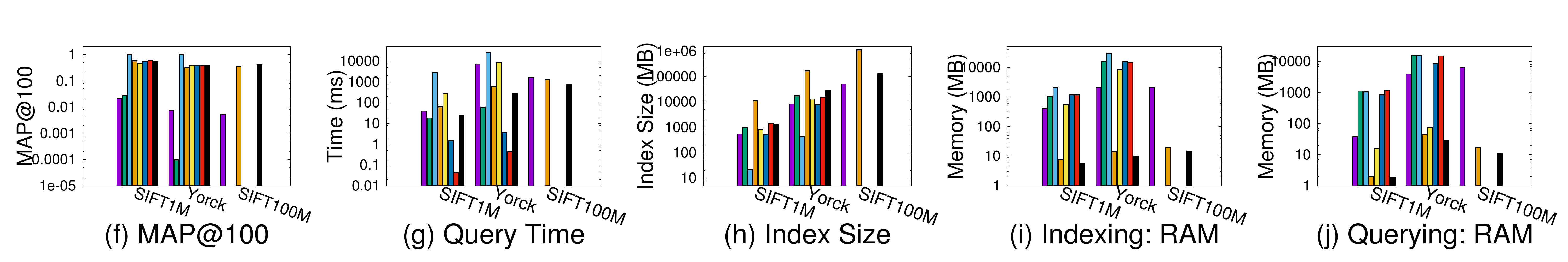}
	\\
	\vspace*{-3mm}	
	\includegraphics[width=1.01\linewidth]{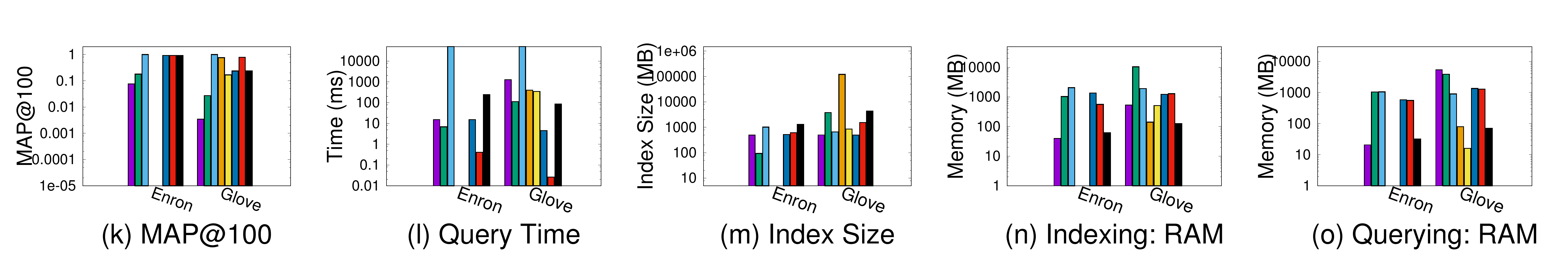}
	\vspace*{-3mm}
	\moveups
	\moveups
	\figcaption{\textbf{Comparison on different parameters for $k=100$ using
	small datasets (a-e), larger datasets (f-j), and text datasets (k-o).} }
\label{fig:techniques}
\vspace*{1mm}
\end{figure*}

\subsection{Quality Metrics}
\label{subsec:map_vs_ratio}

With the \name parameters fine-tuned, we evaluate the two quality metrics,
MAP@$k$ and approximation ratio vis-\`a-vis the other methods. It is evident
from Fig.~\ref{fig:map_ratio}a-\ref{fig:map_ratio}e that even for good
approximation ratios ($c\leq1.5$), MAP values can be surprisingly and
significantly low (MAP $\leq0.15$). Moreover, with increase in dimensionality
$\nu$, this effect becomes more and more prominent. For instance, in
Fig.~\ref{fig:map_ratio}c, $c=1.15$ and $1.39$ for C2LSH and SRS respectively,
while MAP are $0.04$ and $0.01$ respectively.

This analysis highlights a notable limitation of the techniques that
possess theoretical guarantees on the approximation ratio as it empirically
proves that the ratio loses its significance.
Thus, henceforth, we use only MAP@k.

\subsection{Comparative Study}
\label{subsec:comparison}

In this section, we perform an in-depth comparison of the different methods with
respect to three aspects: (a)~quality of results, (b)~efficiency of retrieval in
terms of running time, and (c)~scalability in terms of space usage for both main
memory and on disk.

\subsubsection{Quality}

Figs.~\ref{fig:techniques}a,~\ref{fig:techniques}f,~\ref{fig:techniques}k show
a comparison of MAP@100 of the different techniques.  Being an exact technique,
iDistance possesses the best quality, with a MAP of $1$. However, it is neither
efficient nor scalable.  Table~\ref{tab:compareOverall} shows that \name
significantly outperforms all the other techniques except HNSW and OPQ in terms
of quality.  The empty bars in the figures indicate that the methods could not
be run for the corresponding datasets (e.g., in Fig.~\ref{fig:techniques}a,
Multicurves was unable to run on SUN).

\subsubsection{Efficiency}

Figs.~\ref{fig:techniques}b,~\ref{fig:techniques}g,~\ref{fig:techniques}l show a
comparison of the running times of \name with the different methods.  iDistance
possesses the highest running times and was almost as slow as the linear scan
method and is, therefore, not efficient at all.  OPQ and HNSW, being
memory-based techniques, run extremely fast.  As indicated later, this is
possible since they consume and utilize a large amount of main memory.  Compared
to purely disk-based methods that aim to scale for large datasets, these methods
enjoy an unfair advantage in terms of running time efficiency.  C2LSH is the
most efficient disk-based technique; it, however, crashed on SIFT100M.
\name is comparatively not very efficient for smaller datasets, as is clear from
Fig.~\ref{fig:techniques}b. However, its running time scales gracefully with
dataset size and it becomes increasingly efficient with larger data sizes.

\subsubsection{Scalability}

The final set of results deal with the scalability of the various techniques.
Figs.~\ref{fig:techniques}c,~\ref{fig:techniques}h,~\ref{fig:techniques}m show a
comparison of the index sizes of the different techniques.  Multicurves
possesses the largest sized index and consumes up to 1.2TB of disk-space for the
SIFT100M dataset. Owing to this, it is unable to scale to the SIFT1B dataset on
our machine, as its projected index size in this case is $\approx$12TB. HNSW,
although extremely fast and fairly accurate, requires a very large amount of
main memory.  For SIFT1M, it consumes 1.43GB of main memory.  Hence, its
projected main memory requirement for SIFT100M is beyond the 32GB capacity of
our machine.  Not surprisingly, it also crashes for this dataset.  iDistance
possesses the smallest index size.  However, in addition to index size, the
memory consumption during the index construction step is also significant.
iDistance consumes a lot of main memory (RAM) during the index construction step
as shown in Figs.~\ref{fig:techniques}d,~\ref{fig:techniques}i, and
~\ref{fig:techniques}n.  Thus, it crashed for SIFT100M during the index
construction step. C2LSH, QALSH, and OPQ also suffer from the same problem and,
thus, they also crashed on SIFT100M. 

With this, the only two \emph{scalable} techniques left are \name and SRS, with
the index size of SRS being 2-3 times smaller when compared to that of \name.
The implementation of SRS possesses a memory leak, which is evident from the
increasing RAM consumption with dataset sizes. To enable SRS to run on larger
datasets, as mentioned in Sec.~\ref{sec:comments}, the memory leak was solved
using chunking of the large datasets. Due to this chunking effect, SRS exhibits
a stable memory consumption of 2GB for Yorck and larger datasets.  The results
depict that \name possesses the smallest memory-footprint during the index
construction step, and is as low as 100MB across the different datasets.  This
establishes the fact that \name is the \emph{most scalable} technique.

\begin{table*}[t]
\centering
\scalebox{0.80}{
\begin{tabular}{l|rrrrrrr|rrrrrrr}
\hline
\multicolumn{1}{c|}{\multirow{2}{*}{\bf Dataset}} & \multicolumn{1}{c}{\bf Query} & \multicolumn{6}{c|}{\bf Gain of \name in Query Time over} & \bf MAP &
\multicolumn{6}{c}{\bf Gain of \name in MAP@100 over} \\
\cline{3-8}
\cline{10-15}
& \bf Time (ms) & \bf C2LSH & \bf SRS & \bf Multicurves & \bf QALSH & \bf OPQ & \bf HNSW & \bf @100 & \bf C2LSH & \bf SRS & \bf Multicurves & \bf QALSH & \bf OPQ & \bf HNSW \\
\hline
\textbf{SIFT10K} & 19.46 & 0.06x & 0.17x & \bf 2.88x & 0.20x & 0.10x & 0.004x  & 0.98 & \bf 2.44x & \bf 4.45x & 0.98x & \bf 1.81x &  0.99x & 0.98x   \\
\textbf{Audio} & 44.18 & 0.02x & 0.88x & \bf 2.44x & 0.21x & 0.02x & 0.0007x  & 0.86 & \bf 14.33x & \bf 6.61x & \bf 3.05x & \bf 1.28x &  0.98x & 0.99x   \\
\textbf{SUN} & 105.78 & 0.12x & 0.22x & \bf NP & 0.15x & 0.02x & 0.007x  & 0.69 & \bf 3.83x & \bf 23.00x & \bf NP & \bf 1.72x &  1.00x & 0.88x   \\
\textbf{SIFT1M} & 25.10 & \bf 5.30x & \bf 1.56x & \bf 22.98x & \bf 11.27x & 0.05x & 0.002x  & 0.56 & \bf 2.80x & \bf 28.00x & 0.97x & \bf 1.19x &  1.00x & 0.92x   \\
\textbf{Yorck} & 262.29 & 0.27x & \bf 27.56x & \bf 2.21x & \bf 33.54x & 0.01x & 0.002x  & 0.39 & \bf 1542.51x & \bf 39.18x & \bf 1.24x & \bf 1.01x  &  1.00x & \bf 1.02x  \\
\textbf{SIFT100M} & 732.04 & \bf CR & \bf 2.17x & \bf 1.73x & \bf CR  & \bf CR & \bf CR  & 0.40 & \bf CR & \bf 75.72x & \bf 1.13x & \bf CR  & \bf CR & \bf CR  \\
\textbf{SIFT1B} & 4855.20 & \bf CR & \bf DNF & \bf CR & \bf CR  & \bf CR &  \bf CR  & 0.25 & \bf CR & \bf DNF & \bf CR & \bf CR  & \bf CR & \bf CR  \\
\textbf{Enron} & 242.69 & 0.02x & 0.06x & \bf NP & \bf NP & 0.06x & 0.002x  & 0.92 & \bf 5.12x & \bf 12.07x & \bf NP &  \bf NP &  0.99x & 0.99x  \\
\textbf{Glove} & 85.20 & 0.75x & 0.06x & \bf 2.6x & \bf 2.28x & 0.05x & 0.0003x  & 0.24 & \bf 8.87x & \bf 80.00x & \bf 1.26x & \bf 1.43x &  0.99x & 0.31x  \\
\hline
\end{tabular}
}
\tabcaption{{\bf Comparison of \emph{\name} with other techniques. DNF
indicates that index construction did not terminate even after running for 20
times the duration of the slowest among the other techniques. CR indicates that
index construction crashed due to running out of resources. NP indicates that
index construction is not at all possible due to an inherent limitation of the
technique.}}
\label{tab:compareOverall}
\vspace*{1mm}
\end{table*}

We also analyze the main memory consumption during the querying stage.
Figs.~\ref{fig:techniques}e,~\ref{fig:techniques}j,~\ref{fig:techniques}o show
that apart from \name, Multicurves and QALSH (each consuming less than 40MB),
all the other techniques consume a significant amount of RAM.  Utilizing a
large amount of RAM during the querying phase puts these methods at an unfair
advantageous position vis-\`a-vis the strictly disk-based methods when running
times are concerned.

\subsubsection{Billion Scale Dataset}

We could not compare the results for the largest SIFT1B dataset with any other
technique, as none of them were able to run on it. Since C2LSH, iDistance,
QALSH, OPQ and HNSW crashed on SIFT100M (we use a commodity machine with 32GB
of RAM and 2TB of HDD), they could not be run.  Multicurves consumed 1.2TB of
disk space on SIFT100M and, hence, could not scale to the 10 times larger
SIFT1B dataset on our machine. Although the memory leak in the index
construction step of SRS was solved using chunking of the datasets, even after
running for more than 25 days, it was unable to complete the index construction
for SIFT1B.

\name required around 10 days and 1.2TB of space to construct the index.
Querying produced a MAP@100 of 0.25 in 4.8s per query by consuming 30MB of main
memory.

\subsection{Real-World Application: Image Search}
\label{subsec:image_search}

We finally highlight the importance of kANN searches and the MAP measure
through an actual image retrieval application.  We use the Yorck
dataset \cite{yorck} that consists of SURF descriptors from 10,000 art images.
We mention only the quality results here and do not repeat the running times.

All the descriptors from the query image are searched for $100$ nearest
neighbors.  The retrieved results, i.e., the database descriptors are then
aggregated using the Borda count \cite{borda}.  In Borda count, a score is
given to each image $i$ based on their depth in the kANN result of each of the
$Q$ descriptors in the query image $q$ (details are in
\ifarxiv
Appendix~\ref{app:borda}).
\else
Appendix~D in the extended version \cite{full}).
\fi
The $k$ images with the \emph{highest} Borda counts are retrieved as the top-$k$
\emph{similar} images to the query image.
The top-$k$ image results for a representative query for the different methods are
shown in
\ifarxiv
Appendix~\ref{app:image}.
\else
Appendix~E in the extended version \cite{full}.
\fi

In general, \name, QALSH, OPQ and HNSW possess the highest overlap with the
ground truth produced by the linear scan method.  While linear scan has
impractical running times, \name, OPQ, HNSW and C2LSH have better running times
than SRS and QALSH. However, C2LSH has a poor image retrieval quality, while SRS
exhibits moderate quality.

\name is faster than QALSH but slower than both OPQ and HNSW.  However, it has a
much smaller memory footprint when compared to OPQ and HNSW while both indexing
and querying.

\subsection{Summary}
\label{sec:expsummary}

\begin{figure}[t]
	\centering	
	\vspace*{-2mm}
	\includegraphics[width=0.80\linewidth]{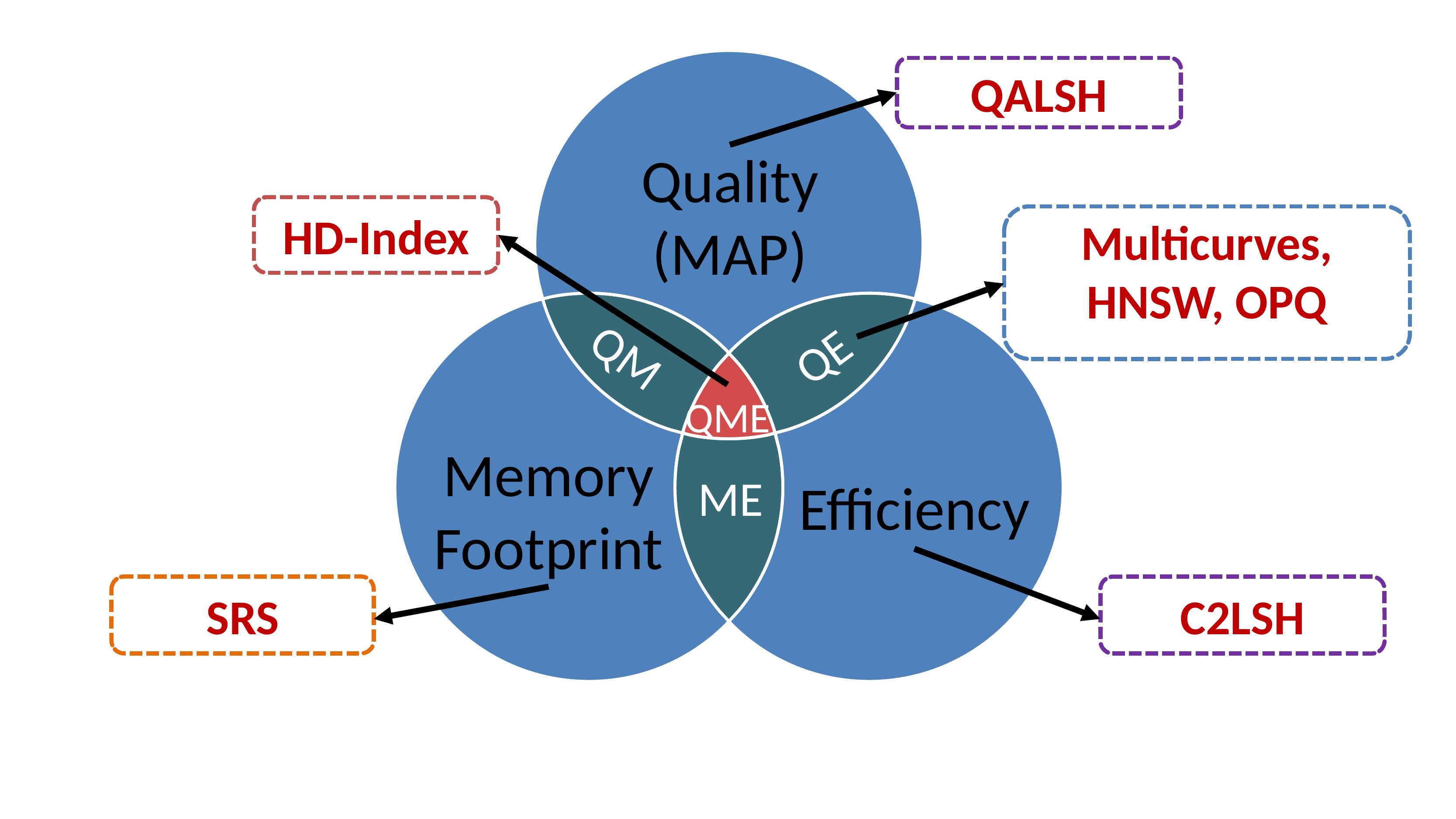}
	\vspace*{-5mm}
	\moveups
	\figcaption{\textbf{Summarizing the spectrum of approximate nearest neighbor search techniques.}}
	\label{fig:summary}
	\moveups
\end{figure}

A good indexing technique for approximate nearest neighbor search stands on
three pillars: \emph{quality} of retrieved objects, running time
\emph{efficiency}, and \emph{memory} footprint (both external memory while
storing the index and main memory while querying).  Our experimental analyses
compared \name with $6$ state-of-the-art techniques across all of these
features.  A quantitative summary of the results is presented in
Table~\ref{tab:compareOverall} while a broad qualitative summary is depicted in
Fig.~\ref{fig:summary}.

QALSH \cite{qalsh}, SRS \cite{srs} and C2LSH \cite{c2lsh} are the
state-of-the-art techniques with respect to quality, memory footprint and
running time efficiency respectively. However, just optimizing one category is
not sufficient. For instance, it is useless to produce spurious results
extremely fast with a high memory overhead.  Therefore, for a technique to be
useful, it should be classified at the intersection of at least two classes,
that too preferably ``QE'' or ``QM'', since poor quality results is of no
importance.  If quality is sacrificed, even a technique producing random
results without any computation would fall in the class ``ME''.  To this end,
Multicurves \cite{valle}, HNSW \cite{hnsw} and OPQ \cite{opq} (classified as
``QE'') are useful; however, they possess several disadvantages. First, owing
to its large index space requirements, Multicurves finds difficulty in scaling
to massive datasets.  Next, due to its inherent design, it cannot scale to
extremely high-dimensional datasets such as the $512$-dimensional SUN dataset.
HNSW and OPQ, being memory-based techniques, are quite fast and fairly
accurate; however, their memory requirements are impractically large.  They,
therefore, fail to run on large datasets.  Thus, as verified empirically, none
of these methods could run on the billion-scale dataset.

In sum, \name (classified as ``QME'') stands strong on all the three pillars,
and is an \emph{effective}, \emph{efficient} and \emph{scalable} index
structure which scales gracefully to \emph{massive high-dimensional} datasets
even on commodity hardware.

\section{Conclusions and Future Work}
\label{sec:conc}

In this paper, we addressed the problem of approximate nearest neighbor
searching in massive high-dimensional datasets within practical compute times
while simultaneously exhibiting high quality.  We designed an efficient
indexing scheme, \name, based on a novel index structure, \tree.  Experiments
portrayed the effectiveness, efficiency and scalability of \name.

In future, we would like to attempt a parallel implementation to cater to
higher dimensional datasets.  Also, we would like to analyze the effect of
parameters from a theoretical point of view.

\bibliographystyle{abbrv}
\balance
\bibliography{papers}

\pagebreak

\appendix

\section{Reference Object Selection}
\label{app:reference}

\begin{figure}[t]
\moveup
\moveups
\centering	
	\subfloat[Time]
	{
		\scalebox{0.48}{
			\includegraphics[width=0.82\linewidth]{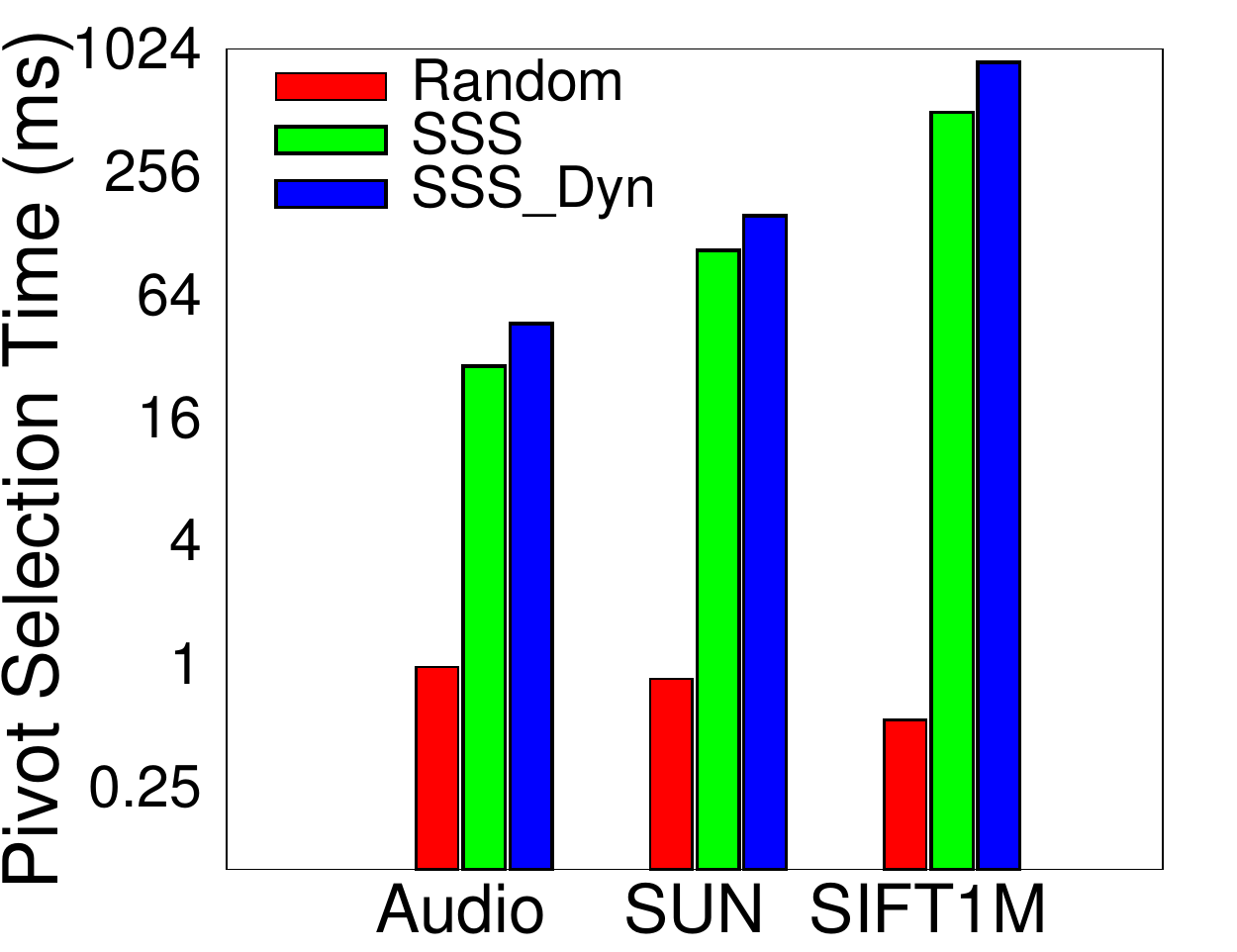}
		}
		\label{fig:ref_algo_time}
	}
	\subfloat[MAP@k]
	{
		\scalebox{0.48}{
			\includegraphics[width=0.82\linewidth]{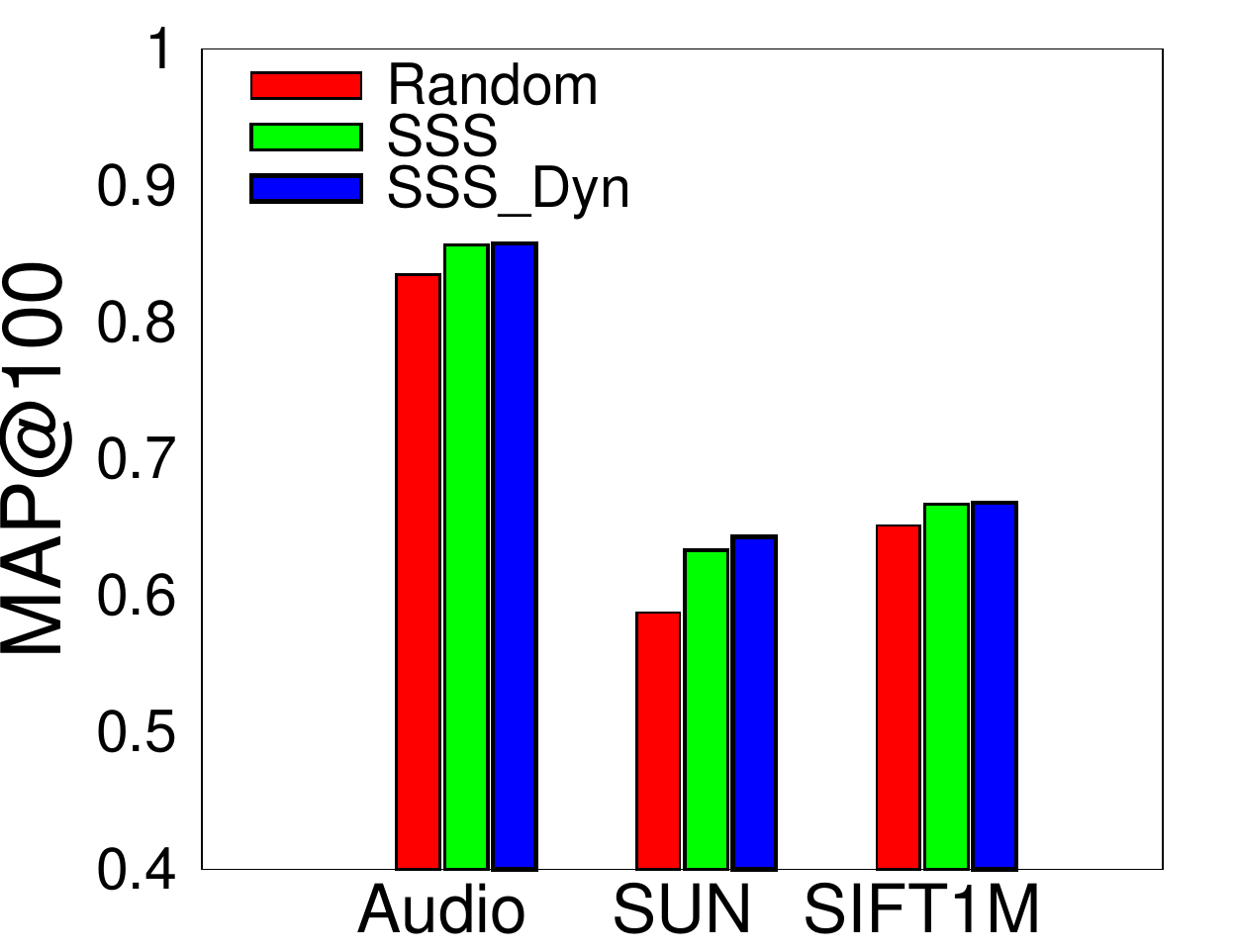}
		}
		\label{fig:ref_algo_map}
	}
\figcaption{\textbf{Comparison of different reference object selection algorithms.}
}
\label{fig:ref_algo}
\vspace*{1mm}
\end{figure}

The comparison of different reference object selection methods is shown in
Fig.~\ref{fig:ref_algo}.  The method selected is \pivot.

\section{Ptolemaic Inequality}
\label{app:ptolemaic}

\begin{figure*}[t]
\centering
	\subfloat[Query Time (SIFT10K)]
	{
		\scalebox{0.24}{
			\includegraphics[width = \filterwidth]{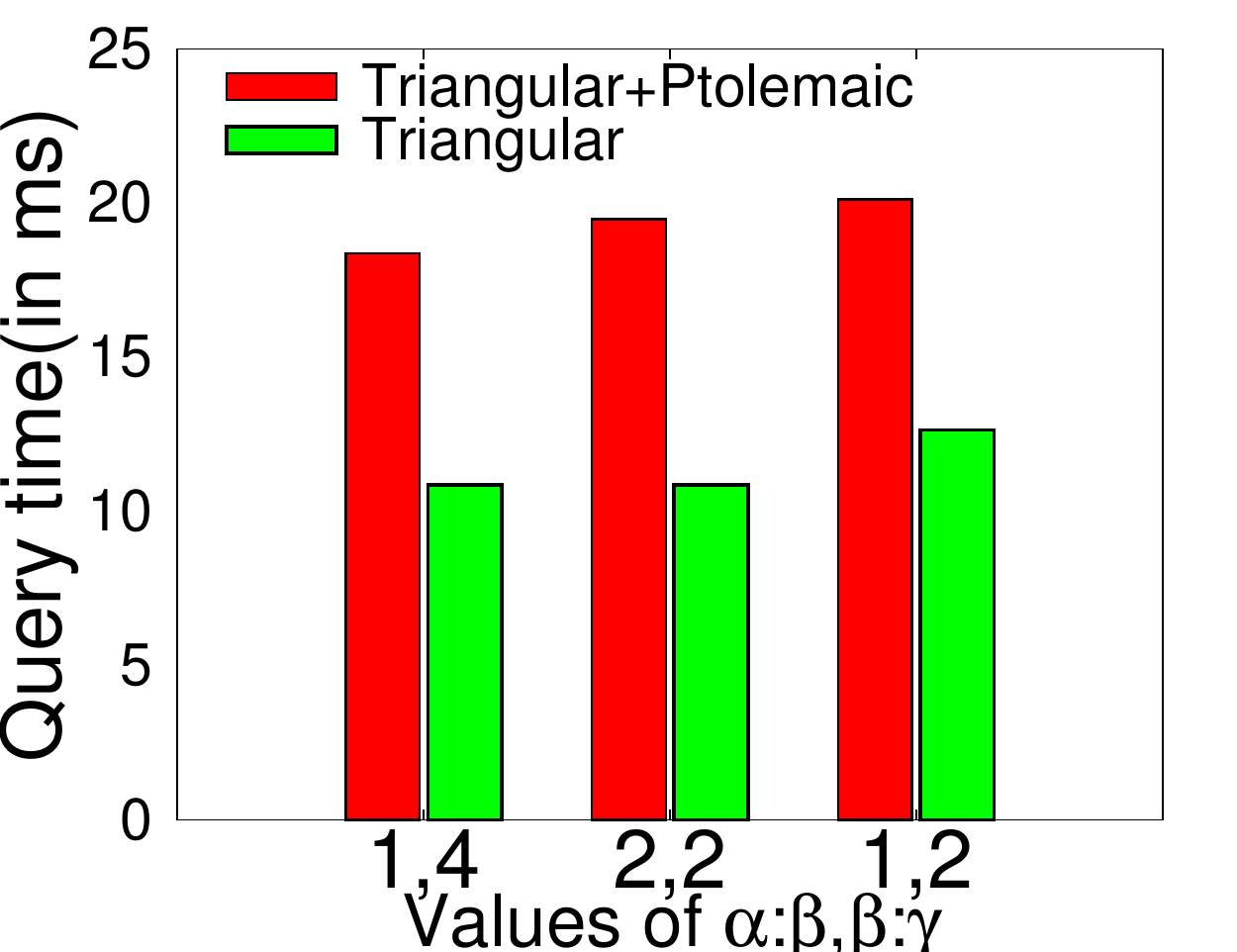}
		}
		\label{fig:ptolemaic2_time1}
	}
	\subfloat[Query Time (AUDIO)]
	{
		\scalebox{0.24}{
			\includegraphics[width = \filterwidth]{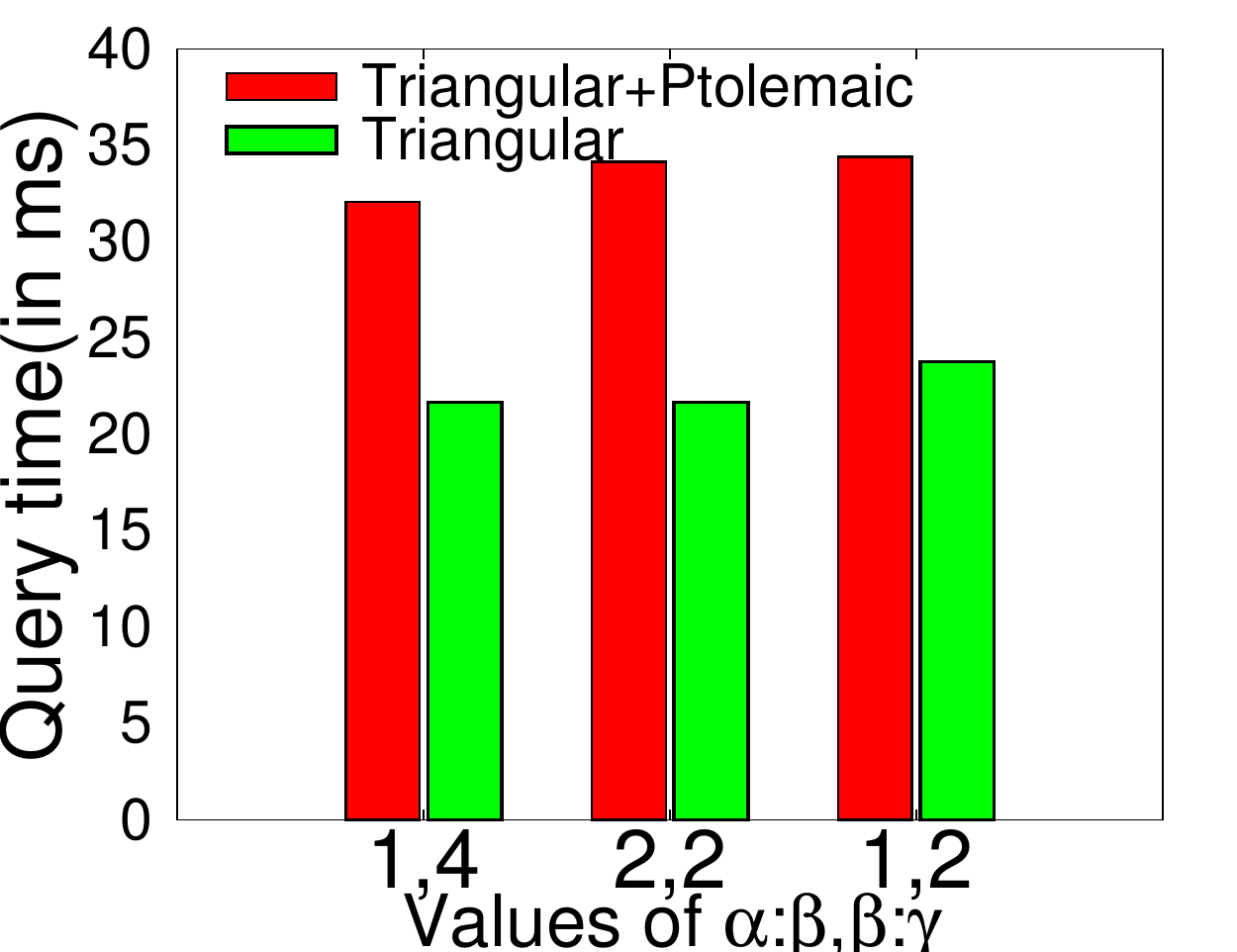}
		}
		\label{fig:ptolemaic2_time2}
	}
	\subfloat[Query Time (SUN)]
	{
		\scalebox{0.24}{
			\includegraphics[width = \filterwidth]{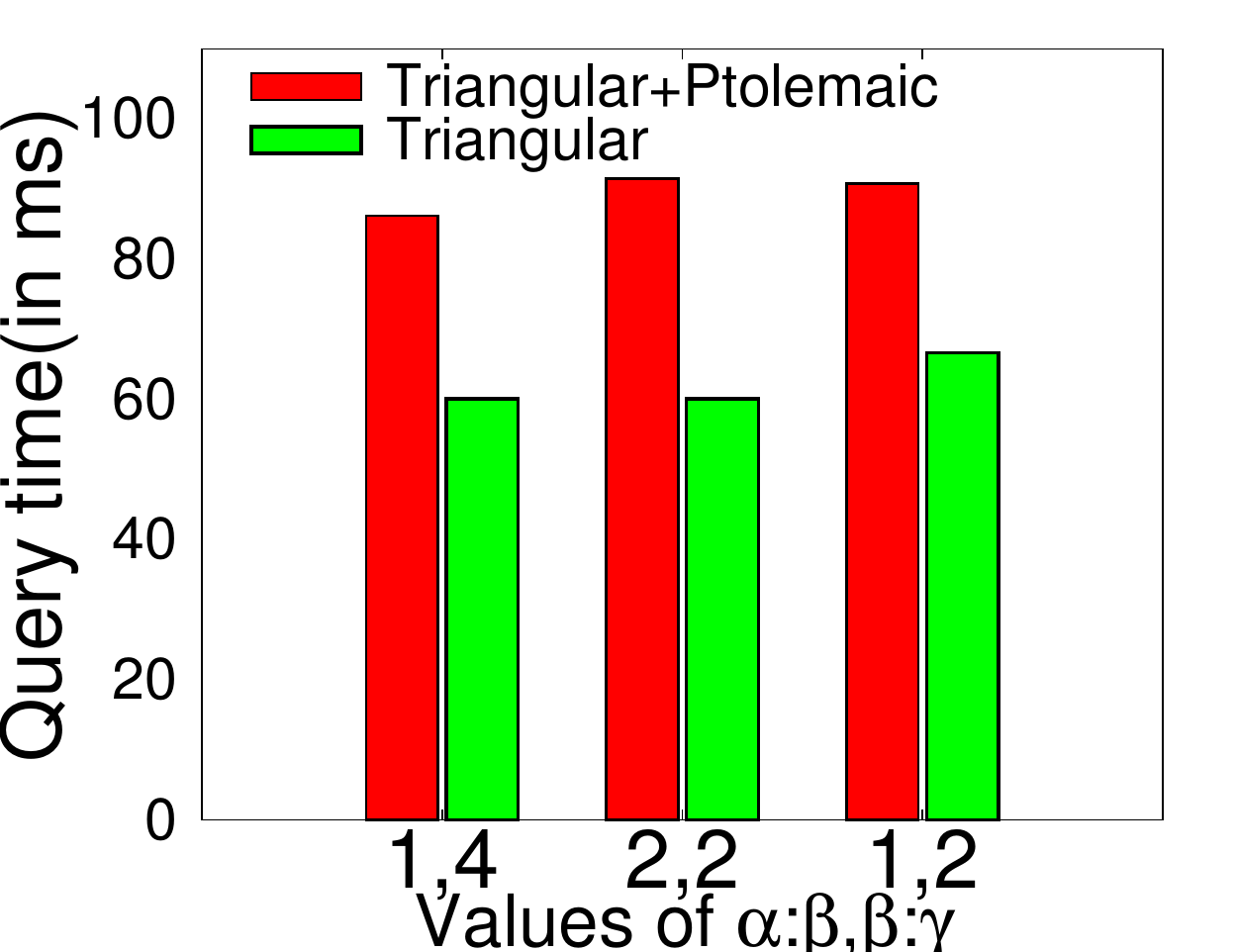}
		}
		\label{fig:ptolemaic2_time3}
	}
	\subfloat[Query Time (SIFT1M)]
	{
		\scalebox{0.24}{
			\includegraphics[width=\filterwidth]{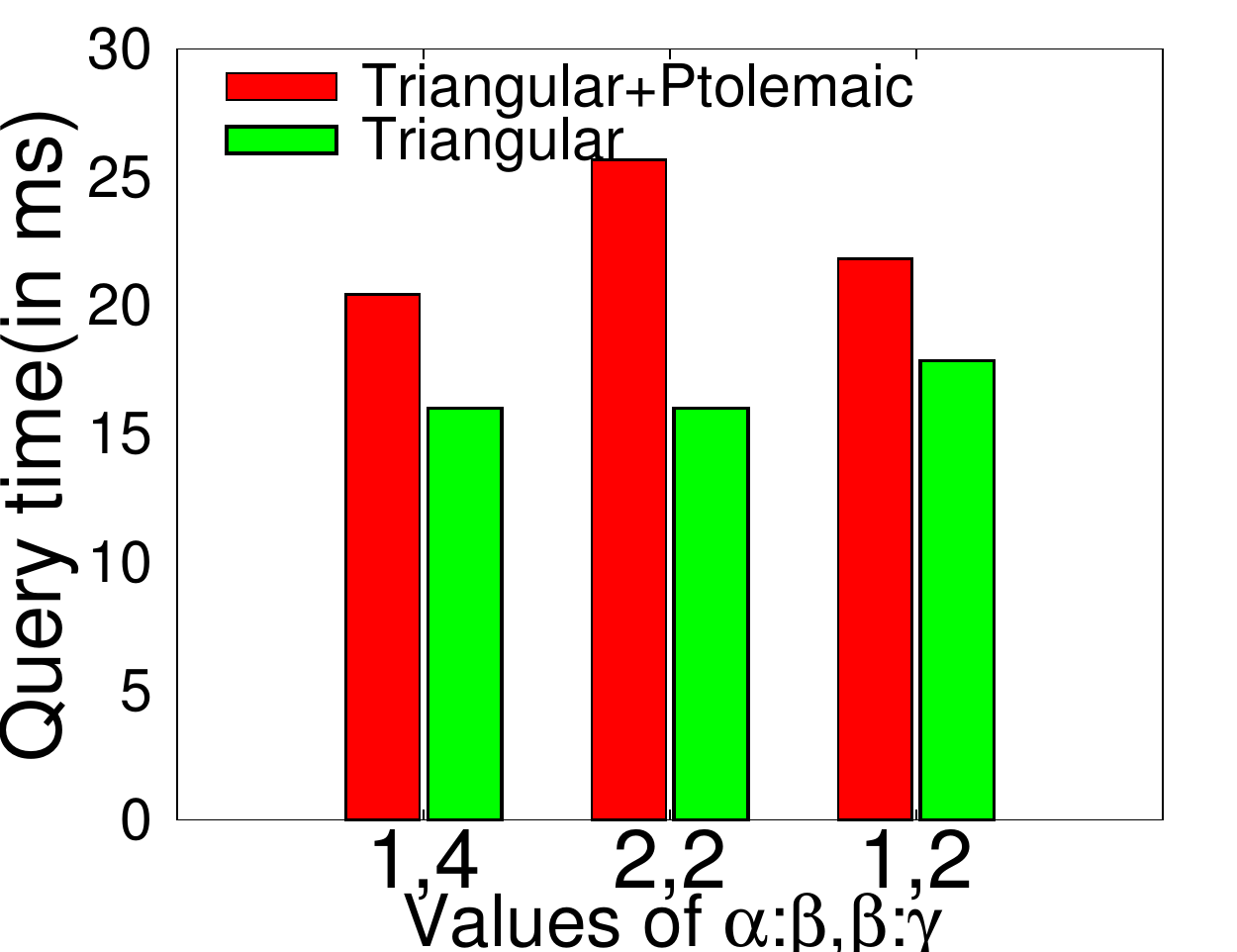}
		}
		\label{fig:ptolemaic2_time4}
	}
	\\
	\vspace{-3.5mm}	
	\subfloat[MAP@10 (SIFT10K)]
	{
		\scalebox{0.24}{
			\includegraphics[width = \filterwidth]{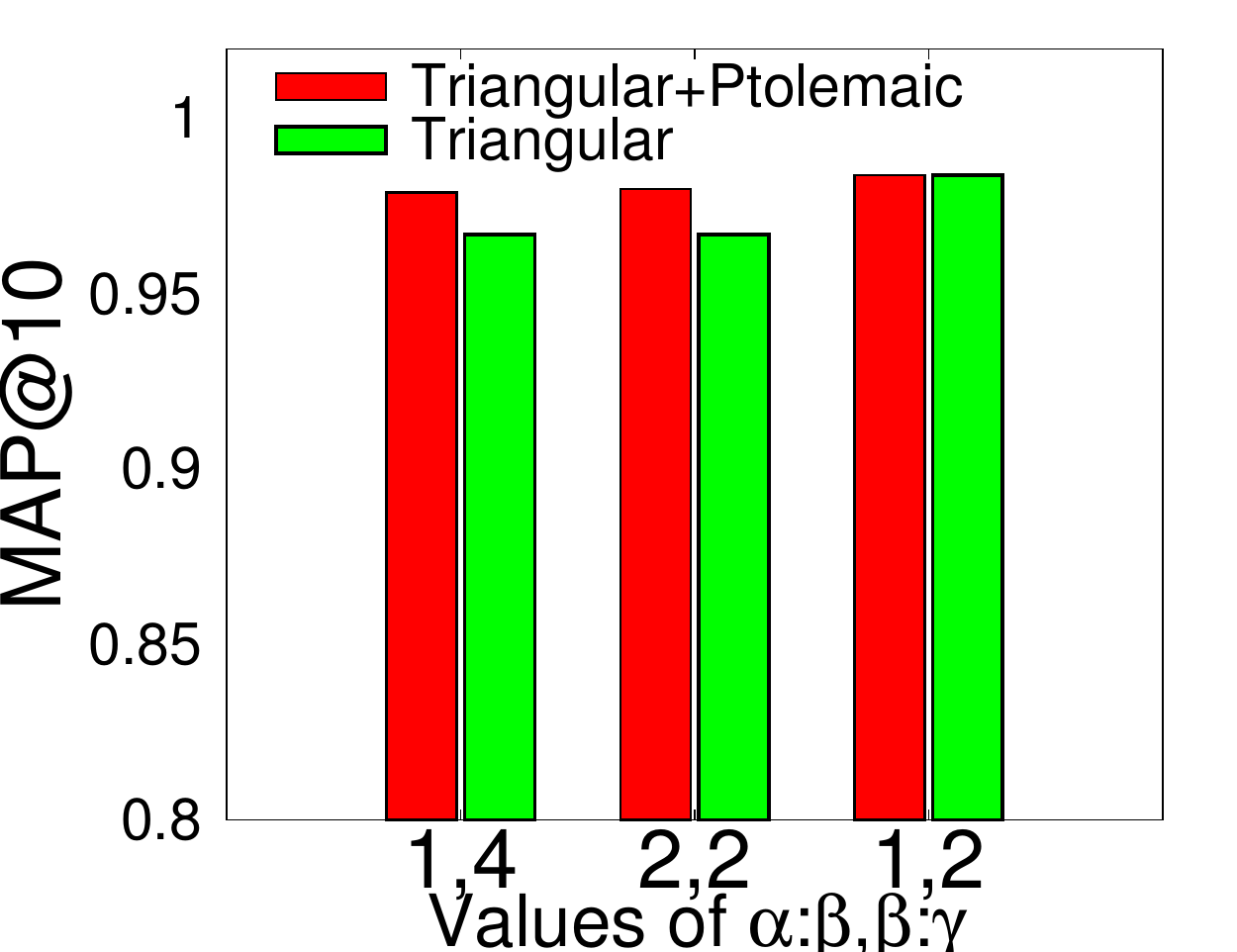}
		}
		\label{fig:ptolemaic2_map1}
	}
	\subfloat[MAP@10 (AUDIO)]
	{
		\scalebox{0.24}{
			\includegraphics[width = \filterwidth]{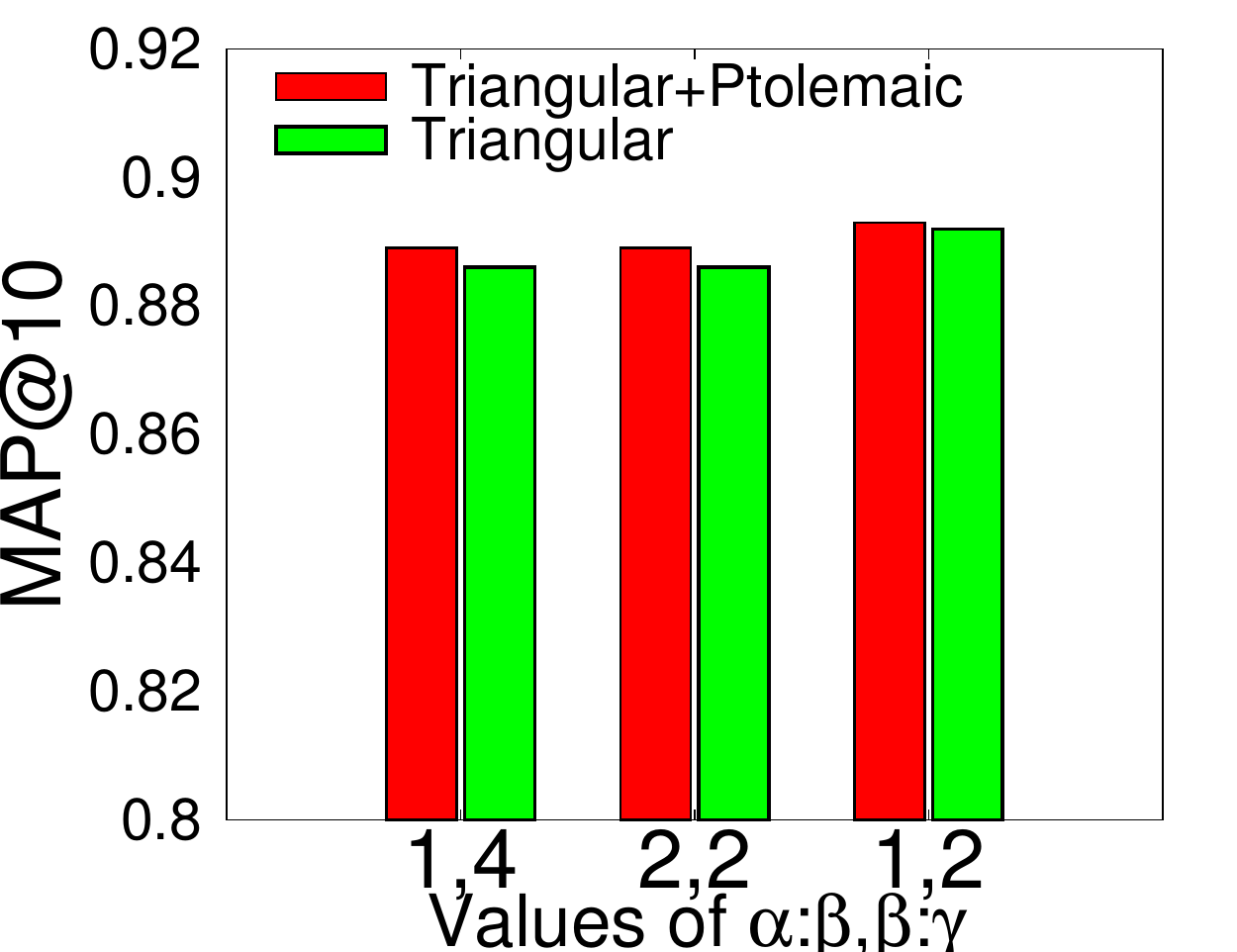}
		}
		\label{fig:ptolemaic2_map2}
	}
	\subfloat[MAP@10 (SUN)]
	{
		\scalebox{0.24}{
			\includegraphics[width = \filterwidth]{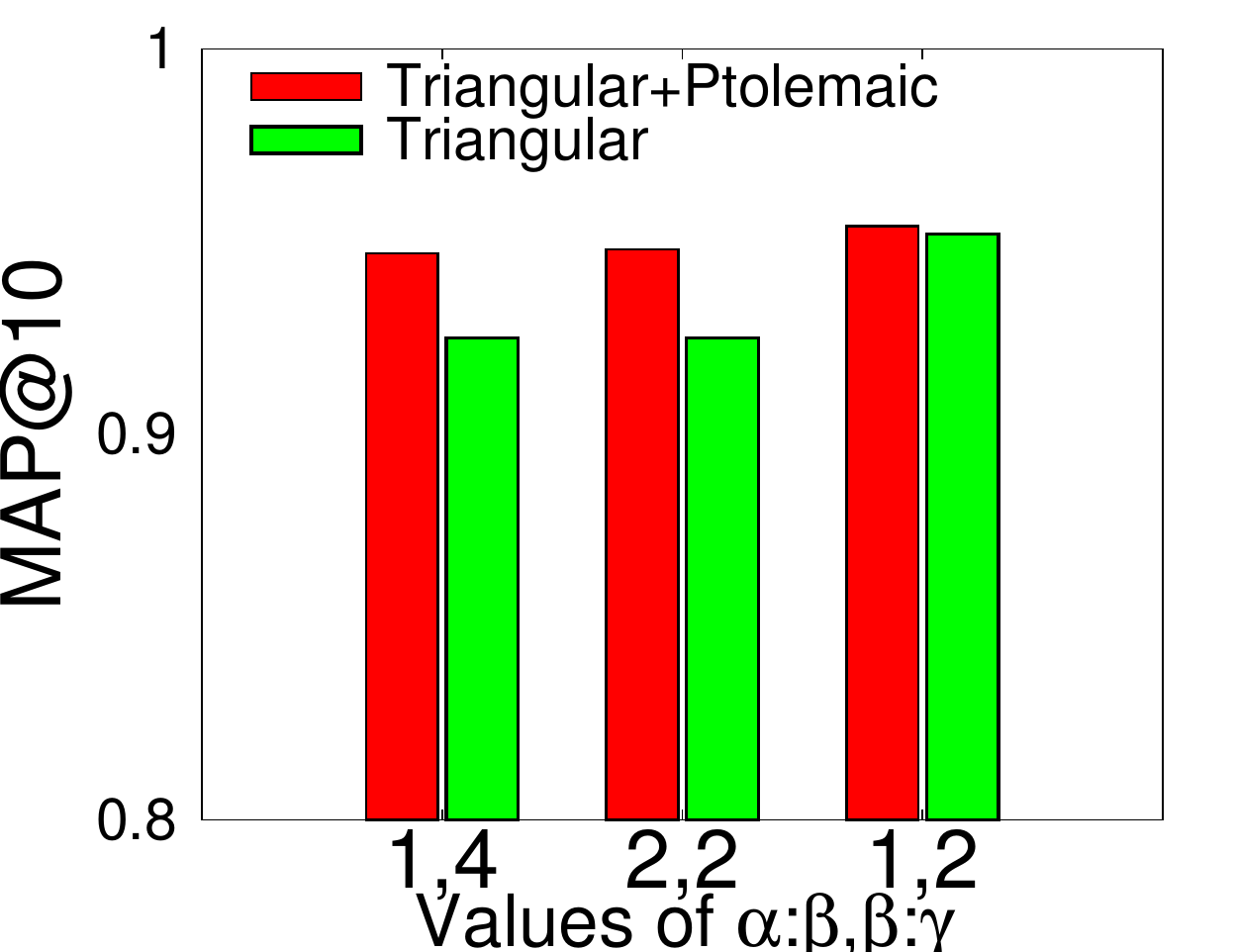}
		}
		\label{fig:ptolemaic2_map3}
	}
	\subfloat[MAP@10 (SIFT1M)]
	{
		\scalebox{0.24}{
			\includegraphics[width=\filterwidth]{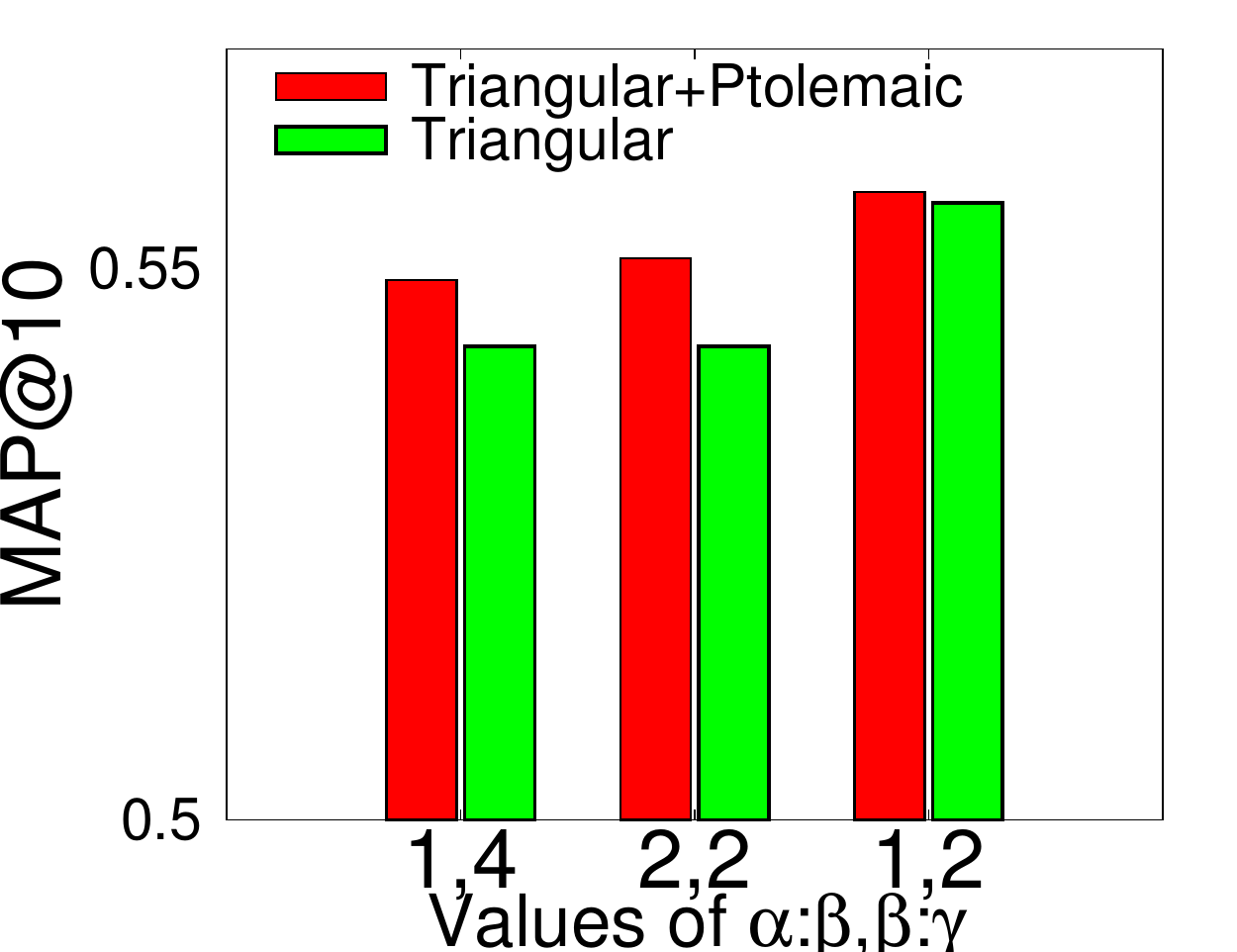}
		}
		\label{fig:ptolemaic2_map4}
	}
	\figcaption{\textbf{Effect of the filtering mechanism (either triangular
	inequality alone or a combination of triangular and Ptolemaic inequalities)
on query time (a-d) and MAP@10 (e-h) with $\alpha=2048$.}
}
\label{fig:ptolemaic2}
\moveups
\end{figure*}

\begin{figure*}[t]
\moveup
\centering
	\subfloat[Query Time (SIFT10K)]
	{
		\scalebox{0.24}{
			\includegraphics[width = \filterwidth]{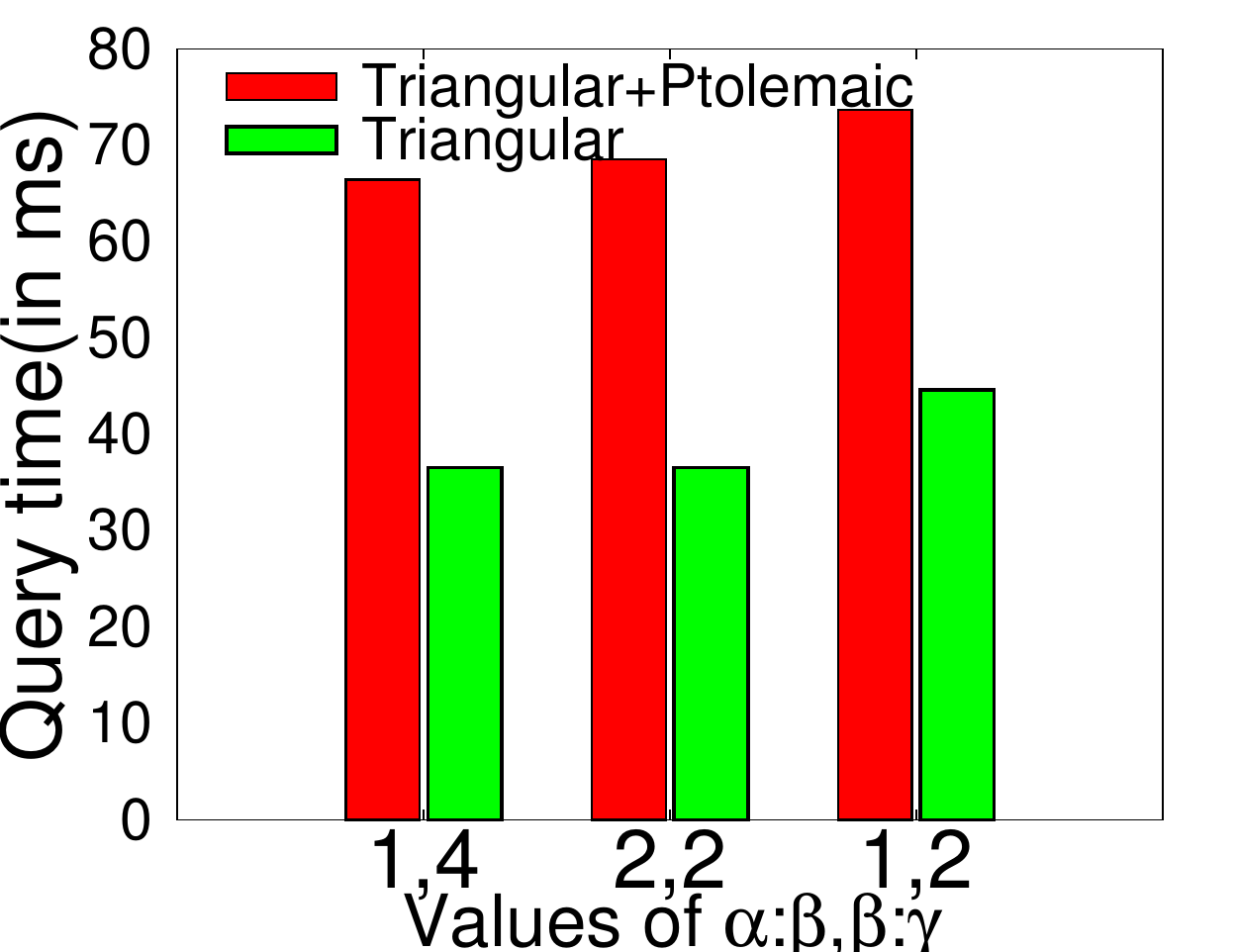}
		}
		\label{fig:ptolemaic3_time1}
	}
	\subfloat[Query Time (AUDIO)]
	{
		\scalebox{0.24}{
			\includegraphics[width = \filterwidth]{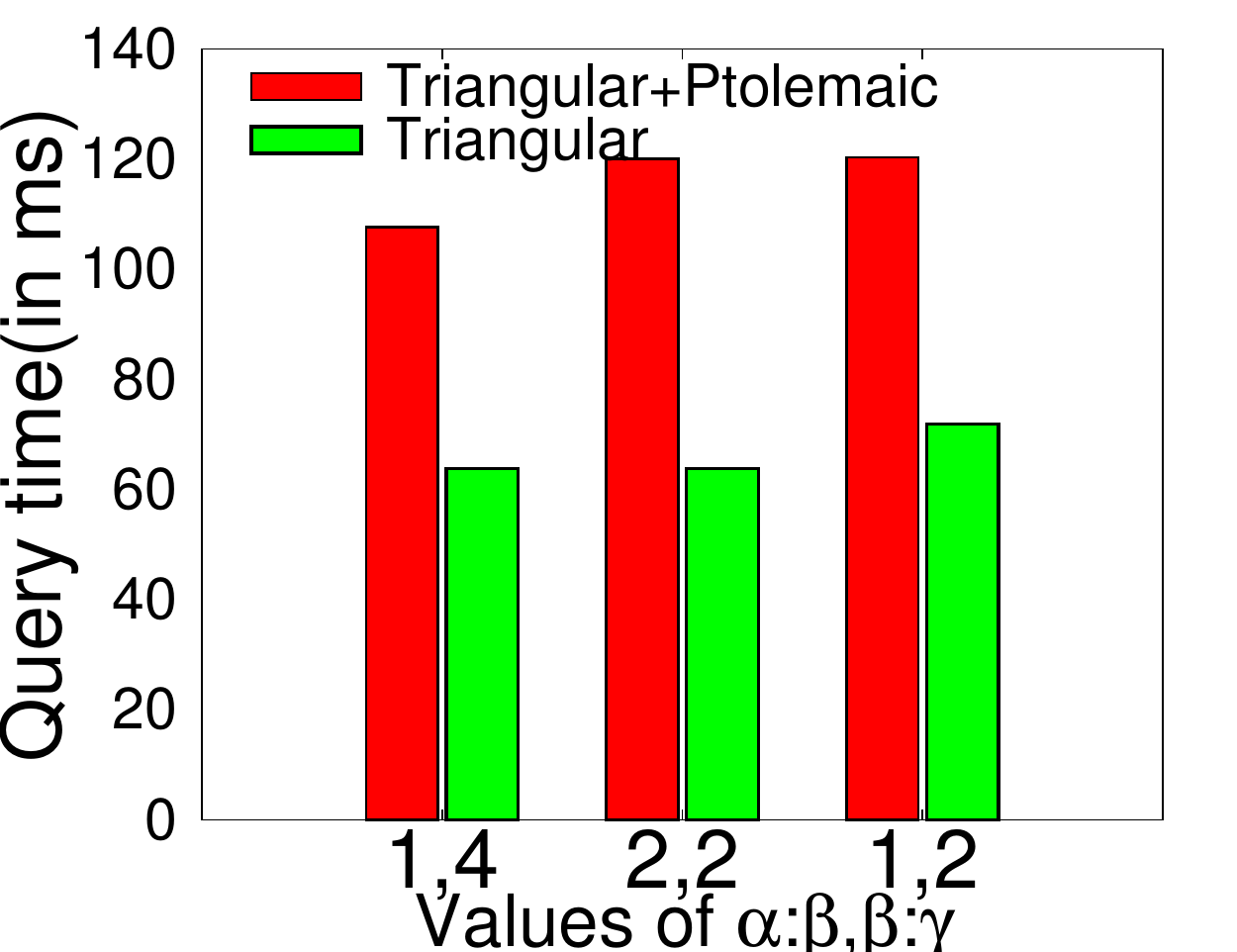}
		}
		\label{fig:ptolemaic3_time2}
	}
	\subfloat[Query Time (SUN)]
	{
		\scalebox{0.24}{
			\includegraphics[width = \filterwidth]{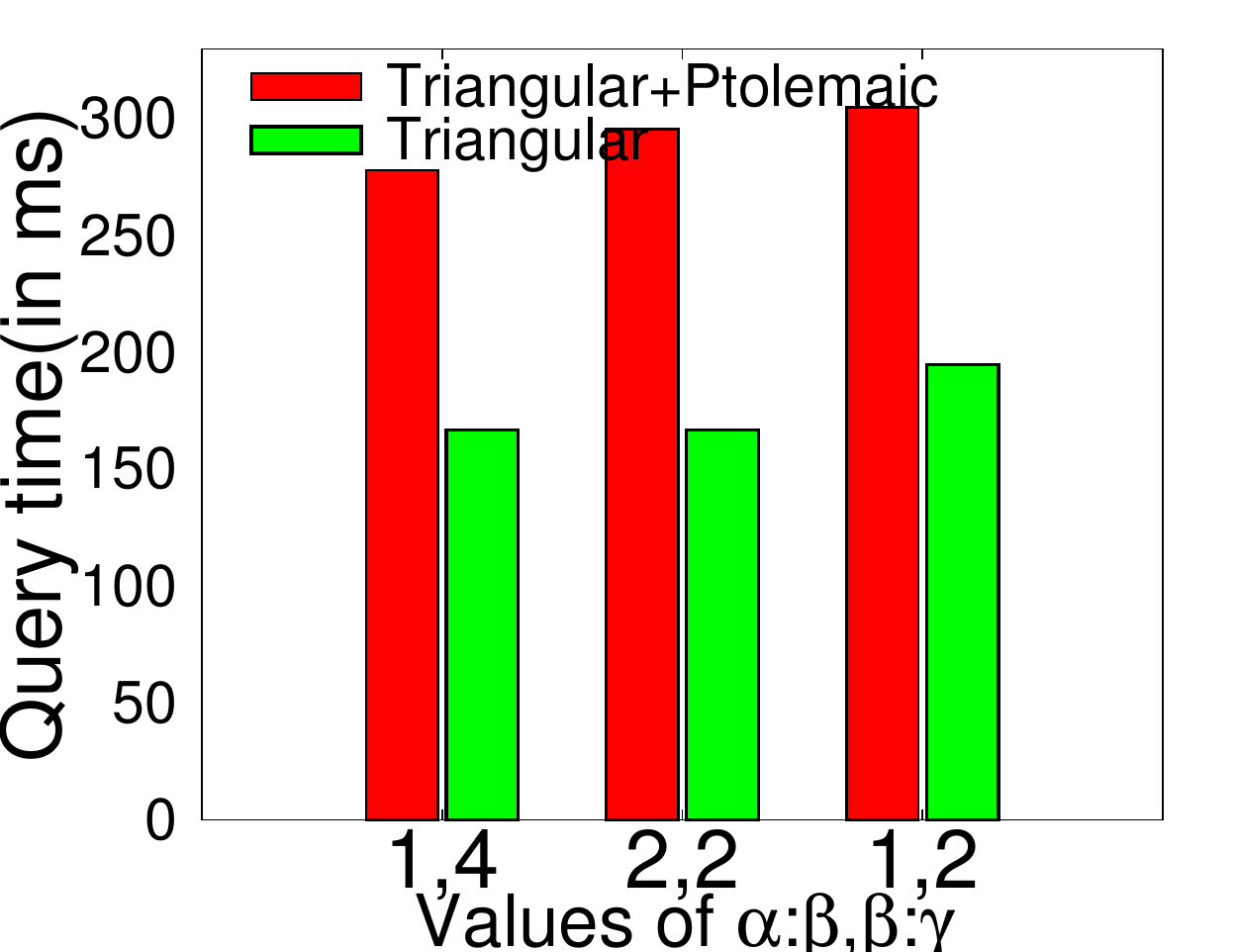}
		}
		\label{fig:ptolemaic3_time3}
	}
	\subfloat[Query Time (SIFT1M)]
	{
		\scalebox{0.24}{
			\includegraphics[width=\filterwidth]{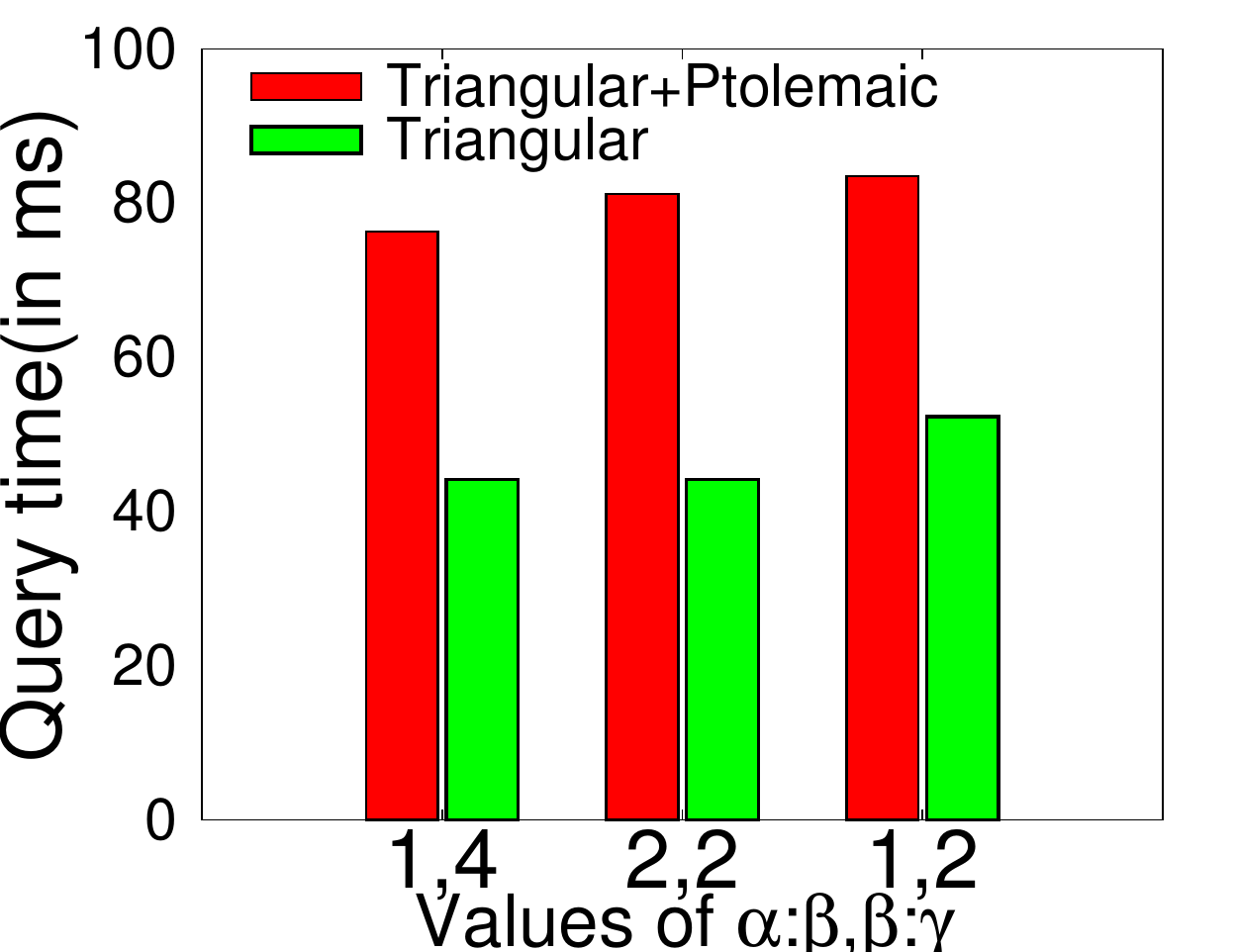}
		}
		\label{fig:ptolemaic3_time4}
	}
	\\
	\vspace{-3.5mm}	
	\subfloat[MAP@10 (SIFT10K)]
	{
		\scalebox{0.24}{
			\includegraphics[width = \filterwidth]{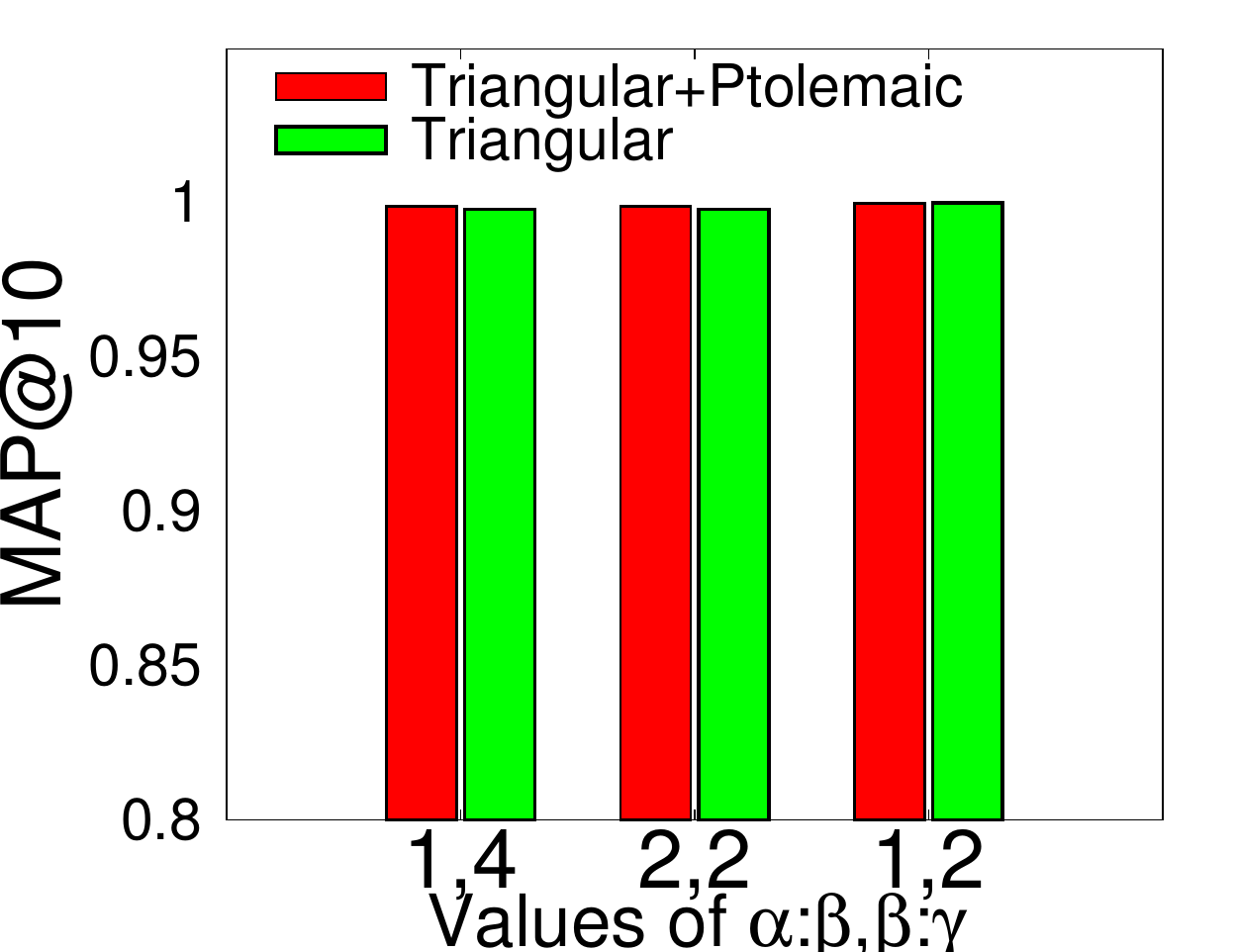}
		}
		\label{fig:ptolemaic3_map1}
	}
	\subfloat[MAP@10 (AUDIO)]
	{
		\scalebox{0.24}{
			\includegraphics[width = \filterwidth]{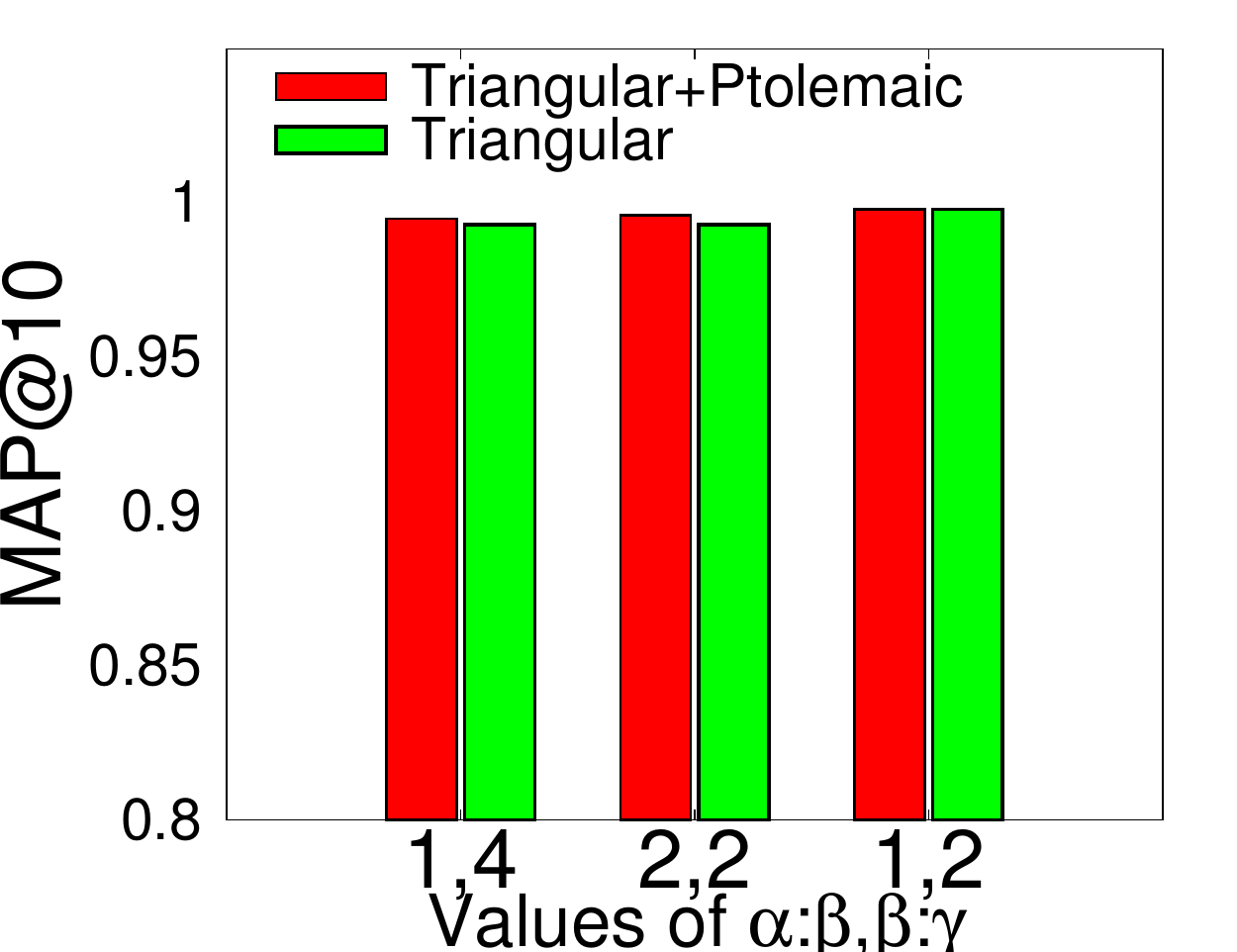}
		}
		\label{fig:ptolemaic3_map2}
	}
	\subfloat[MAP@10 (SUN)]
	{
		\scalebox{0.24}{
			\includegraphics[width = \filterwidth]{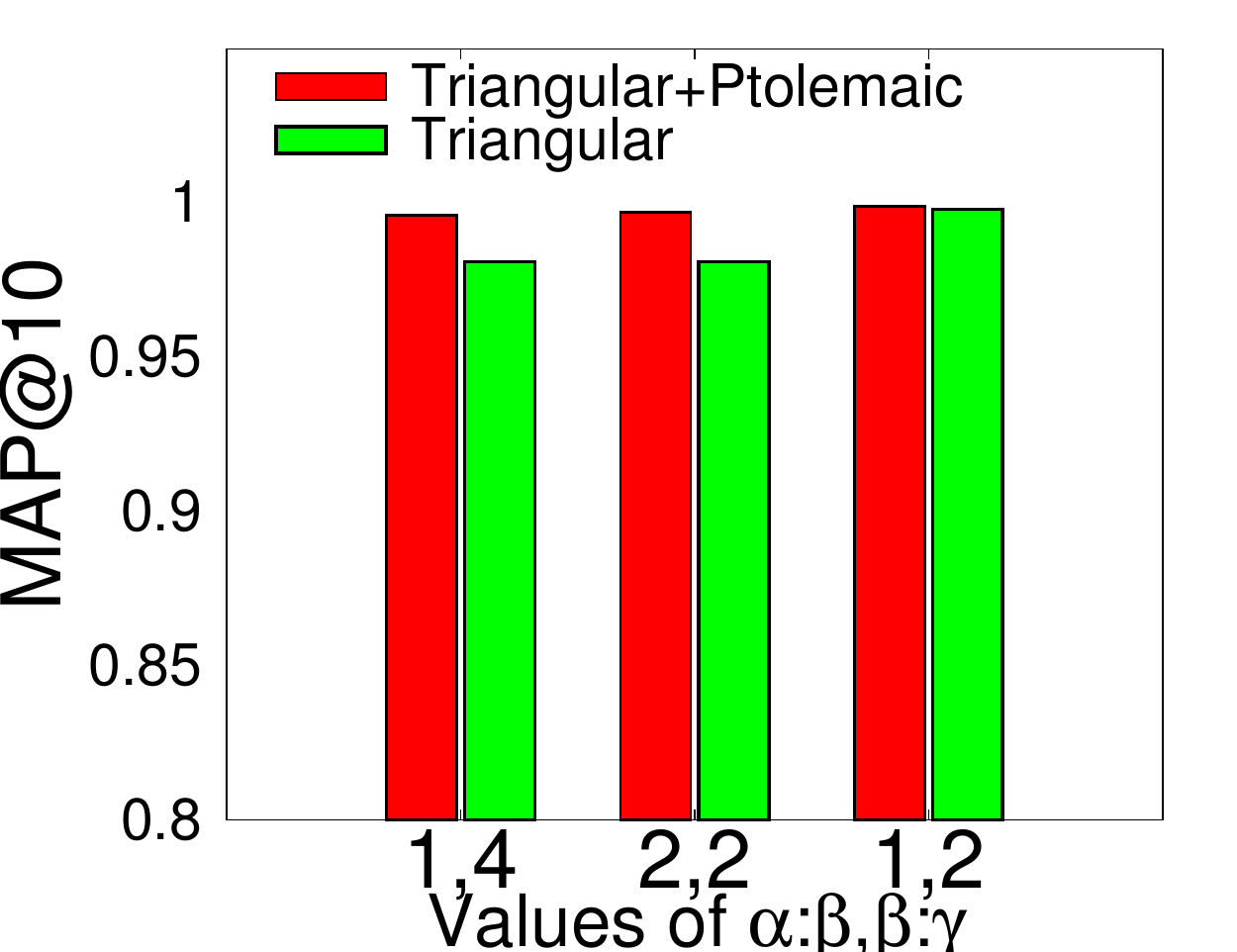}
		}
		\label{fig:ptolemaic3_map3}
	}
	\subfloat[MAP@10 (SIFT1M)]
	{
		\scalebox{0.24}{
			\includegraphics[width=\filterwidth]{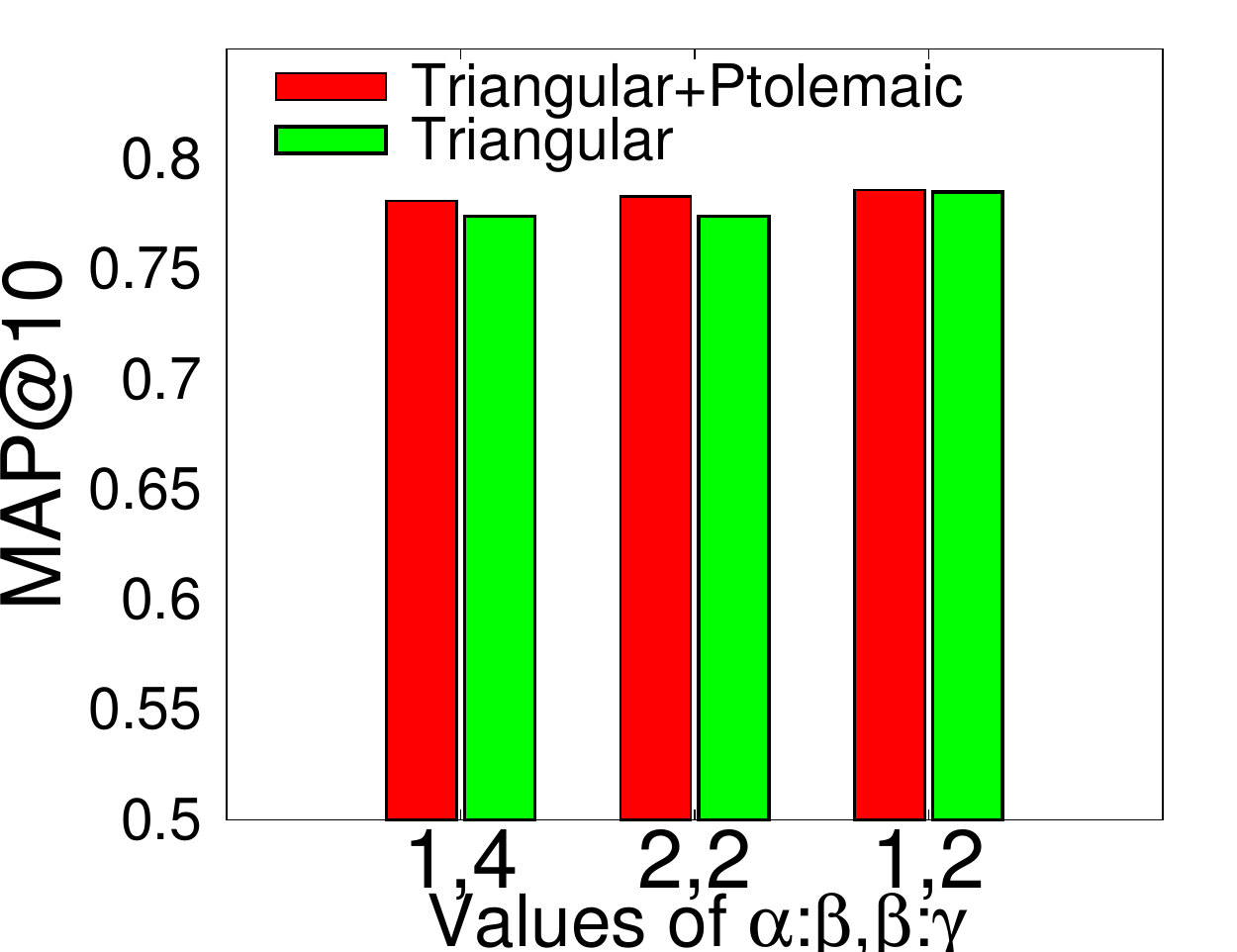}
		}
		\label{fig:ptolemaic3_map4}
	}
	\figcaption{\textbf{Effect of the filtering mechanism (either triangular
	inequality alone or a combination of triangular and Ptolemaic inequalities)
on query time (a-d) and MAP@10 (e-h) with $\alpha=8192$.}
}
\label{fig:ptolemaic3}
\moveups
\end{figure*}

Fig.~\ref{fig:ptolemaic2}  and Fig.~\ref{fig:ptolemaic3} describe the detailed
results of using the triangular inequality alone versus a combination of
triangular and Ptolemaic inequalities.  For a fixed value of $\alpha$ and the
same amount of reduction through the filtering process, using only the
triangular inequality provides the fastest response times, being $1.5$-$2$
times faster than when both the inequalities are used.  However, it is clear
from the highlighted entries in the 5$^\text{th}$ column that usually the best
MAP@10 is obtained when the Ptolemaic inequality is employed either alone or in
conjunction with the triangular inequality. Since Ptolemaic inequality provides
tighter bounds, the improvements in MAP values are prominent where the amount of
reduction using the filters is large.

\section{Number of Nearest Neighbors}
\label{app:k}

\begin{figure*}[t]
\centering
	\subfloat[SIFT10K]
	{
		\scalebox{0.19}{
			\includegraphics[width=0.99\linewidth]{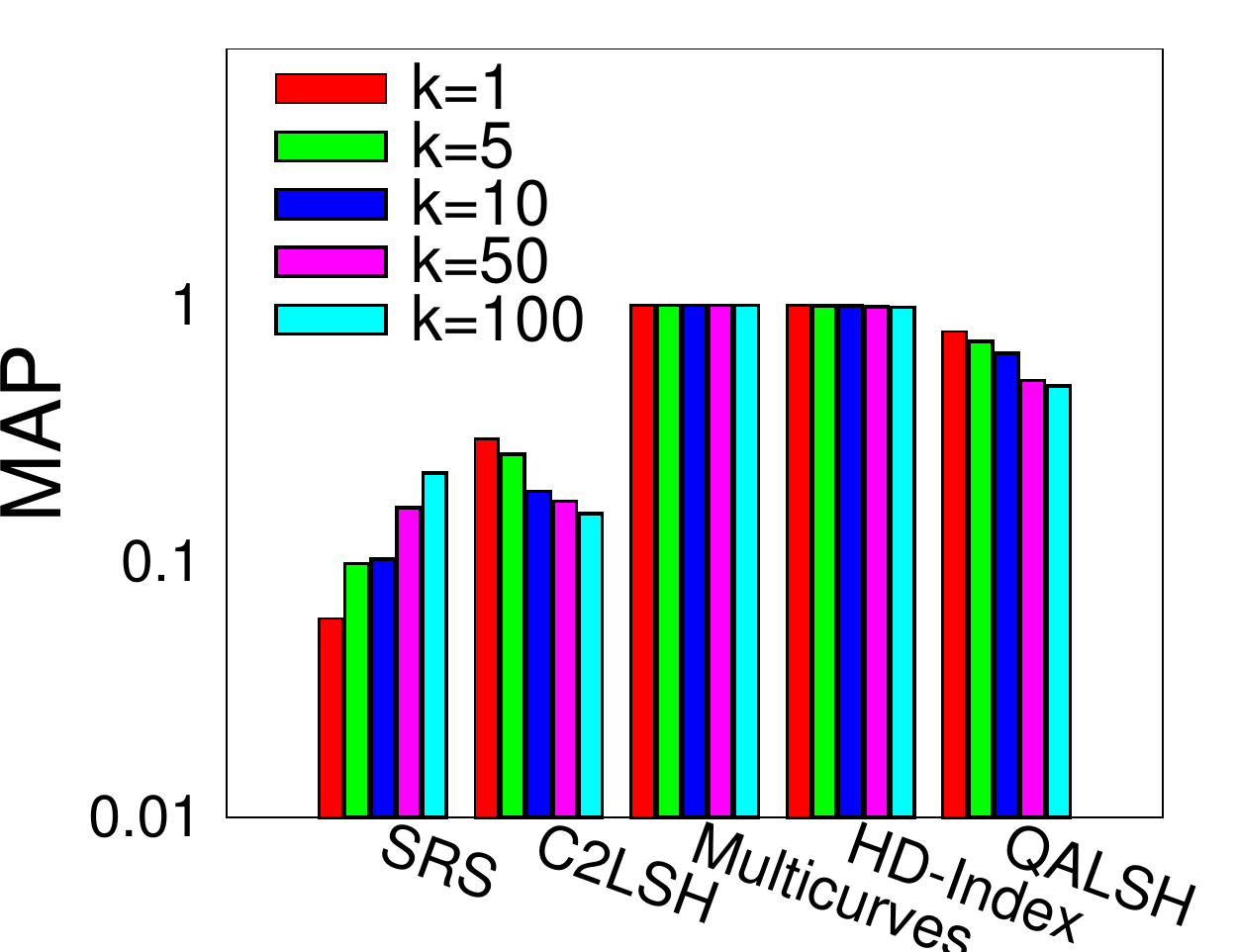}
		}
		\label{fig:map_k1}
	}
	\subfloat[Audio]
	{
		\scalebox{0.19}{
			\includegraphics[width=0.99\linewidth]{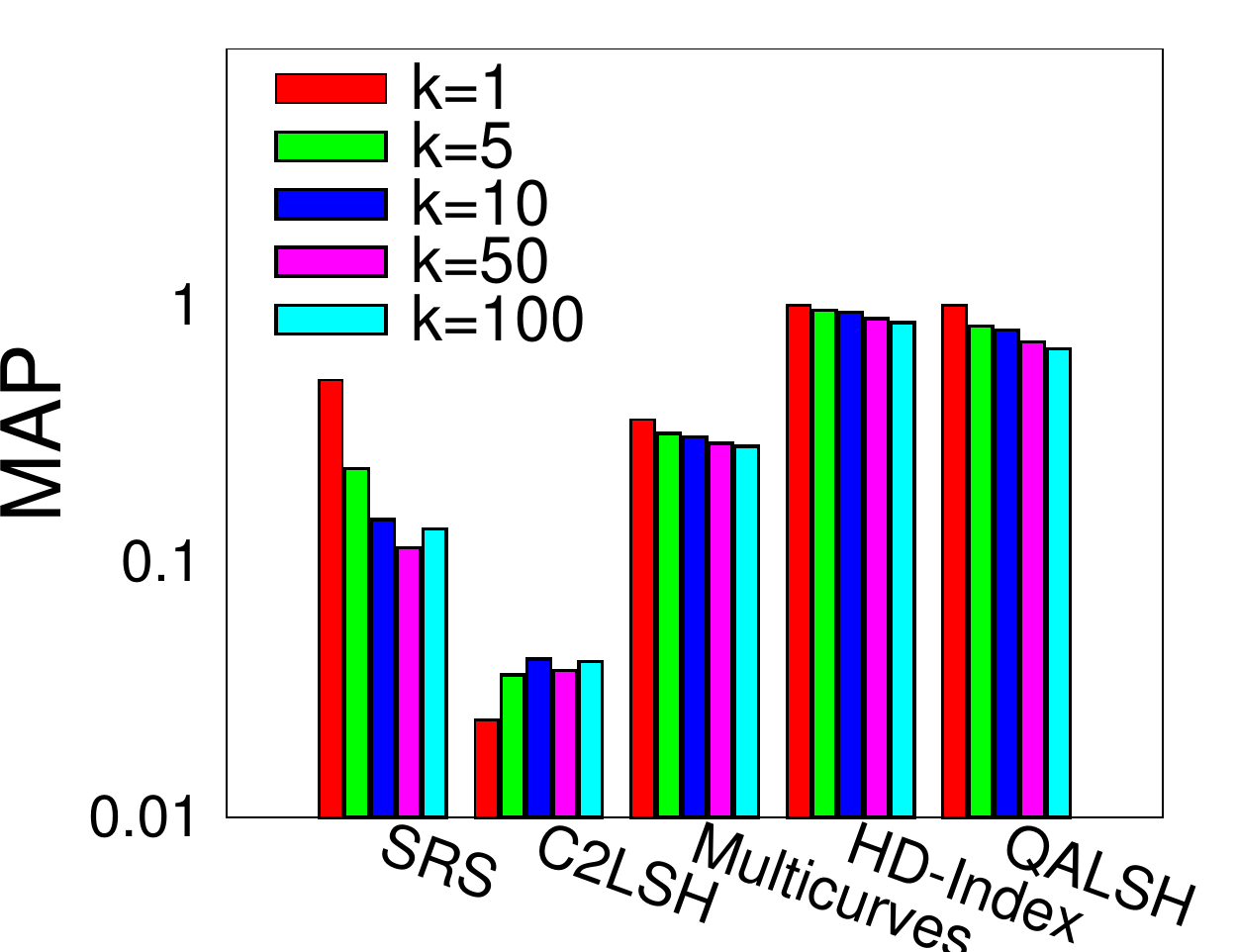}
		}
		\label{fig:map_k2}
	}
	\subfloat[SUN]
	{
		\scalebox{0.19}{
			\includegraphics[width=0.99\linewidth]{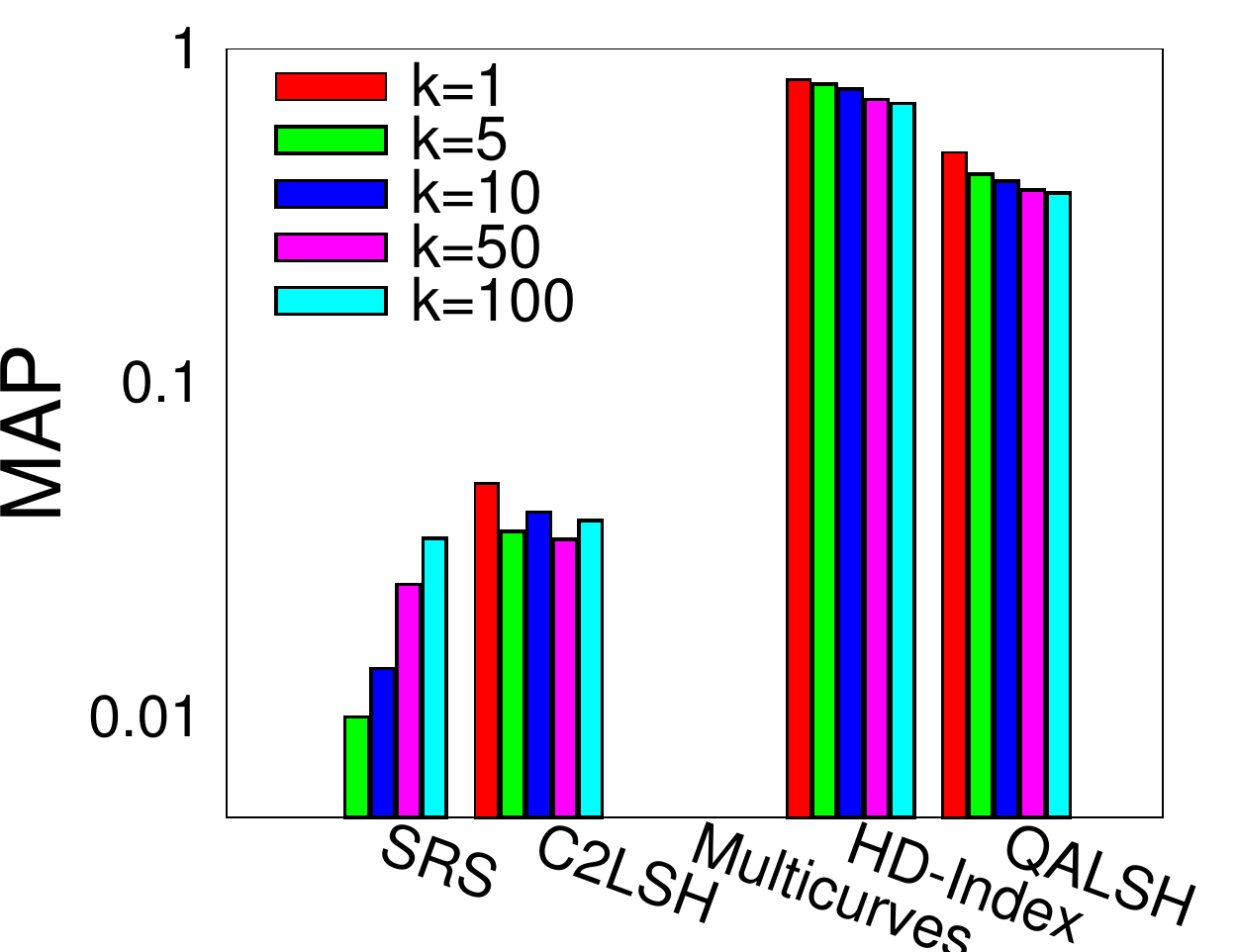}
		}
		\label{fig:map_k3}
	}
	\subfloat[SIFT1M]
	{
		\scalebox{0.19}{
			\includegraphics[width=0.99\linewidth]{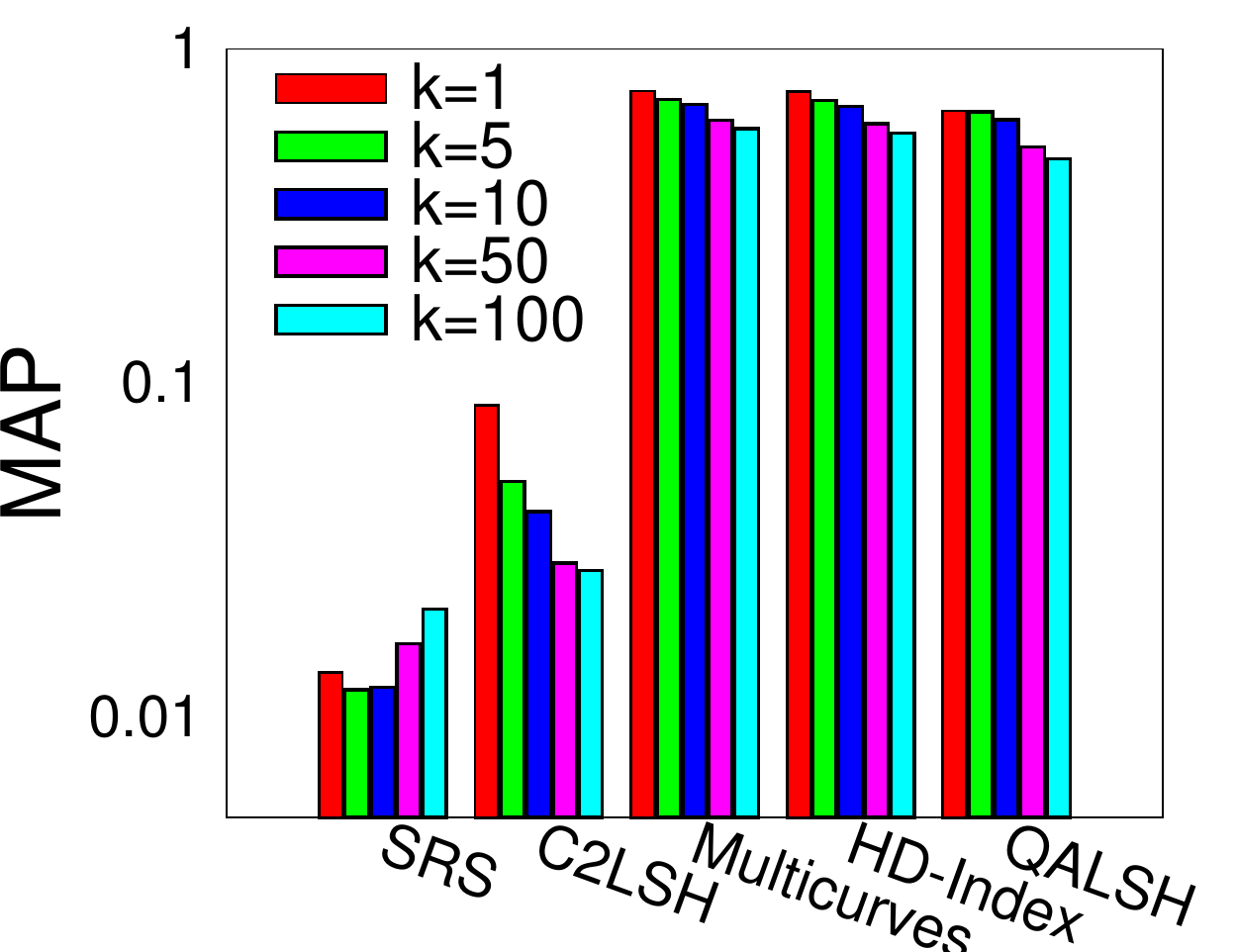}
		}
		\label{fig:map_k4}
	}
	\subfloat[Yorck]
	{
		\scalebox{0.19}{
			\includegraphics[width=0.99\linewidth]{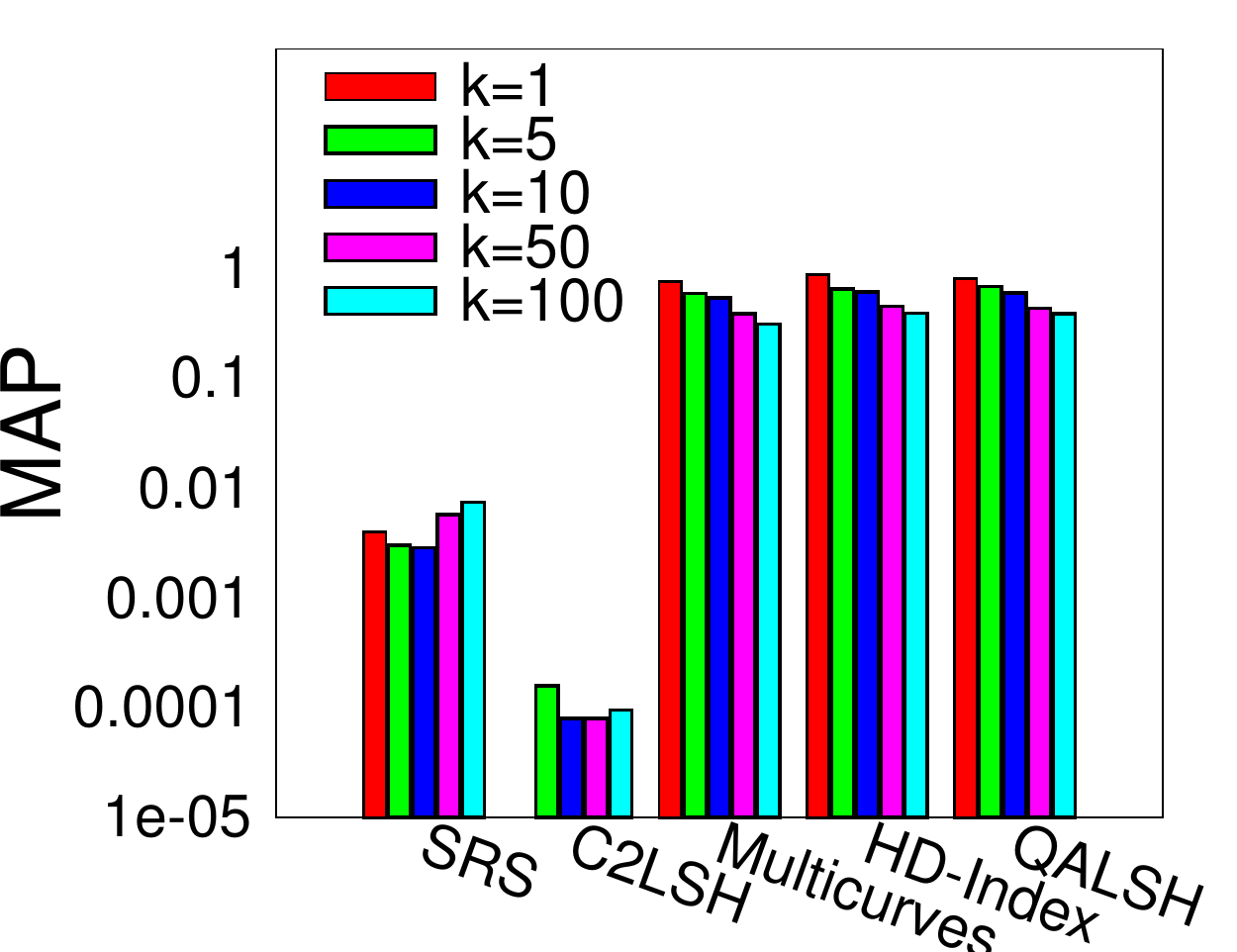}
		}
		\label{fig:map_k5}
	}
	\\
	\vspace{-3.5mm}	
	\subfloat[SIFT10K]
	{
		\scalebox{0.19}{
			\includegraphics[width=0.99\linewidth]{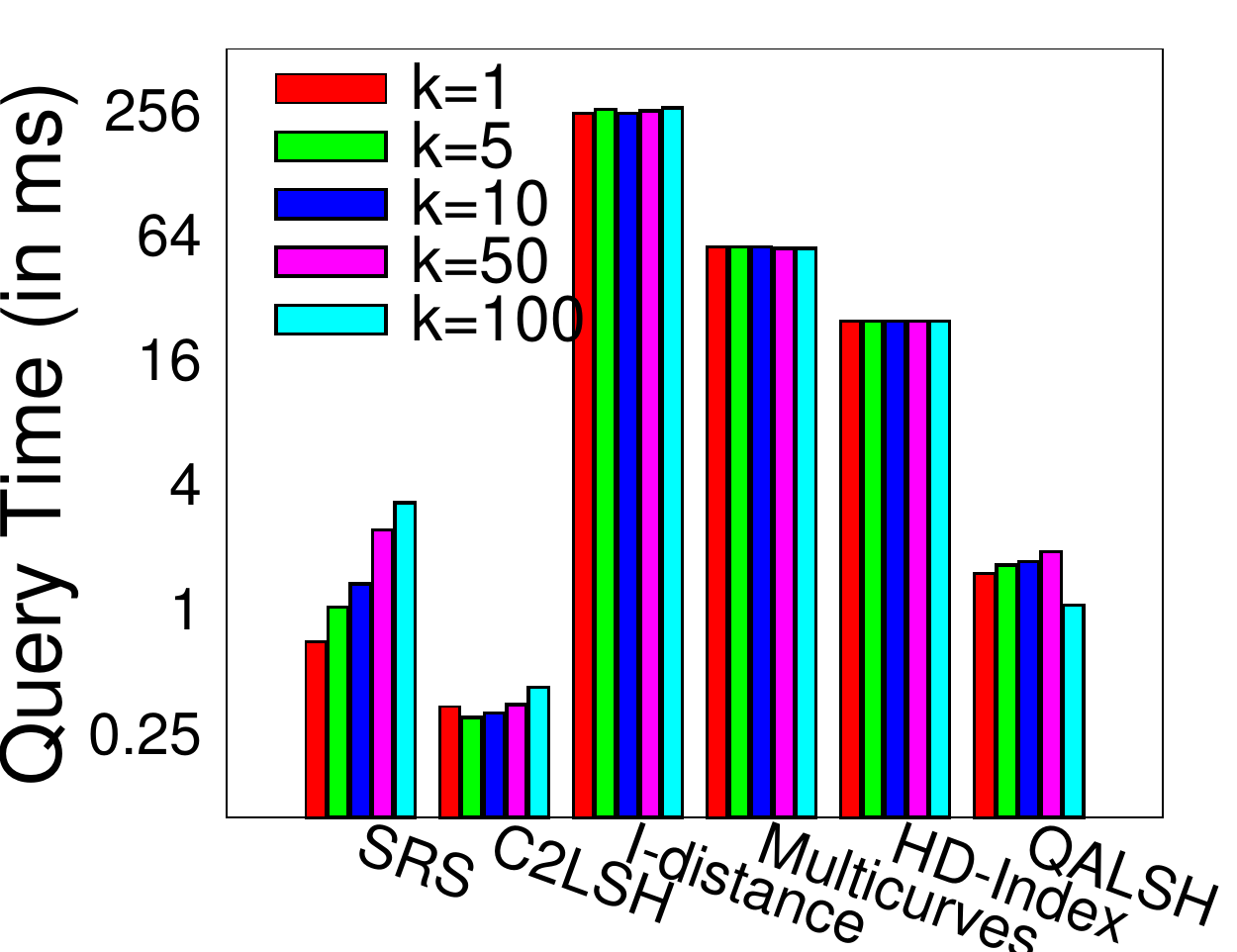}
		}
		\label{fig:time_k1}
	}
	\subfloat[Audio]
	{
		\scalebox{0.19}{
			\includegraphics[width=0.99\linewidth]{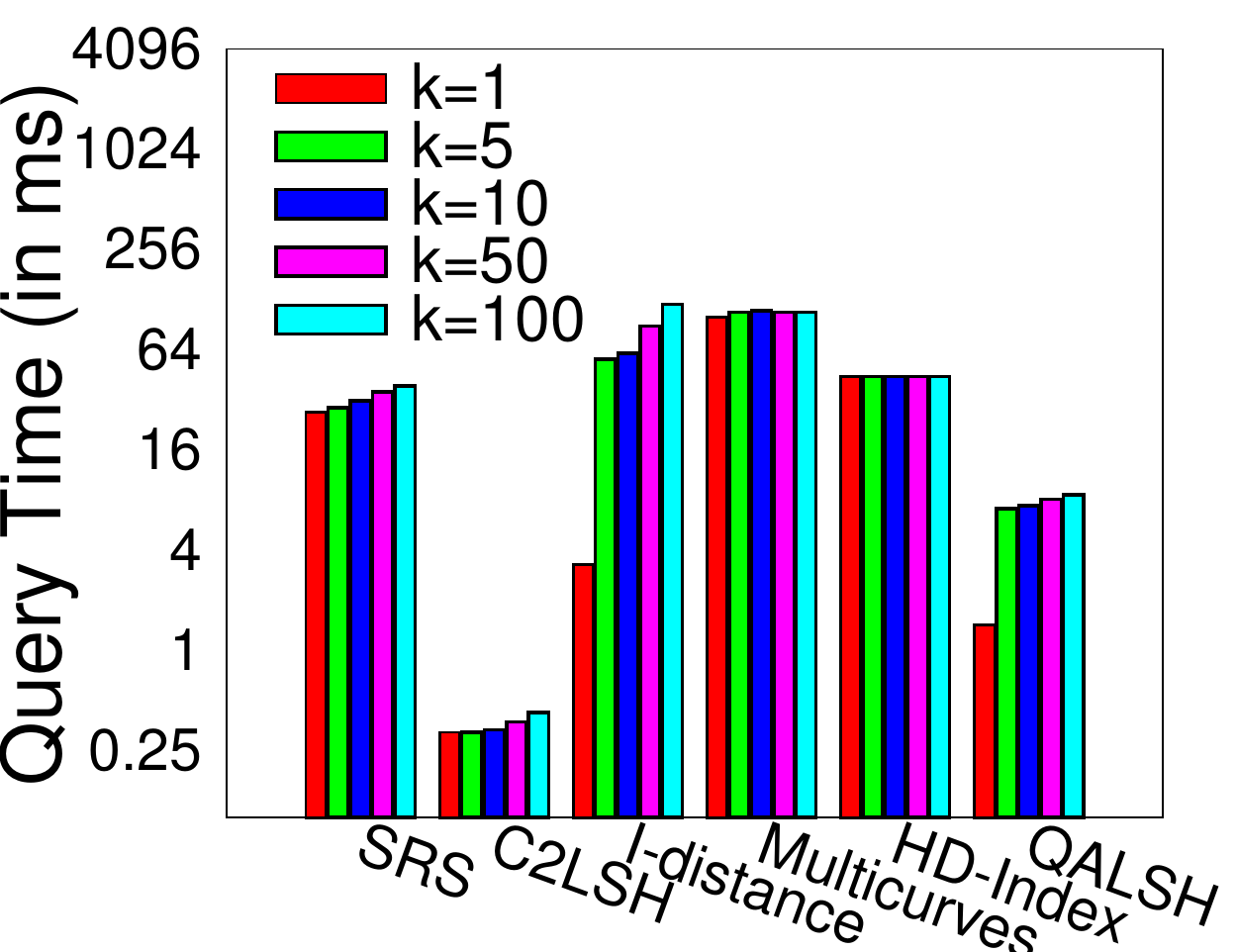}
		}
		\label{fig:time_k2}
	}
	\subfloat[SUN]
	{
		\scalebox{0.19}{
			\includegraphics[width=0.99\linewidth]{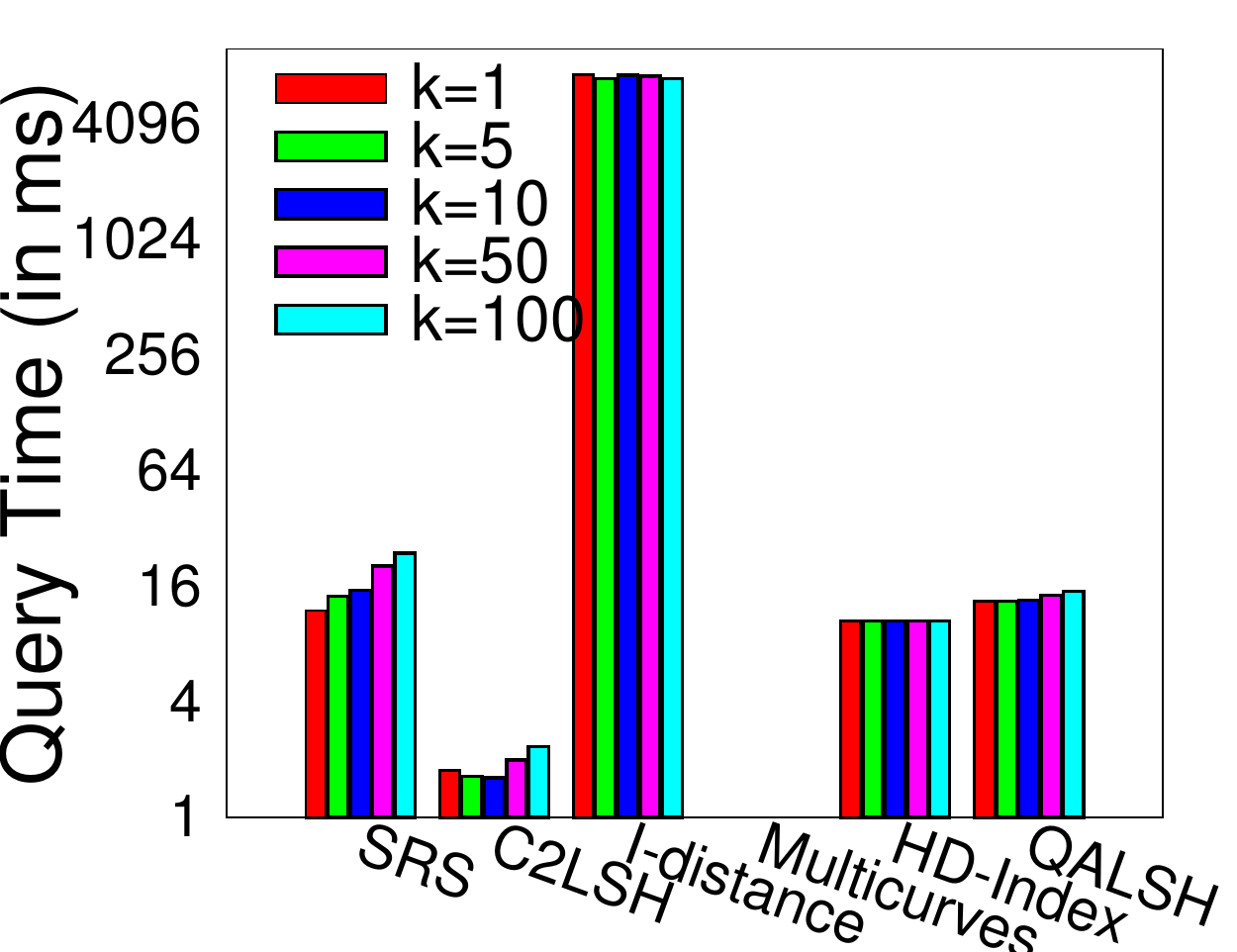}
		}
		\label{fig:time_k3}
	}
	\subfloat[SIFT1M]
	{
		\scalebox{0.19}{
			\includegraphics[width=0.99\linewidth]{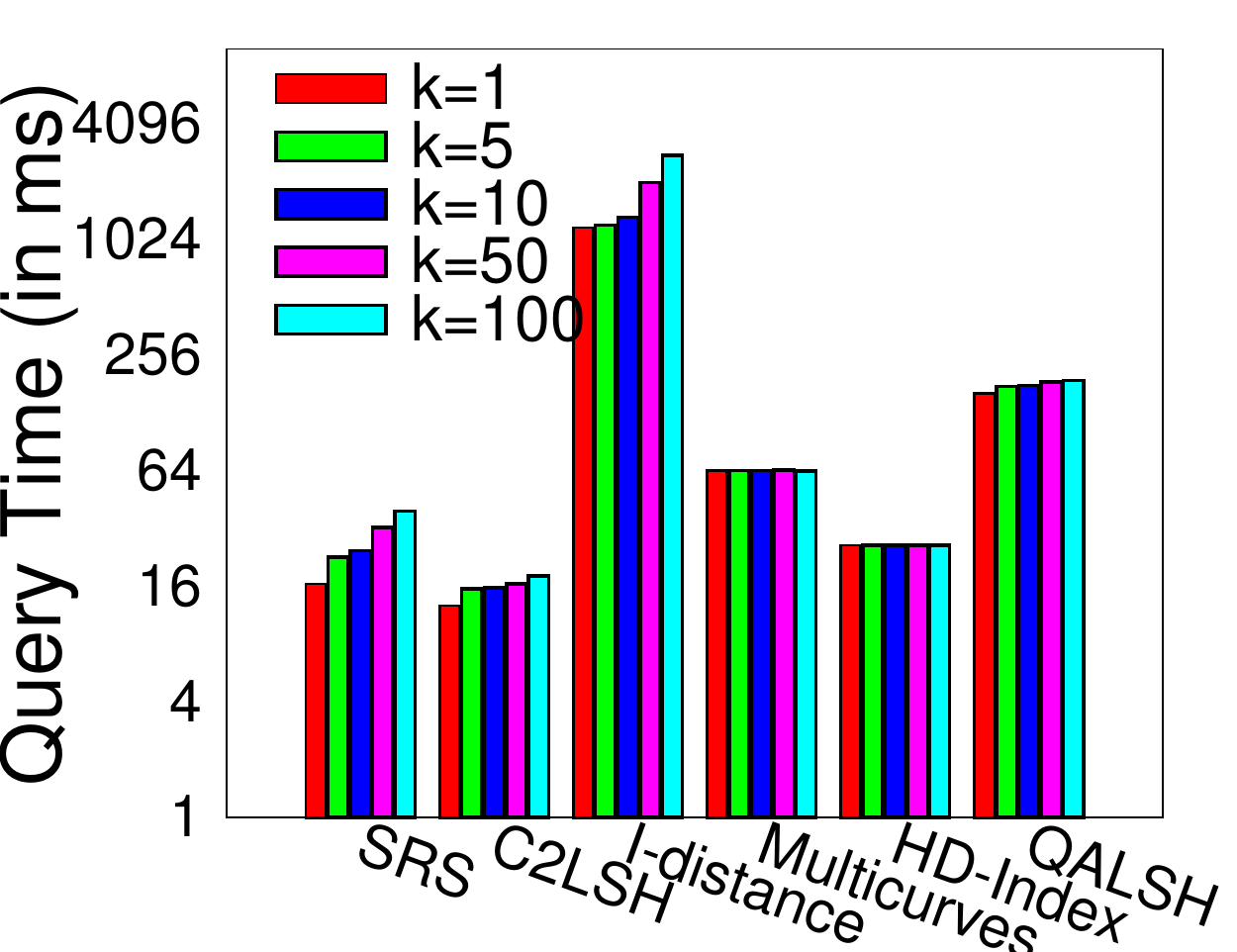}
		}
		\label{fig:time_k4}
	}
	\subfloat[Yorck]
	{
		\scalebox{0.19}{
			\includegraphics[width=0.99\linewidth]{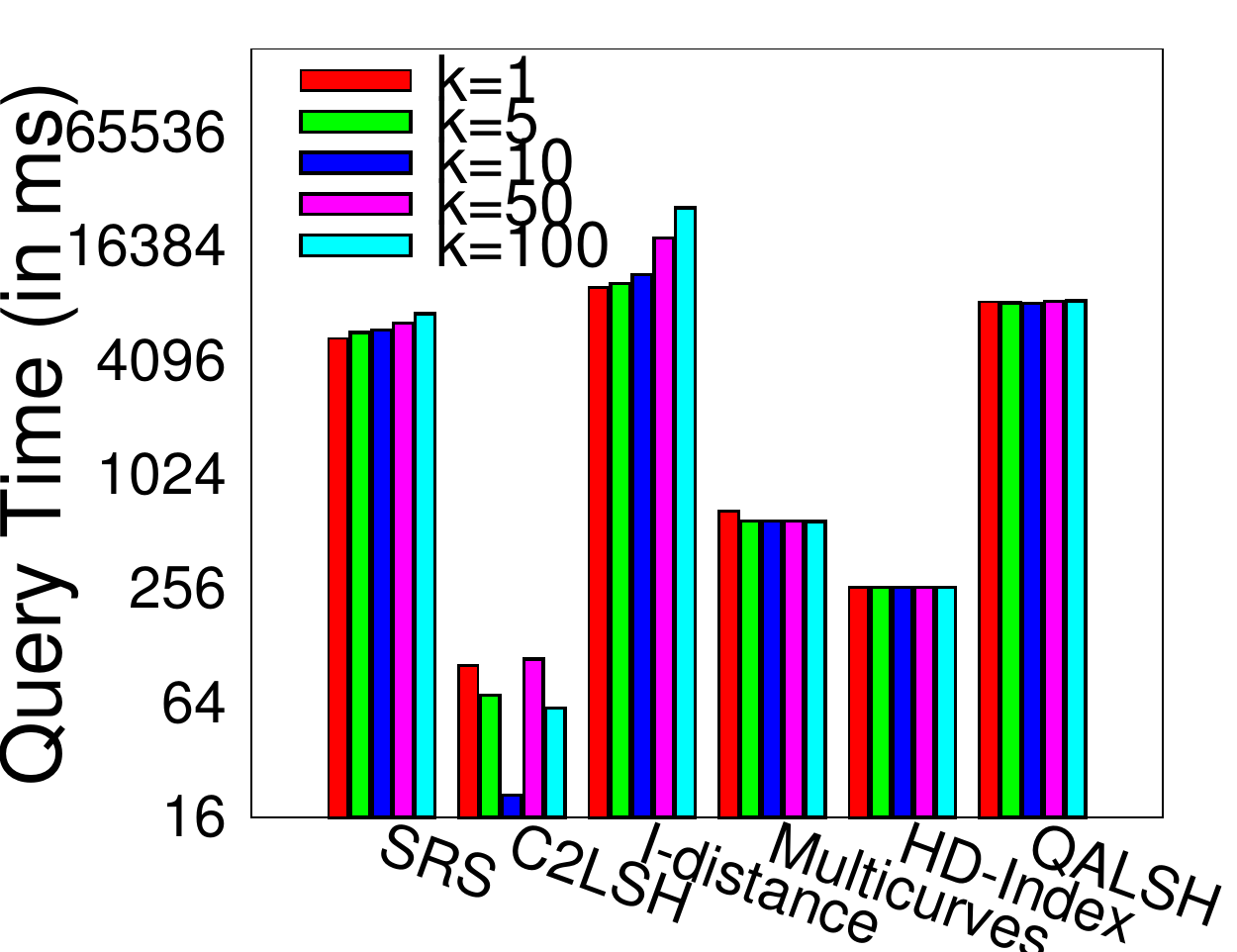}
		\label{fig:time_k5}
		}
	}
	\figcaption{\textbf{Comparison of MAP@$k$ and query time for varying $k$ on
different datasets.}
}
\label{fig:maps_times_varying_k}
\end{figure*}

The results of varying the number of nearest neighbors, $k$, for all the
methods on the tiny, small and medium datasets are shown in
Fig.~\ref{fig:maps_times_varying_k}.  Since iDistance is an exact method, it
always returns correct answers and, hence, possesses a MAP@k of $1$.
Consequently, we do not show its MAP results.  Moreover, the results of the
Multicurves technique is not shown for SUN (Figs.~\ref{fig:map_k3}
and~\ref{fig:time_k3}) due to reasons explained in
Sec.~\ref{subsec:comparison}.

The running time remains almost constant for \name and Multicurves, while for
the other techniques, it increase with $k$.  This is mainly due to the design
of \name and Multicurves as explained in Sec.~\ref{sec:k}.

Moreover, as shown in Figs.~\ref{fig:map_k1} to~\ref{fig:map_k5}, even the
MAP@k for \name and Multicurves almost remain constant with varying $k$.  For
the other techniques, MAP@k is unstable with varying $k$.  This is due to the
fact that when these techniques are allowed to examine more objects, i.e., $k$
is increased, some actual answers are sometimes retrieved.  Since the answers
are returned in a sorted manner, the true answers find their correct places and
MAP increases.

Overall, \name reports the best MAP@$k$ and scales gracefully with dataset size.

\section{Borda Count}
\label{app:borda}

Consider a query image $q$ with $N$ descriptors.  Each descriptor $desc(q)$ is
queried against the database and its kANN result set $r(j,q)$ with $|r(j,q)| =
k$ descriptors, sorted according to their distance to $desc(q)$ are retrieved.
The Borda count method \cite{borda} is then employed to get the $k$ nearest
neighbor images of $q$.

Consider a database image $i$.  A score is given to $i$ based on the depth of
its descriptors in the kANN result sets $r(j,q)$ of $q$.  For each result set
$r(j,q)$, if a descriptor of image $i$ is found at position $l$, then the image
$i$ accumulates a score $k+1-l$.  These scores are then summed over all the $N$
results sets of $q$ to obtain the Borda count of $i$ against $q$:
\begin{align}
	\label{eq:borda}
	\text{BC}(i,q) =
		\sum_{j=1}^{N} \left[ \sum_{l=1}^{k} \left( (k+1-l) . I(i,j,l) \right) \right]
\end{align}
where $I$ is the indicator function that is $1$ only when a descriptor from
image $i$ is found at position $l$ of result $r(j,q)$:
\begin{align*}
    I(i,j,l)= 
		\begin{cases}
		    1 & \text{if descriptor from $i$ occurs at position $l$ of $r(j,q)$} \\
		    0 & \text{otherwise}
		\end{cases}
\end{align*}

The $k$ images with the largest Borda counts are returned in order as the $k$
nearest neighbors of the query image $q$.

\section{Image Search Results}
\label{app:image}

The top-$3$ results for a representative query for the different methods is
shown in Table~\ref{tab:image}.

\begin{table}[t]
	\centering
	{\scriptsize
	\begin{tabular}{|c||c||c|c|c|c|}
		\hline
		{\bf Linear} &
		{\bf \name} &
		{\bf SRS} &
		{\bf C2LSH} &
		{\bf Multicurves} &
		{\bf QALSH} \\
		\hline
		\hline
		\includegraphics[width=\imagewidth]{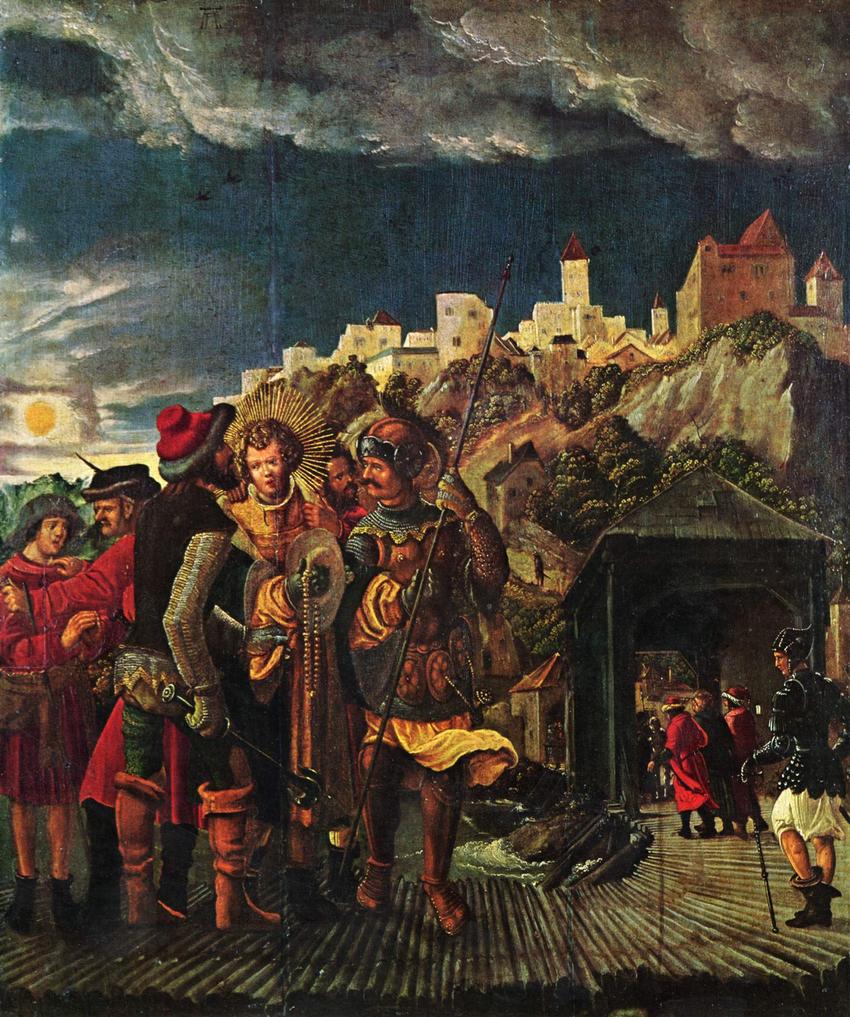} &
		\includegraphics[width=\imagewidth]{scaled/l1} &
		\includegraphics[width=\imagewidth]{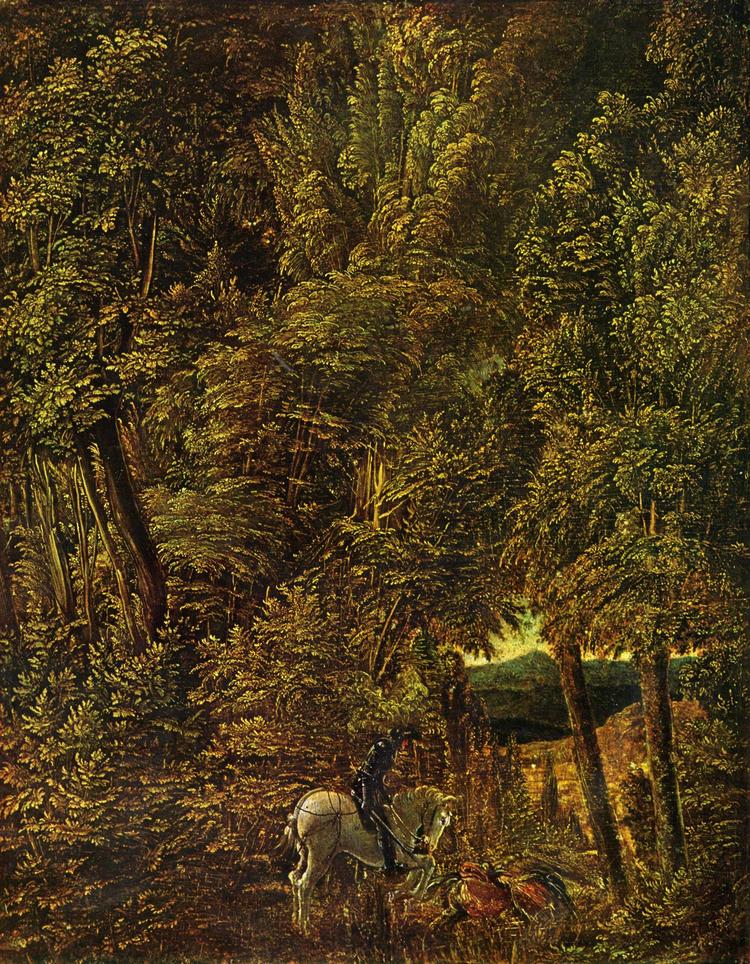} &
		\includegraphics[width=\imagewidth]{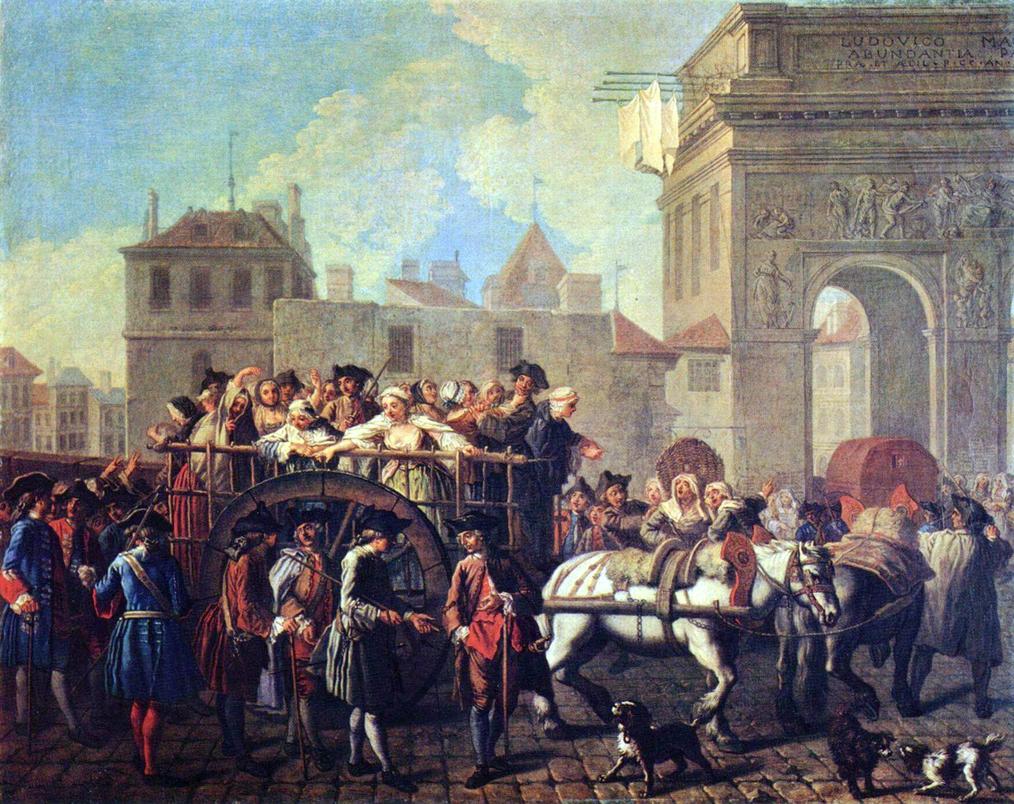} &
		\includegraphics[width=\imagewidth]{scaled/l1} &
		\includegraphics[width=\imagewidth]{scaled/l1} \\
		\hline
		\includegraphics[width=\imagewidth]{scaled/l2} &
		\includegraphics[width=\imagewidth]{scaled/l2} &
		\includegraphics[width=\imagewidth]{scaled/l1} &
		\includegraphics[width=\imagewidth]{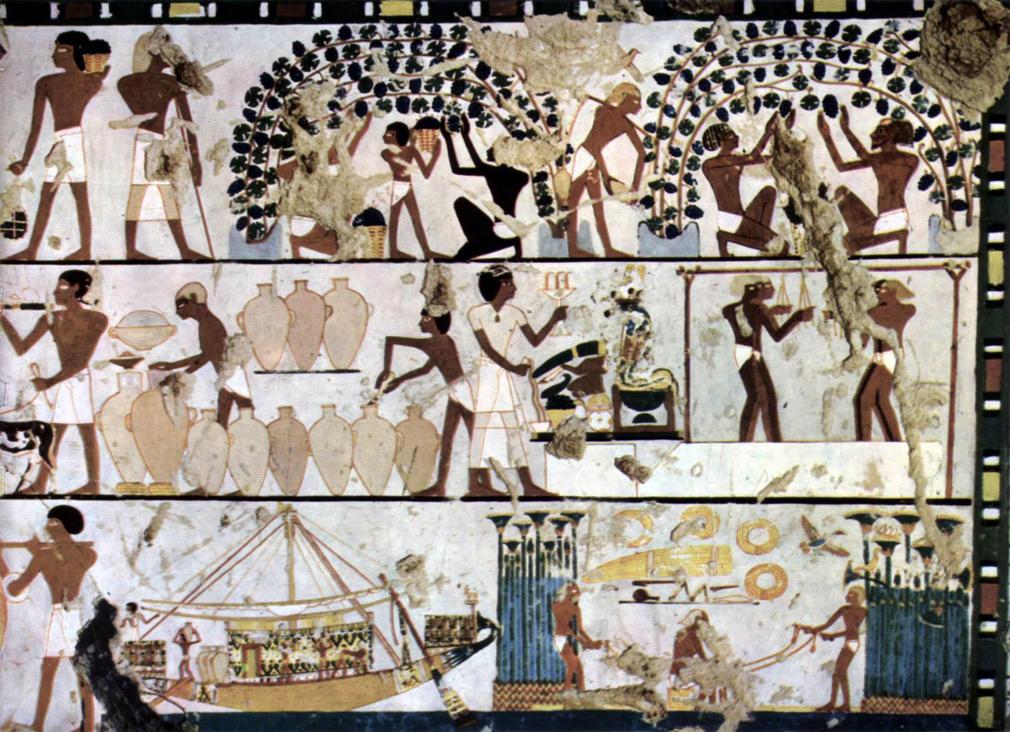} &
		\includegraphics[width=\imagewidth]{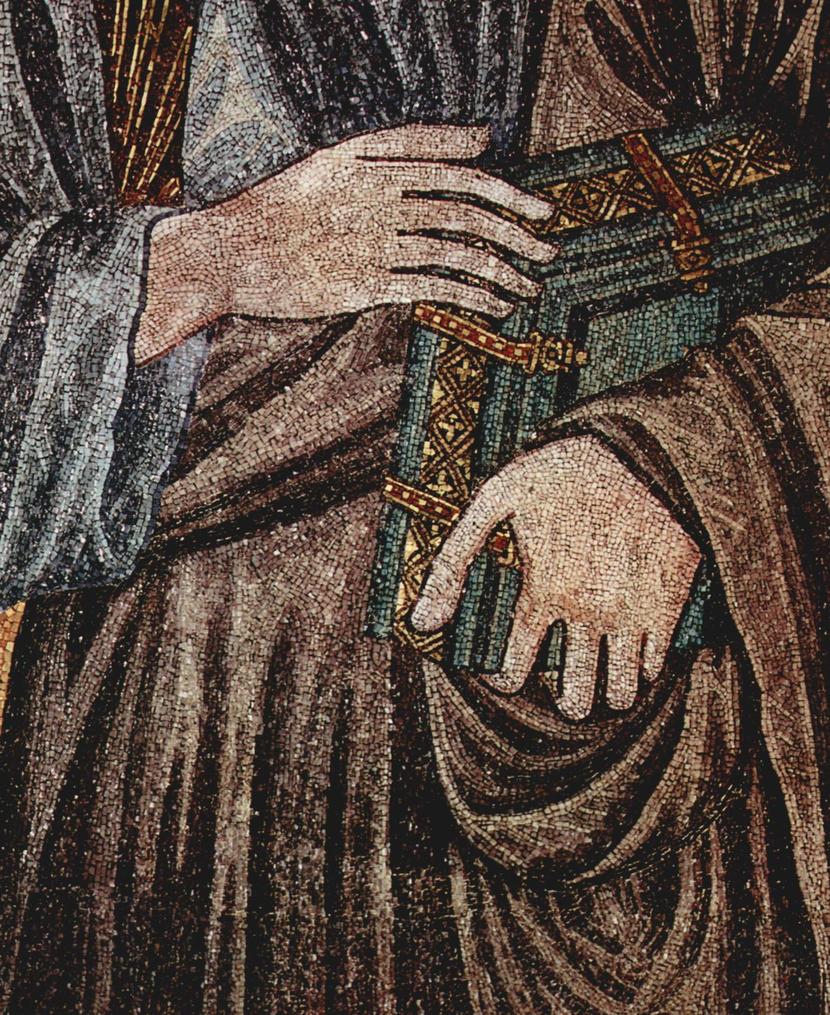} &
		\includegraphics[width=\imagewidth]{scaled/l2} \\
		\hline
		\includegraphics[width=\imagewidth]{scaled/l3} &
		\includegraphics[width=\imagewidth]{scaled/l3} &
		\includegraphics[width=\imagewidth]{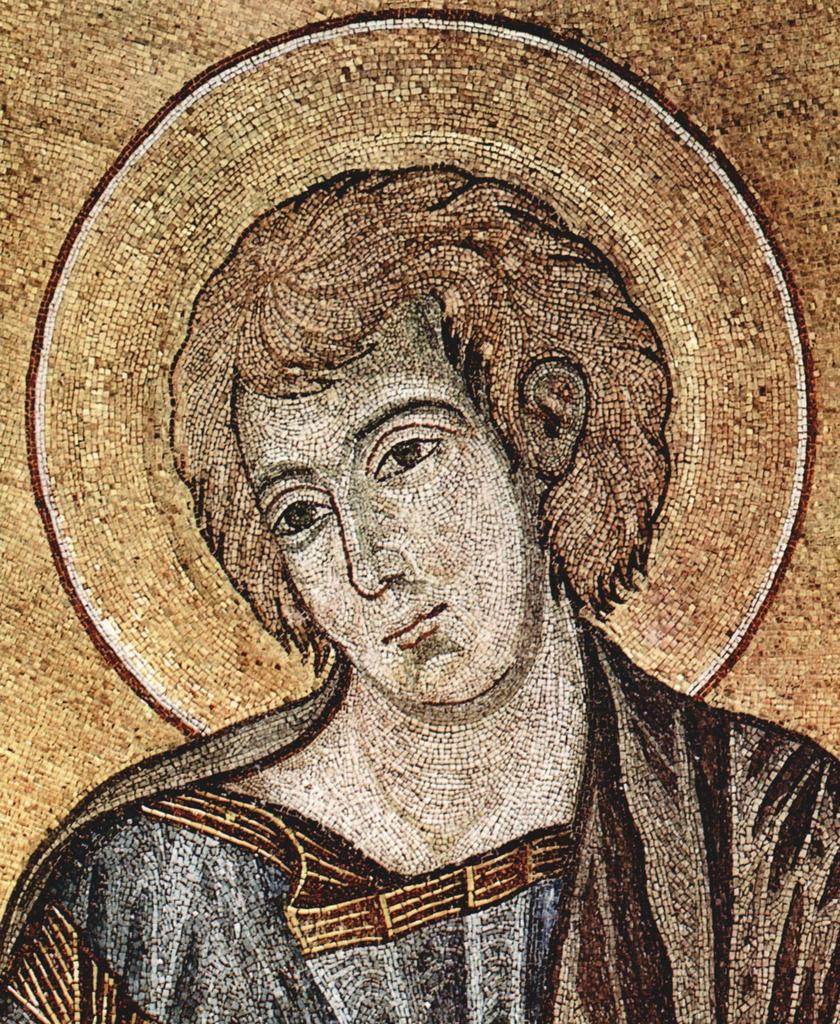} &
		\includegraphics[width=\imagewidth]{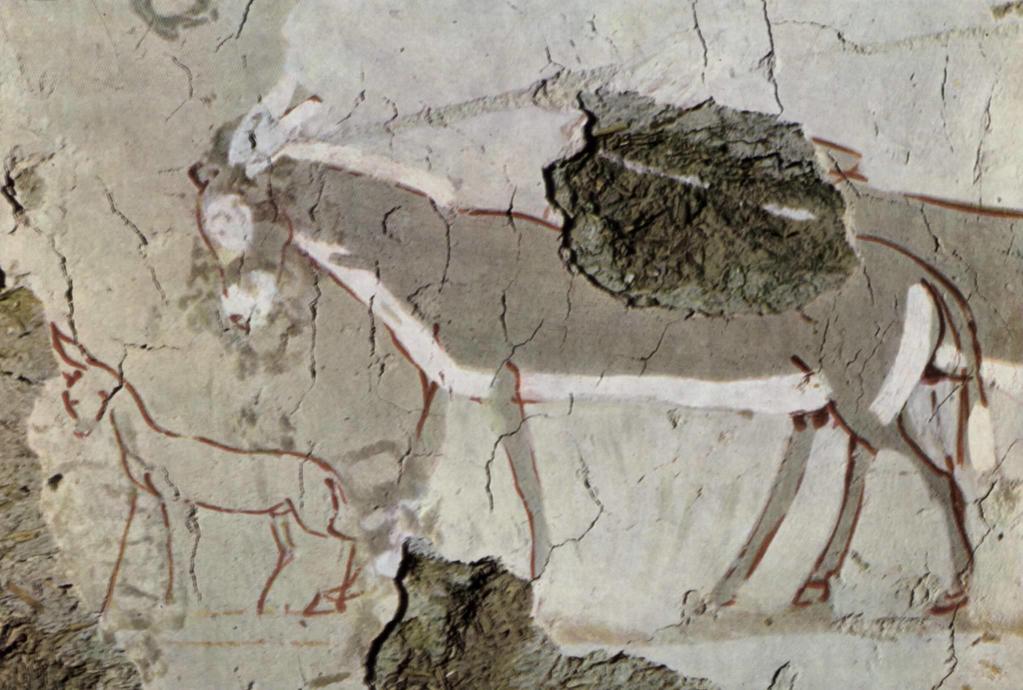} &
		\includegraphics[width=\imagewidth]{scaled/l2} & 
		\includegraphics[width=\imagewidth]{scaled/l3} \\
		\hline
	\end{tabular}
	}
	\tabcaption{\textbf{Image search results.}}
	\label{tab:image}
	\vspace*{1mm}
\end{table}

\end{document}